\documentclass[11pt]{article}

\usepackage{amsfonts,amsmath,amssymb,chet,dsfont,graphicx,mathrsfs,pgfplots,tikz}
\usetikzlibrary{calc}
\pgfplotsset{compat=1.14}
\allowdisplaybreaks

\newcommand{\1}{\mathds{1}}

\newcommand{\Op}[2]{\mathcal{O}_{#1}(\eta_{#2})}

\newcommand{\ee}[3]{(\eta_{#1}\cdot\eta_{#2})^{#3}}
\newcommand{\e}[3]{\eta_{#1}^{#2_{#3}}}
\newcommand{\D}{\mathcal{D}}
\newcommand{\A}{\mathcal{A}}
\newcommand{\cOPE}[4]{{}_{#1}c_{#2#3}^{\phantom{#2#3}#4}}

\newcommand{\tOPE}[6]{{}_{#1}t_{#2#3}^{#5#6#4}}
\newcommand{\cCF}[4]{{}_{#1}c_{#2#3#4}}
\newcommand{\tCF}[6]{{}_{#1}t_{#2#3#4}^{#5#6}}

\newcommand{\aCF}[4]{{}_{#1}\alpha_{#2#3#4}}
\newcommand{\FCF}[6]{{}_{#1}F_{#2#3#4}^{#5#6}}
\newcommand{\vev}[1]{\langle{#1}\rangle}
\newcommand{\Vev}[1]{\left\langle{#1}\right\rangle}

\newcommand{\Diag}[8]{
\begin{tikzpicture}[baseline={([yshift=-.5ex]$(C.center)$)},dot/.style={circle,fill=black,minimum size=#1*1pt,inner sep=0pt},every loop/.style={min distance=#1*19pt}]
\node (C) at (0,0) {};
\node (EF) at (0:#1) {};
\node (EE) at (120:#1) {};
\node (FF) at (240:#1) {};
\draw[fill=black] (C) circle (#1*1pt);
\draw[thick,dashed](0,0)--(EF);
\draw[thick](0,0)--(EE);
\draw[thick,dotted](0,0)--(FF);
\ifnum #2>0 \foreach \lEF in {1,...,#2}{\node[dot](\lEF) at (0:{#1*\lEF/(#2+#5+#8+1)}) {};\draw[thick,dashed](\lEF) to [out=120-20,in=60+20,loop] ();}\fi
\ifnum #5>0 \foreach \ElEF in {1,...,#5}{\node[dot](#2+\ElEF) at (0:{#1*(#2+\ElEF)/(#2+#5+#8+1)}) {};\draw[thick](#2+\ElEF)--+(90:#1*0.5);}\fi
\ifnum #8>0 \foreach \FlEF in {1,...,#8}{\node[dot](#2+#5+\FlEF) at (0:{#1*(#2+#5+\FlEF)/(#2+#5+#8+1)}) {};\draw[thick,dotted](#2+#5+\FlEF)--+(90:#1*0.5);}\fi
\ifnum #3>0 \foreach \lEE in {1,...,#3}{\node[dot](#2+#5+#8+\lEE) at (120:{#1*\lEE/(#3+#4+1)}) {};\draw[thick](#2+#5+#8+\lEE) to [out=240-20,in=180+20,loop] ();}\fi
\ifnum #4>0 \foreach \ElEE in {1,...,#4}{\node[dot](#2+#5+#8+#3+\ElEE) at (120:{#1*(#3+\ElEE)/(#3+#4+1)}) {};\draw[thick](#2+#5+#8+#3+\ElEE)--+(210:#1*0.5);}\fi
\ifnum #6>0 \foreach \lFF in {1,...,#6}{\node[dot](#2+#5+#8+#3+#4+\lFF) at (240:{#1*\lFF/(#6+#7+1)}) {};\draw[thick,dotted](#2+#5+#8+#3+#4+\lFF) to [out=0-20,in=300+20,loop] ();}\fi
\ifnum #7>0 \foreach \FlFF in {1,...,#7}{\node[dot](#2+#5+#8+#3+#4+#6+\FlFF) at (240:{#1*(#6+\FlFF)/(#6+#7+1)}) {};\draw[thick,dotted](#2+#5+#8+#3+#4+#6+\FlFF)--+(330:#1*0.5);}\fi
\end{tikzpicture}
}

\title{Efficient Rules for All Conformal Blocks}

\author{Jean-Fran\c{c}ois Fortin$^{\ast,}$\email{jean-francois.fortin@phy.ulaval.ca}, Wen-Jie Ma$^{\ast,}$\email{wenjie.ma.1@ulaval.ca}, 
Valentina Prilepina$^{\ast,}$\email{valentina.prilepina.1@ulaval.ca}, and Witold Skiba$^{\dagger,}$\email{witold.skiba@yale.edu}}

\affiliation{
$^\ast$D\'epartement de Physique, de G\'enie Physique et d'Optique,\\Universit\'e Laval, Qu\'ebec, QC G1V 0A6, Canada\\
$^\dagger$Department of Physics, Yale University, New Haven, CT 06520, USA
}

\abstract{We formulate a set of general rules for computing $d$-dimensional four-point global conformal blocks of operators in arbitrary Lorentz representations in the context of the embedding space operator product expansion formalism \cite{Fortin:2019dnq}.  With these rules, the procedure for determining any conformal block of interest is reduced to (1) identifying the relevant projection operators and tensor structures and (2) applying the conformal rules to obtain the blocks.  To facilitate the bookkeeping of contributing terms, we introduce a convenient diagrammatic notation.  We present several concrete examples to illustrate the general procedure as well as to demonstrate and test the explicit application of the rules.  In particular, we consider four-point functions involving scalars $S$ and some specific irreducible representations $R$, namely $\vev{SSSS}$, $\vev{SSSR}$, $\vev{SRSR}$ and $\vev{SSRR}$ (where, when allowed, $R$ is a vector or a fermion), and determine the corresponding blocks for all possible exchanged representations.}

\date{February 2020} 

\begin{document}

\maketitle

\toc


\section{Introduction}\label{SecIntro}

Conformal field theories (CFTs) are special quantum field theories that enjoy an enhanced symmetry, namely invariance under the conformal group $SO(2,d)$.  They describe the intriguing universal physics of critical scale invariant fixed points and also lie at the core of our understanding of the space of all quantum field theories (QFTs).  CFTs represent fixed points of renormalization group flows and describe second order phase transitions of statistical physics systems.  Strikingly, they shed light on the structure of the space of all QFTs, furnish concrete implementations of quantum gravity theories via the AdS/CFT correspondence and holography, and illuminate problems in black hole physics.  It is evident that the urge for a profound understanding of the landscape of CFTs cannot be overemphasized.

In recent years, this field has experienced a veritable explosion of results, largely owing to the success of the conformal bootstrap, a program which seeks to systematically apply symmetries and consistency conditions to carve out the allowed space of CFTs.  The vast bootstrap literature has been summarized in several comprehensive reviews and lectures (see for example \cite{Rychkov:2016iqz,Simmons-Duffin:2016gjk,Poland:2018epd,Chester:2019wfx} and references therein).  This profusion of progress spans a wide range of high-precision numerical results as well as many remarkable analytic advances, in addition to contributions involving global symmetries and higher-spin fields.

An implementation of the bootstrap program calls for a determination of the complete set of so-called conformal blocks, which are the building blocks of four-point correlation functions that capture contributions of particular exchanged representations in the operator product expansion (OPE).  To date, only a handful of these objects have been worked out in $d>2$, due to the challenging nature of the computations involved \cite{Dolan:2000ut,Dolan:2003hv} (see also \cite{Ferrara:1973vz,Ferrara:1974nf,Dobrev:1977qv,Exton_1995} for earlier work).  With renewed interest in the bootstrap, a host of novel approaches and revisions of old methods have been proposed, including further developments of the shadow and the weight shifting operator formalisms, adding to the ever growing variety of methods \cite{Costa:2011mg,Dolan:2011dv,Costa:2011dw,SimmonsDuffin:2012uy,Costa:2014rya,Echeverri:2015rwa,Iliesiu:2015qra,Hijano:2015zsa,Penedones:2015aga,Iliesiu:2015akf,Alkalaev:2015fbw,Echeverri:2016dun,Isachenkov:2016gim,Costa:2016hju,Costa:2016xah,Chen:2016bxc,Nishida:2016vds,Cordova:2016emh,Schomerus:2016epl,Kravchuk:2016qvl,Gliozzi:2017hni,Castro:2017hpx,Dyer:2017zef,Sleight:2017fpc,Chen:2017yia,Pasterski:2017kqt,Cardoso:2017qmj,Karateev:2017jgd,Kravchuk:2017dzd,Schomerus:2017eny,Isachenkov:2017qgn,Faller:2017hyt,Chen:2017xdz,Sleight:2018epi,Costa:2018mcg,Kobayashi:2018okw,Bhatta:2018gjb,Lauria:2018klo,Liu:2018jhs,Gromov:2018hut,Rosenhaus:2018zqn,Zhou:2018sfz,Kazakov:2018gcy,Li:2019dix,Goncalves:2019znr,Jepsen:2019svc,Rejon-Barrera:2015bpa}.

An alternative technique was recently suggested in \cite{Fortin:2019fvx,Fortin:2019dnq}.  This method hinges on exploiting the OPE directly in the embedding space, where the conformal group acts linearly \cite{Dirac:1936fq,Mack:1969rr,Weinberg:2010fx,Weinberg:2012mz}.  The embedding space OPE framework was originally proposed in \cite{Ferrara:1971vh,Ferrara:1971zy,Ferrara:1972cq,Ferrara:1973eg,Dobrev:1975ru,Mack:1976pa} and later expanded further in \cite{Fortin:2016lmf,Fortin:2016dlj,Comeau:2019xco}.  Subsequent work has established this framework on a firm footing, starting with \cite{Fortin:2019fvx,Fortin:2019dnq}, where the formalism was fully expounded for general $M$-point correlation functions, and later followed up by \cite{Fortin:2019xyr,Fortin:2019pep,Fortin:2019gck}, where it was tested and exemplified for two-, three-, and four-point functions, respectively.\footnote{Scalar $M$-point correlation functions in the comb channel were also obtained with this method in \cite{Fortin:2019zkm}.  See also \cite{Parikh:2019dvm} for an independent computation using AdS/CFT.}  In this formalism, operators in arbitrary Lorentz representations are uplifted to the embedding space in a uniform fashion by building general representations solely out of products of spinor representations.  A key advantage of this approach is that arbitrary operators, whether they are fermions or bosons, are treated democratically, so that from the perspective of the Dynkin indices, all representations effectively look the same.  The method is designed to work at the fundamental level of the OPE and therefore applies to arbitrary correlation functions.  A crucial aspect is the appropriate definition of an optimal embedding space OPE differential operator, which is symmetric and traceless in the embedding space indices by construction.  This feature renders the operator exceptionally useful, due to a variety of nice properties and identities, which enable one to readily generalize the scalar case to the tensorial ones.  The action of this operator on any quantity which may potentially crop up in an arbitrary $M$-point function has been explicitly worked out in \cite{Fortin:2019fvx,Fortin:2019dnq}.  This computation subsequently led to the tensorial generalization of the scalar Exton-$G$ function for $M$-point correlators.  The reader interested in the details of the general method is referred to \cite{Fortin:2019fvx,Fortin:2019dnq}.

It turns out that further refinement of this approach enables one to compute the conformal blocks for arbitrary four-point functions quite efficiently.  The method yields the infinite towers of blocks in a compact form.  The blocks are expressed as specific linear combinations of Gegenbauer polynomials in a special variable $X$, with a unique substitution rule ascribed to each polynomial piece.  Once each relevant rule is applied to its associated Gegenbauer term, we directly generate the complete conformal block in terms of a four-point tensorial generalization of the Exton $G$-function.  As detailed in \cite{Fortin:2019gck}, in the context of this formalism, the procedure for determining a given block comes down to (1) writing down the relevant group theoretic quantities, namely the projection operators and tensor structures (which effectively serve as intertwiners among the respective external and exchanged representations), and (2) identifying the specific linear combination of Gegenbauer polynomials along with the corresponding substitution rules for each piece.

While this approach is complete and clearly formulated as it stands, it is rather cumbersome to apply in practice for infinite towers of exchanged quasi-primary operators in irreducible representations $\boldsymbol{N}_m+\ell\boldsymbol{e}_1$.  In the analysis \cite{Fortin:2019gck}, it was apparent that various parts involved in the derivation of the substitution rules recurred, suggesting that the procedure could be made completely systematic for any $\ell$.  Further, while the determination of the appropriate linear combination of Gegenbauer polynomials for a given case was straightforward, a systematic approach for the identification of these combinations was lacking.  Moreover, the nature of the substitution rules themselves seemed somewhat mysterious.  A careful inspection of the form of the various rules for different blocks, \textit{e.g.}\ $\vev{SVSV}$ and $\vev{SSVV}$, revealed that combinations of these rules were related to each other and that in some cases, one could map certain rules to others via a set of integer shifts, implying a deep relationship among the different rules.  However, the origin of such shifts remained unclear.

In this work, we seek to cast our prescription for obtaining the blocks into a systematic form.  In particular, we wish to understand the underlying structure of the substitution rules as well as how to methodically generate the relevant linear combination of Gegenbauer polynomials for each case of interest, \textit{i.e.}\ for any $\ell$ for some exchanged representation $\boldsymbol{N}_m+\ell\boldsymbol{e}_1$.  Our general philosophy is to formulate the procedure in terms of parameters which depend entirely on the projection operators and the tensor structures, so that effectively, all that needs to be done for a given case is to determine these objects.  The remainder would be subsequently handled by a set of conformal rules, designed to be easy to apply for arbitrary $\ell$.  These conformal rules (which are reminiscent of the Feynman rules but are non-perturbative as they lead to the exact blocks) would directly use the information about the structure of the projection operators and the tensor structures as well as the identity of the external operators to generate the complete tower of blocks for a given case.  With this machinery in place, the calculation of the blocks would become essentially effortless.  It is the purpose of this work to formulate these conformal rules, check their validity, and demonstrate how to apply them in practice.

This paper is organized as follows: Section \ref{SecReview} provides a brief review of the method for $(M\leq4)$-point functions and summarizes the general results.  This section also includes a discussion of various bases for the blocks and proposes a specific preferred basis, the mixed basis, in which the expressions for the blocks assume the simplest possible form.  One can transform between the various bases using special rotation matrices, which can be constructed explicitly.  In Section \ref{SecInput}, the necessary ingredients of the formalism are described.  In particular, here we discuss the input data required for the determination of the conformal blocks, namely the tensor structures and projection operators.  We detail how to determine these objects.  This section also includes a decomposition of the projection operators in terms of shifted symmetric traceless projectors as well as a general index separation algorithm necessary for the determination of the complete set of independent terms, along with their substitution rules, appearing in the conformal blocks.  We also introduce a convenient diagrammatic notation (somewhat reminiscent of Feynman diagrams) which serves to encode the index separation in a compact form.

The following two sections, Sections \ref{SecRM} and \ref{SecCB}, expound the general algorithm for determining the rotation matrices and the conformal blocks, respectively.  The rotation matrices entail a careful analysis of the three-point functions, while the conformal blocks necessitate an expansion in Gegenbauer polynomials in a special variable $X$, coupled with associated substitution rules.  In these two sections, we consider infinite towers of exchanged quasi-primary operators in some irreducible representations $\boldsymbol{N}_m+\ell\boldsymbol{e}_1$.  We provide proofs on how to handle the universal $\ell$-dependent parts of these exchanged representations, which leads to some simple $\ell$-independent rules.  The reader not interested in the proofs of Sections \ref{SecRM} and \ref{SecCB} can skip directly to Section \ref{SecSum}, where we present a summary of the results along with a dictionary of the notation.  In Section \ref{SecEx}, we illustrate how to apply these rules in practice.  We analyze several interesting examples.  We begin by revisiting all the conformal blocks obtained in \cite{Fortin:2019gck}, namely $\vev{SSSS}$, $\vev{SSS\boldsymbol{e}_2}$, $\vev{SVSV}$ and $\vev{SSVV}$.  We demonstrate that the application of the conformal rules allows us to effortlessly rederive these results.  The diagrammatic notation is also illustrated through these examples.  We next proceed to treat the remaining cases of the type $\vev{SSSR}$, $\vev{SRSR}$ and $\vev{SSRR}$ when $R$ is a vector or a fermion.

Finally, Section \ref{SecConc} concludes with a summary of the results, a preview of future work, and questions of interest raised by this analysis, while Appendix \ref{SecProj} provides details on the projection operators needed to compute the various conformal blocks presented in this paper.


\section{Review of the Embedding Space OPE Method}\label{SecReview}

This section presents a quick review of the embedding space OPE method and its implications for correlation functions up to four points.  The reader interested in the details is refereed to \cite{Fortin:2019fvx,Fortin:2019dnq,Fortin:2019xyr,Fortin:2019pep,Fortin:2019gck}.


\subsection{Embedding Space OPE}

The form of the embedding space OPE that is most convenient for the determination of $M$-point correlation functions was found in \cite{Fortin:2019fvx,Fortin:2019dnq}.  It is given by
\eqn{
\begin{gathered}
\Op{i}{1}\Op{j}{2}=(\mathcal{T}_{12}^{\boldsymbol{N}_i}\Gamma)(\mathcal{T}_{21}^{\boldsymbol{N}_j}\Gamma)\cdot\sum_k\sum_{a=1}^{N_{ijk}}\frac{\cOPE{a}{i}{j}{k}\tOPE{a}{i}{j}{k}{1}{2}}{\ee{1}{2}{p_{ijk}}}\cdot\D_{12}^{(d,h_{ijk}-n_a/2,n_a)}(\mathcal{T}_{12\boldsymbol{N}_k}\Gamma)*\Op{k}{2},\\
p_{ijk}=\frac{1}{2}(\tau_i+\tau_j-\tau_k),\qquad h_{ijk}=-\frac{1}{2}(\chi_i-\chi_j+\chi_k),\\
\tau_\mathcal{O}=\Delta_\mathcal{O}-S_\mathcal{O},\qquad\chi_\mathcal{O}=\Delta_\mathcal{O}-\xi_\mathcal{O},\qquad\xi_\mathcal{O}=S_\mathcal{O}-\lfloor S_\mathcal{O}\rfloor,
\end{gathered}
}[EqOPE]
where $\Delta_\mathcal{O}$ and $S_\mathcal{O}$ are the conformal dimension and spin of the quasi-primary operator $\mathcal{O}$, respectively, while $\cOPE{a}{i}{j}{k}$ are the OPE coefficients.  The remaining quantities appearing in the OPE \eqref{EqOPE} are described below.

The first quantity of interest here is the OPE differential operator $\D_{12}^{(d,h_{ijk}-n_a/2,n_a)}$.  It is given by
\eqn{
\begin{gathered}
\D_{ij}^{(d,h,n)A_1\cdots A_n}=\frac{1}{\ee{1}{2}{\frac{n}{2}}}\D_{ij}^{2(h+n)}\eta_j^{A_1}\cdots\eta_j^{A_n},\\
\D_{ij}^2=\ee{i}{j}{}\partial_j^2-(d+2\eta_j\cdot\partial_j)\eta_i\cdot\partial_j.
\end{gathered}
}
The explicit action of this operator on arbitrary functions of the embedding space coordinates and cross-ratios was found in \cite{Fortin:2019fvx,Fortin:2019dnq}.  A useful consequence is that in the context of the computation of conformal blocks, the action of this operator can be taken care of by simple substitution rules on specific quantities.

The remaining objects of interest are fundamental group theoretic quantities, including the projection operators, the half-projection operators, and the tensor structures.  In the OPE \eqref{EqOPE}, we require their embedding space analogs, and these are readily obtained from the corresponding position space quantities.  We therefore first determine the position space objects and then translate them into their embedding space counterparts via some simple substitutions detailed below.

We begin with a brief discussion of the projection and half-projection operators in position space.  The projection operators are central ingredients in the construction of $M$-point correlation functions in the context of the present framework.  The position space projectors are defined as operators that satisfy the following properties:
\begin{enumerate}
\item the projection property
\eqn{\hat{\mathcal{P}}^{\boldsymbol{N}}\cdot\hat{\mathcal{P}}^{\boldsymbol{N}'}=\delta_{\boldsymbol{N}'\boldsymbol{N}}\hat{\mathcal{P}}^{\boldsymbol{N}},}
\item the completeness relation 
\eqn{\sum_{\boldsymbol{N}|n_v\text{ fixed}}\hat{\mathcal{P}}^{\boldsymbol{N}}=\1-\text{traces},}
\item the tracelessness condition
\eqn{g\cdot\hat{\mathcal{P}}^{\boldsymbol{N}}=\gamma\cdot\hat{\mathcal{P}}^{\boldsymbol{N}}=\hat{\mathcal{P}}^{\boldsymbol{N}}\cdot g=\hat{\mathcal{P}}^{\boldsymbol{N}}\cdot\gamma=0,}
\end{enumerate}
where $n_v$ is the total number of vector indices.  They are labeled by the irreducible representations $\boldsymbol{N}$ of $SO(p,q)$.  An arbitrary irreducible representation of $SO(p,q)$ is in turn indexed by a set of nonnegative integers, the Dynkin indices, denoted by $\boldsymbol{N}=\{N_1,\ldots,N_r\}\equiv\sum_iN_i\boldsymbol{e}_i$, where $r$ is the rank of the Lorentz group and $\boldsymbol{e}_i\equiv(\boldsymbol{e}_i)_j=\delta_{ij}$.  There exists a variety of methods for constructing such projectors to general irreducible representations $\boldsymbol{N}$ of the Lorentz group.  Some examples include Young tableaux techniques with the birdtrack notation \cite{Costa:2016hju}, the weight-shifting operator formalism \cite{Karateev:2017jgd}, and an approach based on the tensor product decomposition and the defining properties above \cite{Fortin:2019gck}.  Irrespective of the method employed, the procedure comes down to an application of group theory.

We may build up the projection operators to general irreducible representations from the corresponding operators for the defining irreducible representations.  These act as building blocks for the general operators.  By properly subtracting traces and smaller irreducible representations from appropriately symmetrized products of the defining projectors, we may in principle generate any projection operator of interest.

The hatted projectors to defining irreducible representations in odd spacetime dimensions are given by 
\eqn{
\begin{gathered}
(\hat{\mathcal{P}}^{\boldsymbol{e}_r})_\alpha^{\phantom{\alpha}\beta}=\delta_\alpha^{\phantom{\alpha}\beta},\qquad(\hat{\mathcal{P}}^{\boldsymbol{e}_{i\neq r}})_{\mu_i\cdots\mu_1}^{\phantom{\mu_i\cdots\mu_1}\nu_1\cdots\nu_i}=\delta_{[\mu_1}^{\phantom{[\mu_1}\nu_1}\cdots\delta_{\mu_i]}^{\phantom{\mu_i]}\nu_i},\qquad(\hat{\mathcal{P}}^{2\boldsymbol{e}_r})_{\mu_r\cdots\mu_1}^{\phantom{\mu_r\cdots\mu_1}\nu_1\cdots\nu_r}=\delta_{[\mu_1}^{\phantom{[\mu_1}\nu_1}\cdots\delta_{\mu_r]}^{\phantom{\mu_r]}\nu_r},
\end{gathered}
}
while in even dimensions they are
\eqn{
\begin{gathered}
(\hat{\mathcal{P}}^{\boldsymbol{e}_{r-1}})_\alpha^{\phantom{\alpha}\beta}=\delta_\alpha^{\phantom{\alpha}\beta},\qquad(\hat{\mathcal{P}}^{\boldsymbol{e}_r})_{\tilde{\alpha}}^{\phantom{\tilde{\alpha}}\tilde{\beta}}=\delta_{\tilde{\alpha}}^{\phantom{\tilde{\alpha}}\tilde{\beta}},\qquad(\hat{\mathcal{P}}^{\boldsymbol{e}_{i\neq r-1,r}})_{\mu_i\cdots\mu_1}^{\phantom{\mu_i\cdots\mu_1}\nu_1\cdots\nu_i}=\delta_{[\mu_1}^{\phantom{[\mu_1}\nu_1}\cdots\delta_{\mu_i]}^{\phantom{\mu_i]}\nu_i},\\
(\hat{\mathcal{P}}^{\boldsymbol{e}_{r-1}+\boldsymbol{e}_r})_{\mu_{r-1}\cdots\mu_1}^{\phantom{\mu_{r-1}\cdots\mu_1}\nu_1\cdots\nu_{r-1}}=\delta_{[\mu_1}^{\phantom{[\mu_1}\nu_1}\cdots\delta_{\mu_{r-1}]}^{\phantom{\mu_{r-1}]}\nu_{r-1}},\\
(\hat{\mathcal{P}}^{2\boldsymbol{e}_{r-1}})_{\mu_r\cdots\mu_1}^{\phantom{\mu_r\cdots\mu_1}\nu_1\cdots\nu_r}=\frac{1}{2}\delta_{[\mu_1}^{\phantom{[\mu_1}\nu_1}\cdots\delta_{\mu_r]}^{\phantom{\mu_r]}\nu_r}+(-1)^r\frac{\mathscr{K}}{2r!}\epsilon_{\mu_1\cdots\mu_r}^{\phantom{\mu_1\cdots\mu_r}\nu_r\cdots\nu_1},\\
(\hat{\mathcal{P}}^{2\boldsymbol{e}_r})_{\mu_r\cdots\mu_1}^{\phantom{\mu_r\cdots\mu_1}\nu_1\cdots\nu_r}=\frac{1}{2}\delta_{[\mu_1}^{\phantom{[\mu_1}\nu_1}\cdots\delta_{\mu_r]}^{\phantom{\mu_r]}\nu_r}-(-1)^r\frac{\mathscr{K}}{2r!}\epsilon_{\mu_1\cdots\mu_r}^{\phantom{\mu_1\cdots\mu_r}\nu_r\cdots\nu_1}.
\end{gathered}
}
Here $\delta_{[\mu_1}^{\phantom{[\mu_1}\nu_1}\cdots\delta_{\mu_i]}^{\phantom{\mu_i]}\nu_i}$ is the totally antisymmetric normalized product of $\delta_\mu^{\phantom{\mu}\nu}$, while $\mathscr{K}$ is the proportionality constant in $\gamma^{\mu_1\cdots\mu_d}=\mathscr{K}\epsilon^{\mu_1\cdots\mu_d}\1$ which satisfies $\mathscr{K}^2=(-1)^{r+q}$ with $r$ the rank of the Lorentz group, $q$ the signature of the Lorentz group, and $\epsilon^{1\cdots d}=1$.  These hatted projectors operate on the ``dummy'' indices that are fully contracted in expressions for correlation functions.  They are in place to restrict the operators to the relevant irreducible representations.

Meanwhile, the half-projection operators encode the transformation properties of operators $\mathcal{O}^{\boldsymbol{N}}$ in general irreducible representations $\boldsymbol{N}$ under Lorentz transformations, $\mathcal{O}^{\boldsymbol{N}}\sim\mathcal{T}^{\boldsymbol{N}}$.  They are aptly named, because they satisfy $\mathcal{T}_{\boldsymbol{N}}*\mathcal{T}^{\boldsymbol{N}}=\hat{\mathcal{P}}^{\boldsymbol{N}}$, where the star product corresponds to contractions of the spinor indices.  These operators play the role of translating the spinor indices carried by each operator to the dummy vector and spinor indices that need to be contracted when constructing correlation functions.

The position space half-projectors to arbitrary irreducible representations $\boldsymbol{N}$ are given by
\eqna{
(\mathcal{T}^{\boldsymbol{N}})_{\alpha_1\cdots\alpha_n}^{\mu_1\cdots\mu_{n_v}\delta}&=\left((\mathcal{T}^{\boldsymbol{e}_1})^{N_1}\cdots(\mathcal{T}^{\boldsymbol{e}_{r-1}})^{N_{r-1}}(\mathcal{T}^{2\boldsymbol{e}_r})^{\lfloor N_r/2\rfloor}(\mathcal{T}^{\boldsymbol{e}_r})^{N_r-2\lfloor N_r/2\rfloor}\right)_{\alpha_1\cdots\alpha_n}^{\mu_1'\cdots\mu_{n_v}'\delta'}\\
&\phantom{=}\qquad\times(\hat{\mathcal{P}}^{\boldsymbol{N}})_{\delta'\mu_{n_v}'\cdots\mu_1'}^{\phantom{\delta'\mu_{n_v}'\cdots\mu_1'}\mu_1\cdots\mu_{n_v}\delta},
}[EqTPS]
where $n=2S=2\sum_{i=1}^{r-1}N_i+N_r$ is twice the spin $S$ of the irreducible representation $\boldsymbol{N}$, $n_v=\sum_{i=1}^{r-1}iN_i+r\lfloor N_r/2\rfloor$ is the number of vector indices of the irreducible representation $\boldsymbol{N}$, and $\delta$ is the spinor index which appears only if $N_r$ is odd (in odd spacetime dimensions).  In \eqref{EqTPS}, the spinor indices $\alpha_1,\ldots,\alpha_n$ match the free indices on the corresponding quasi-primary operator, while the remaining indices $\mu_1,\ldots,\mu_{n_v},\delta$ are dummy indices that are contracted.

Further, in \eqref{EqTPS} the corresponding half-projectors to the defining representations are given by
\eqn{(\mathcal{T}^{\boldsymbol{e}_{i\neq r}})_{\alpha\beta}^{\mu_1\cdots\mu_i}=\frac{1}{\sqrt{2^ri!}}(\gamma^{\mu_1\cdots\mu_i}C^{-1})_{\alpha\beta},\qquad(\mathcal{T}^{\boldsymbol{e}_r})_{\alpha}^{\beta}=\delta_\alpha^{\phantom{\alpha}\beta},\qquad(\mathcal{T}^{2\boldsymbol{e}_r})_{\alpha\beta}^{\mu_1\cdots\mu_r}=\frac{1}{\sqrt{2^rr!}}(\gamma^{\mu_1\cdots\mu_r}C^{-1})_{\alpha\beta},}[EqPositionHP]
where
\eqn{\gamma^{\mu_1\cdots\mu_n}=\frac{1}{n!}\sum_{\sigma\in S_n}(-1)^\sigma\gamma^{\mu_{\sigma(1)}}\cdots\gamma^{\mu_{\sigma(n)}},}
is the totally antisymmetric product of $\gamma$-matrices.  Lastly, the operator $\hat{\mathcal{P}}^{\boldsymbol{N}}$ in \eqref{EqTPS} contracts with the dummy indices of the half-projector.  It is present to ensure projection onto the proper irreducible representation $\boldsymbol{N}$.  We note that the definitions above extend straightforwardly to even dimensions.

The final objects of central interest here are the tensor structures.  These are purely group theoretic quantities that are entirely determined by the irreducible representations of the quasi-primary operators in question.  In a three-point function $\vev{\mathcal{O}^{\boldsymbol{N}_i} \mathcal{O}^{\boldsymbol{N}_j} \mathcal{O}^{\boldsymbol{N}_k} }$, the objects $\tCF{a}{i}{j}{k}{1}{2}$ serve to intertwine three irreducible representations of the Lorentz group into a symmetric traceless representation.  In fact, the number $N_{ijk}$ of symmetric irreducible representations appearing in $\boldsymbol{N}_i\otimes\boldsymbol{N}_j\otimes\boldsymbol{N}_k$ precisely corresponds to the number of such independent tensor structures and OPE coefficients.  Moreover, for fixed $\boldsymbol{N}_i$, $\boldsymbol{N}_j$, and $\boldsymbol{N}_k$, the set of all tensor structures forms a basis for a vector space.

Equivalently, these structures may be viewed as contracting four irreducible representations together into a singlet, with the fourth representation corresponding to the symmetric traceless differential operator.  As such, owing to the OPE, the corresponding embedding space quantities can be made to satisfy the following identity \cite{Fortin:2019dnq}
\eqn{\tCF{a}{i}{j}{k}{1}{2}=(\hat{\mathcal{P}}_{12}^{\boldsymbol{N}_i})(\hat{\mathcal{P}}_{21}^{\boldsymbol{N}_j})(\hat{\mathcal{P}}_{12}^{\boldsymbol{N}_k})(\hat{\mathcal{P}}_{21}^{n_a\boldsymbol{e}_1})\cdot\tCF{a}{i}{j}{k}{1}{2},}[EqPTS]
where the order of the contractions is self-evident.  The purpose of this condition is to restrict the tensor structures onto the appropriate irreducible representations for the three quasi-primary operators and the symmetric traceless differential operator.

It is straightforward to obtain the embedding space projection operators, half-projectors, and tensor structures from their position space counterparts by making the following substitutions:
\eqn{
\begin{gathered}
g^{\mu\nu}\to\A_{12}^{AB}\equiv g^{AB}-\frac{\eta_1^A\eta_2^B}{\ee{1}{2}{}}-\frac{\eta_1^B\eta_2^A}{\ee{1}{2}{}},\\
\epsilon^{\mu_1\cdots\mu_d}\to\epsilon_{12}^{A_1\cdots A_d}\equiv\frac{1}{\ee{1}{2}{}}\eta_{1A_0'}\epsilon^{A_0'A_1'\cdots A_d'A_{d+1}'}\eta_{2A_{d+1}'}\A_{12A_d'}^{\phantom{12A_d'}A_d}\cdots\A_{12A_1'}^{\phantom{12A_1'}A_1},\\
\gamma^{\mu_1\cdots\mu_n}\to\Gamma_{12}^{A_1\cdots A_n}\equiv\Gamma^{A_1'\cdots A_n'}\A_{12A_n'}^{\phantom{12A_n'}A_n}\cdots\A_{12A_1'}^{\phantom{12A_1'}A_1}\qquad\forall\,n\in\{0,\ldots,r\}.
\end{gathered}
}[EqSubs]
By construction, these exhibit all the requisite properties (\textit{e.g.}\ trace, number of vector indices, \textit{etc.}) to guarantee proper contraction with the corresponding irreducible representations in position space.

In the embedding space, the appropriate counterparts of the half-projectors \eqref{EqPositionHP} are given by
\eqna{
(\mathcal{T}_{ij}^{\boldsymbol{N}}\Gamma)&\equiv\left(\left(\frac{\sqrt{2}}{\ee{i}{j}{\frac{1}{2}}}\mathcal{T}^{\boldsymbol{e}_2}\eta_i\A_{ij}\right)^{N_1}\cdots\left(\frac{\sqrt{r}}{\ee{i}{j}{\frac{1}{2}}}\mathcal{T}^{\boldsymbol{e}_{r_E-1}}\eta_i\A_{ij}\cdots\A_{ij}\right)^{N_{r-1}}\right.\\
&\phantom{=}\qquad\times\left.\left(\frac{\sqrt{r+1}}{\ee{i}{j}{\frac{1}{2}}}\mathcal{T}^{2\boldsymbol{e}_{r_E}}\eta_i\A_{ij}\cdots\A_{ij}\right)^{\lfloor N_r/2\rfloor}\left(\frac{1}{\sqrt{2}\ee{i}{j}{}}\mathcal{T}^{\boldsymbol{e}_{r_E}}\eta_i\cdot\Gamma\eta_j\cdot\Gamma\right)^{N_r-2\lfloor N_r/2\rfloor}\right)\cdot\hat{\mathcal{P}}_{ij}^{\boldsymbol{N}},
}
where
\eqn{(\mathcal{T}^{\boldsymbol{e}_{n+1}}\eta_i\A_{ij}\cdots\A_{ij})_{ab}^{A_1\cdots A_n}\equiv(\mathcal{T}^{\boldsymbol{e}_{n+1}})_{ab}^{A_0'\cdots A_n'}\A_{ijA_n'}^{\phantom{ijA_n'}A_n}\cdots\A_{ijA_1'}^{\phantom{ijA_1'}A_1}\eta_{iA_0'}.}
The definition of the embedding space half-projectors to the defining representations is the direct analog of the position space definition with the substitutions \eqref{EqSubs} for the projectors and the rank of the Lorentz group $r\to r_E=r+1$, as expected.

With the notation established, we now discuss the two-, three-, and four-point correlation functions from the perspective of the embedding space OPE.  Before proceeding, we first present the identities (these can be proven from the identities in Appendix B of \cite{Fortin:2019dnq})
\eqn{
\begin{gathered}
\eta_j\cdot\Gamma\,\hat{\mathcal{P}}_{ji}^{\boldsymbol{N}}=\eta_j\cdot\Gamma\,(\A_{ji})^{n_v}\hat{\mathcal{P}}_{jk}^{\boldsymbol{N}}(\A_{ji})^{n_v},\\
\hat{\mathcal{P}}_{ij}^{\boldsymbol{N}}\,\eta_j\cdot\Gamma=(\A_{ji})^{n_v}\hat{\mathcal{P}}_{kj}^{\boldsymbol{N}}(\A_{ji})^{n_v}\,\eta_j\cdot\Gamma,
\end{gathered}
}[EqetaP]
valid for an arbitrary irreducible representation $\boldsymbol{N}$.  These identities are powerful in simplifying the computations of correlation functions, as we will show below.


\subsection{Two-Point Functions}

From the OPE \eqref{EqOPE}, it is easy to see that the two-point correlation functions are given by \cite{Fortin:2019xyr}
\eqn{\Vev{\Op{i}{1}\Op{j}{2}}=(\mathcal{T}_{12}^{\boldsymbol{N}_i}\Gamma)(\mathcal{T}_{21}^{\boldsymbol{N}_j}\Gamma)\cdot\frac{\lambda_{\boldsymbol{N}_i}\cOPE{}{i}{j}{\1}\hat{\mathcal{P}}_{12}^{\boldsymbol{N}_i}}{\ee{1}{2}{\tau_i}}=(\mathcal{T}_{12}^{\boldsymbol{N}_i}\Gamma)\cdot(\mathcal{T}_{21}^{\boldsymbol{N}_j}\Gamma)\frac{\lambda_{\boldsymbol{N}_i}\cOPE{}{i}{j}{\1}}{\ee{1}{2}{\tau_i}},}[Eq2pt]
where the sole tensor structures are
\eqn{\tOPE{}{i}{j}{\1}{1}{2}=\lambda_{\boldsymbol{N}_i}\hat{\mathcal{P}}_{12}^{\boldsymbol{N}_i}\to\lambda_{\boldsymbol{N}_i},}
without loss of generality.

As expected, the two-point correlation functions vanish unless the quasi-primary operators are in irreducible representations that are contragredient-reflected with respect to each other, \textit{i.e.}\ $\boldsymbol{N}_i=\boldsymbol{N}_j^{CR}$,\footnote{In Lorentzian signature, the contragredient-reflected representation corresponds to the conjugate representation, \textit{i.e.}\ $\boldsymbol{N}^{CR}=\boldsymbol{N}^C$.} and their conformal dimensions are the same, \textit{i.e.}\ $\tau_i=\tau_j$.  Here $\lambda_{\boldsymbol{N}_i}$ is a normalization constant that can be set to any convenient value.


\subsection{Three-Point Functions}

Applying the OPE \eqref{EqOPE} on the first two quasi-primary operators and then using the form of the two-point functions \eqref{Eq2pt} on the result, we find that the three-point correlation functions are given by \cite{Fortin:2019pep}
\eqna{
\Vev{\Op{i}{1}\Op{j}{2}\Op{m}{3}}&=\frac{(\mathcal{T}_{12}^{\boldsymbol{N}_i}\Gamma)(\mathcal{T}_{21}^{\boldsymbol{N}_j}\Gamma)(\mathcal{T}_{31}^{\boldsymbol{N}_m}\Gamma)}{\ee{1}{2}{\frac{1}{2}(\tau_i+\tau_j-\chi_m)}\ee{1}{3}{\frac{1}{2}(\chi_i-\chi_j+\tau_m)}\ee{2}{3}{\frac{1}{2}(-\chi_i+\chi_j+\chi_m)}}\\
&\phantom{=}\qquad\cdot\sum_{a=1}^{N_{ijm}}\cCF{a}{i}{j}{m}\mathscr{G}_{(a|}^{ij|m},
}[Eq3pt]
where the quantities $\mathscr{G}_{(a|}^{ij|m}$ are defined as
\eqn{\mathscr{G}_{(a|}^{ij|m}=\lambda_{\boldsymbol{N}_m}\bar{J}_{12;3}^{(d,h_{ijm},n_a,\Delta_m,\boldsymbol{N}_m)}\cdot\tCF{a}{i}{j}{m}{1}{2}.}
We refer to these as the three-point conformal blocks in the OPE tensor structure basis.

Further, the relevant three-point correlation function quantities are defined as
\eqn{\cCF{a}{i}{j}{m}=\sum_n\cOPE{a}{i}{j}{n}\cOPE{}{n}{m}{\1},\qquad\tCF{a}{i}{j}{m}{1}{2}=\tOPE{a}{i}{j}{m^{CR}}{1}{2}[(C_\Gamma^{-1})]^{2\xi_m}(g)^{n_v^m}(g)^{n_a}.}[EqCoeff]
We begin by considering the definition of the three-point $\bar{J}$-function in terms of the three-point conformal substitution \cite{Fortin:2019pep}, namely
\eqna{
\bar{J}_{12;3}^{(d,h,n,\Delta,\boldsymbol{N})}&=(\bar{\eta}_3\cdot\Gamma\,\hat{\mathcal{P}}_{32}^{\boldsymbol{N}}\cdot\hat{\mathcal{P}}_{12}^{\boldsymbol{N}}\,\bar{\eta}_2\cdot\Gamma)_{cs_3}\\
&\equiv\left.\bar{\eta}_3\cdot\Gamma\,\hat{\mathcal{P}}_{32}^{\boldsymbol{N}}\cdot\hat{\mathcal{P}}_{12}^{\boldsymbol{N}}\,\bar{\eta}_2\cdot\Gamma\right|_{\substack{(g)^{s_0}(\bar{\eta}_1)^{s_1}(\bar{\eta}_2)^{s_2}(\bar{\eta}_3)^{s_3}\to(g)^{s_0}(\bar{\eta}_1)^{s_1}(\bar{\eta}_3)^{s_3}\\\times\bar{I}_{12}^{(d,h-n/2-s_2,n+s_2;\chi-s_1/2+s_2/2+s_3/2)}}},
}[Eqcs3]
where the three-point homogenized embedding space coordinates are
\eqn{\bar{\eta}_i^A=\frac{\ee{j}{m}{\frac{1}{2}}}{\ee{i}{j}{\frac{1}{2}}\ee{i}{m}{\frac{1}{2}}}\e{i}{A}{},}[Eqetab3]
with $(i,j,m)$ a cyclic permutation of $(1,2,3)$.  The three-point $\bar{I}$-function is given explicitly in the next subsection in \eqref{EqIb3}.  It turns out that the identities \eqref{EqetaP} allow the following simplifications
\eqna{
\bar{J}_{12;3}^{(d,h,n,\Delta,\boldsymbol{N})}&=(\bar{\eta}_3\cdot\Gamma\,\hat{\mathcal{P}}_{32}^{\boldsymbol{N}}(\A_{12})^{n_v}\hat{\mathcal{P}}_{32}^{\boldsymbol{N}}(\A_{12})^{n_v}\,\bar{\eta}_2\cdot\Gamma)_{cs_3}\\
&=(\bar{\eta}_3\cdot\Gamma\,\hat{\mathcal{P}}_{32}^{\boldsymbol{N}}\cdot\hat{\mathcal{P}}_{32}^{\boldsymbol{N}}(\A_{12})^{n_v}\,\bar{\eta}_2\cdot\Gamma)_{cs_3}\\
&=(\bar{\eta}_3\cdot\Gamma\,\hat{\mathcal{P}}_{32}^{\boldsymbol{N}}(\A_{12})^{n_v}\,\bar{\eta}_2\cdot\Gamma)_{cs_3}\\
&=(\bar{\eta}_3\cdot\Gamma\,(\A_{32})^{n_v}\hat{\mathcal{P}}_{31}^{\boldsymbol{N}}(\A_{321})^{n_v}\,\bar{\eta}_2\cdot\Gamma)_{cs_3}.
}[EqJ3barIdentity]
To obtain the first equality here, we applied \eqref{EqetaP} on the second hatted projection operator.  In the second line, the $\A_{12}$ metrics were simplified to $g$ metrics through their contractions with the two hatted projection operators.  In the following line, we invoked the projection property of the hatted projection operator.  Lastly, in the fourth equality, \eqref{EqetaP} was used one more time.

We next remark that if we insert the result \eqref{EqJ3barIdentity} inside \eqref{Eq3pt}, we find that the $(\A_{32})^{n_v^m}$ can be simplified to $g$'s through their contractions with the hatted projection operator $\hat{\mathcal{P}}_{31}^{\boldsymbol{N}_m}$ and the half-projector $(\mathcal{T}_{31}^{\boldsymbol{N}_m}\Gamma)$.  Then the hatted projection operator $\hat{\mathcal{P}}_{31}^{\boldsymbol{N}_m}$ can be commuted through the $\bar{\eta}_3\cdot\Gamma$ and contracted directly with the half-projector $(\mathcal{T}_{31}^{\boldsymbol{N}_m}\Gamma)$, effectively allowing the following rewriting:
\eqna{
\bar{J}_{12;3}^{(d,h,n,\Delta,\boldsymbol{N})}&=(\bar{\eta}_3\cdot\Gamma\,(\A_{321})^{n_v}\,\bar{\eta}_2\cdot\Gamma)_{cs_3}\\
&\equiv\left.\bar{\eta}_3\cdot\Gamma\,(\A_{321})^{n_v}\,\bar{\eta}_2\cdot\Gamma\right|_{\substack{(g)^{s_0}(\bar{\eta}_1)^{s_1}(\bar{\eta}_2)^{s_2}(\bar{\eta}_3)^{s_3}\to(g)^{s_0}(\bar{\eta}_1)^{s_1}(\bar{\eta}_3)^{s_3}\\\times\bar{I}_{12}^{(d,h-n/2-s_2,n+s_2;\chi-s_1/2+s_2/2+s_3/2)}}},
}[EqJ3]
which does not depend explicitly on the irreducible representation, apart from its number of vector indices $n_v$.  Hence, the three-point conformal blocks simplify to
\eqn{\mathscr{G}_{(a|}^{ij|m}=\lambda_{\boldsymbol{N}_m}\left((\A_{321})^{n_v^m}\,\bar{\eta}_3\cdot\Gamma\,\bar{\eta}_2\cdot\Gamma\right)_{cs_3}\cdot\tCF{a}{i}{j}{m}{1}{2},}[Eq3ptCB]
with the proper parameters for the exchanged quasi-primary operator $h=h_{ijm}$, $n=n_a$, $\Delta=\Delta_m$, and $\boldsymbol{N}=\boldsymbol{N}_m$ in the three-point conformal substitution \eqref{Eqcs3}.

As discussed in \cite{Fortin:2019pep,Fortin:2019gck}, although the OPE tensor structure basis is convenient in the context of the OPE, it is not the simplest one to use for the construction of three-point correlation functions.  Rather, the natural optimal basis for three-point correlators is the three-point tensor structure basis.  The two bases, indicated by $(a$ and $[a$ for the OPE basis and the three-point basis, respectively, can be related through rotation matrices as in
\eqn{\mathscr{G}_{(a|}^{ij|m}=\sum_{a'=1}^{N_{ijm}}(R_{ijm}^{-1})_{aa'}\mathscr{G}_{[a'|}^{ij|m},\qquad\qquad\cCF{a}{i}{j}{m}=\sum_{a'=1}^{N_{ijm}}\aCF{a'}{i}{j}{m}(R_{ijm})_{a'a},}[EqRbases]
where the $\aCF{a}{i}{j}{m}$ are the associated three-point function coefficients, implying
\eqn{\sum_{a=1}^{N_{ijm}}\cCF{a}{i}{j}{m}\mathscr{G}_{(a|}^{ij|m}=\sum_{a=1}^{N_{ijm}}\aCF{a}{i}{j}{m}\mathscr{G}_{[a|}^{ij|m}.}[EqRcoeff]
We express the three-point conformal blocks in the three-point tensor structure basis as
\eqn{\mathscr{G}_{[a|}^{ij|m}=\bar{\eta}_3\cdot\Gamma\,\FCF{a}{i}{j}{m}{1}{2}(\A_{12},\Gamma_{12},\epsilon_{12};\A_{12}\cdot\bar{\eta}_3),}[Eq3ptTS]
where it is understood that the factor $\bar{\eta}_3\cdot\Gamma$ on the RHS appears only if $\xi_k=\frac{1}{2}$, \textit{i.e.}\ if the exchanged quasi-primary operator is fermionic.  In this basis, the three-point correlation functions \eqref{Eq3pt} can be effortlessly obtained without the aid of the OPE by simply enumerating the three-point tensor structure basis $\{\FCF{a}{i}{j}{m}{1}{2}\}$ made from $\A_{12}$'s, $\Gamma_{12}$'s, $\epsilon_{12}$'s and $\A_{12}\cdot\bar{\eta}_3$'s.  Note that the $n_a$ factors of $\A_{12}\cdot\bar{\eta}_3$ in the three-point tensor structures $\FCF{a}{i}{j}{m}{1}{2}(\A_{12},\Gamma_{12},\epsilon_{12};\A_{12}\cdot\bar{\eta}_3)$ can contract with any index, including the ones from $\A_{12}$, $\Gamma_{12}$, and $\epsilon_{12}$ originating from the tensor structures.


\subsubsection{Three-Point Tensorial Function}

The three-point tensorial function appearing in the three-point conformal substitution \eqref{EqJ3} was found in \cite{Fortin:2019fvx,Fortin:2019dnq} and is given explicitly by
\eqn{\bar{I}_{12}^{(d,h,n;p)}=\rho^{(d,h;p)}\sum_{\substack{q_0,q_1,q_2,q_3\geq0\\\bar{q}=2q_0+q_1+q_2+q_3=n}}S_{(q_0,q_1,q_2,q_3)}K^{(d,h;p;q_0,q_1,q_2,q_3)},}[EqIb3]
where the totally symmetric $S$-tensor, the $\rho$-function, and the $K$-function are
\eqna{
S_{(q_0,q_1,q_2,q_3)}^{A_1\cdots A_{\bar{q}}}&=g^{(A_1A_2}\cdots g^{A_{2q_0-1}A_{2q_0}}\bar{\eta}_1^{A_{2q_0+1}}\cdots\bar{\eta}_1^{A_{2q_0+q_1}}\\
&\phantom{=}\qquad\times\bar{\eta}_2^{A_{2q_0+q_1+1}}\cdots\bar{\eta}_2^{A_{2q_0+q_1+q_2}}\bar{\eta}_3^{A_{2q_0+q_1+q_2+1}}\cdots\bar{\eta}_3^{A_{\bar{q}})},\\
\rho^{(d,h;p)}&=(-2)^h(p)_h(p+1-d/2)_h,\\
K^{(d,h;p;q_0,q_1,q_2,q_3)}&=\frac{(-1)^{\bar{q}-q_0-q_1-q_2}(-2)^{\bar{q}-q_0}\bar{q}!}{q_0!q_1!q_2!q_3!}\frac{(-h-\bar{q})_{\bar{q}-q_0-q_2}(p+h)_{\bar{q}-q_0-q_1}}{(p+1-d/2)_{-q_0-q_1-q_2}}.
}[EqK3]
In \eqref{EqK3}, $\bar{q}=2q_0+q_1+q_2+q_3=n$, the total number of indices on $\bar{I}_{12}^{(d,h,n;p)}$.  The three-point tensorial function is totally symmetric and satisfies several convenient contiguous relations \cite{Fortin:2019fvx,Fortin:2019dnq}, given by
\eqna{
g\cdot\bar{I}_{12}^{(d,h,n;p)}&=0,\\
\bar{\eta}_1\cdot\bar{I}_{12}^{(d,h,n;p)}&=\bar{I}_{12}^{(d,h+1,n-1;p)},\\
\bar{\eta}_2\cdot\bar{I}_{12}^{(d,h,n;p)}&=\rho^{(d,1;-h-n)}\bar{I}_{12}^{(d,h,n-1;p)},\\
\bar{\eta}_3\cdot\bar{I}_{12}^{(d,h,n;p)}&=\bar{I}_{12}^{(d,h+1,n-1;p-1)}.
}[EqCont3]
These will be of great utility in the determination of rotation matrices.  For future convenience, we also introduce $\widetilde{K}^{(d,h;p;q_0,q_1,q_2,q_3)}=\rho^{(d,h;p)}K^{(d,h;p;q_0,q_1,q_2,q_3)}$ to simplify computations.  Having constructed the three-point functions, we now turn to the four-point correlators.


\subsection{Four-Point Functions}

Together, the OPE \eqref{EqOPE} and the respective results for two- and three-point correlation functions in \eqref{Eq2pt} and \eqref{Eq3pt} lead to the four-point correlation functions \cite{Fortin:2019gck}
\eqna{
&\Vev{\Op{i}{1}\Op{j}{2}\Op{k}{3}\Op{l}{4}}\\
&\qquad=\frac{(\mathcal{T}_{12}^{\boldsymbol{N}_i}\Gamma)(\mathcal{T}_{21}^{\boldsymbol{N}_j}\Gamma)(\mathcal{T}_{34}^{\boldsymbol{N}_k}\Gamma)(\mathcal{T}_{43}^{\boldsymbol{N}_l}\Gamma)}{\ee{1}{2}{\frac{1}{2}(\tau_i-\chi_i+\tau_j+\chi_j)}\ee{1}{3}{\frac{1}{2}(\chi_i-\chi_j+\chi_k-\chi_l)}\ee{1}{4}{\frac{1}{2}(\chi_i-\chi_j-\chi_k+\chi_l)}\ee{3}{4}{\frac{1}{2}(-\chi_i+\chi_j+\tau_k+\tau_l)}}\\
&\qquad\phantom{=}\qquad\cdot\sum_m\sum_{a=1}^{N_{ijm}}\sum_{b=1}^{N_{klm}}\cOPE{a}{i}{j}{m}\aCF{b}{k}{l}{m}\mathscr{G}_{(a|b]}^{ij|m|kl},
}[Eq4pt]
with the (four-point) conformal blocks in the mixed basis (the simplest one, as discussed in \cite{Fortin:2019gck}) given by
\eqn{\mathscr{G}_{(a|b]}^{ij|m|kl}=\sum_{b'=1}^{N_{klm}}(-1)^{2\xi_m}\lambda_{\boldsymbol{N}_m}(R_{klm})_{bb'}\tOPE{a}{i}{j}{m}{1}{2}\cdot\bar{J}_{34;21}^{(d,h_{ijm},n_a,h_{klm},n_b,\Delta_m,\boldsymbol{N}_m)}\cdot\tCF{b'}{k}{l}{m}{3}{4}.}
Here the four-point $\bar{J}$-function expressed in terms of the four-point conformal substitution is 
\eqna{
\bar{J}_{34;21}^{(d,h_1,n_1,h_2,n_2,\Delta,\boldsymbol{N})}&=(x_3^{2\xi}\bar{\eta}_2\cdot\Gamma\,\hat{\mathcal{P}}_{21}^{\boldsymbol{N}}\cdot\hat{\mathcal{P}}_{23}^{\boldsymbol{N}}\,\bar{\eta}_3\cdot\Gamma(\bar{\bar{\eta}}_2\cdot\Gamma\,\hat{\mathcal{P}}_{24}^{\boldsymbol{N}}\cdot\hat{\mathcal{P}}_{34}^{\boldsymbol{N}}\,\bar{\bar{\eta}}_4\cdot\Gamma)_{cs_3})_{cs_4}\\
&\equiv x_3^{2\xi}\bar{\eta}_2\cdot\Gamma\,\hat{\mathcal{P}}_{21}^{\boldsymbol{N}}\cdot\hat{\mathcal{P}}_{23}^{\boldsymbol{N}}\,\bar{\eta}_3\cdot\Gamma\\
&\phantom{=}\qquad\times\left.(\bar{\bar{\eta}}_2\cdot\Gamma\,\hat{\mathcal{P}}_{24}^{\boldsymbol{N}}\cdot\hat{\mathcal{P}}_{34}^{\boldsymbol{N}}\,\bar{\bar{\eta}}_4\cdot\Gamma)_{cs_3}\right|_{(\bar{\eta}_2)^{s_2}x_3^{r_3}x_4^{r_4}\to\bar{I}_{12;34}^{(d,h_1-n_1/2-s_2,n_1+s_2;-h_2+r_3,\chi+h_2+r_4)}},
}[Eqcs4]
where the three- and four-point homogeneized embedding space coordinates\footnote{We note that $\bar{\bar{\eta}}_2=\sqrt{x_3x_4}\bar{\eta}_2$, $\bar{\bar{\eta}}_3=\sqrt{\frac{x_3}{x_4}}\bar{\eta}_3$ and $\bar{\bar{\eta}}_4=\sqrt{\frac{x_4}{x_3}}\bar{\eta}_4$.  Moreover, it is important to realize that the three- and four-point homogeneized embedding space coordinates \eqref{Eqetab3} and \eqref{Eqetab4} are different.  Since the former are used in the computation of the rotation matrices while the latter appear in the four-point conformal blocks, it should be clear from the context which ones are used.} as well as the cross-ratios are
\eqn{
\begin{gathered}
\bar{\bar{\eta}}_2^A=\frac{\ee{3}{4}{\frac{1}{2}}}{\ee{2}{3}{\frac{1}{2}}\ee{2}{4}{\frac{1}{2}}}\eta_2^A,\quad\quad\bar{\bar{\eta}}_3^A=\frac{\ee{2}{4}{\frac{1}{2}}}{\ee{2}{3}{\frac{1}{2}}\ee{3}{4}{\frac{1}{2}}}\eta_3^A,\quad\quad\bar{\bar{\eta}}_4^A=\frac{\ee{2}{3}{\frac{1}{2}}}{\ee{3}{4}{\frac{1}{2}}\ee{3}{4}{\frac{1}{2}}}\eta_4^A,\\
\bar{\eta}_1^A=\frac{\ee{3}{4}{\frac{1}{2}}}{\ee{1}{3}{\frac{1}{2}}\ee{1}{4}{\frac{1}{2}}}\eta_1^A,\qquad\qquad\bar{\eta}_2^A=\frac{\ee{1}{3}{\frac{1}{2}}\ee{1}{4}{\frac{1}{2}}}{\ee{1}{2}{}\ee{3}{4}{\frac{1}{2}}}\eta_2^A,\\
\bar{\eta}_3^A=\frac{\ee{1}{4}{\frac{1}{2}}}{\ee{3}{4}{\frac{1}{2}}\ee{1}{3}{\frac{1}{2}}}\eta_3^A,\qquad\qquad\bar{\eta}_4^A=\frac{\ee{1}{3}{\frac{1}{2}}}{\ee{3}{4}{\frac{1}{2}}\ee{1}{4}{\frac{1}{2}}}\eta_4^A,\\
x_3=\frac{\ee{1}{2}{}\ee{3}{4}{}}{\ee{1}{4}{}\ee{2}{3}{}}=\frac{u}{v},\qquad\qquad x_4=\frac{\ee{1}{2}{}\ee{3}{4}{}}{\ee{1}{3}{}\ee{2}{4}{}}=u,
\end{gathered}
}[Eqetab4]
and the four-point $\bar{I}$-function is discussed below.

As before, we now apply the identities \eqref{EqetaP} to transform $\bar{J}_{34;21}^{(d,h_1,n_1,h_2,n_2,\Delta,\boldsymbol{N})}$.  These imply the simplifications
\eqna{
\bar{J}_{34;21}^{(d,h_1,n_1,h_2,n_2,\Delta,\boldsymbol{N})}&=(x_3^{2\xi}\bar{\eta}_2\cdot\Gamma\,\hat{\mathcal{P}}_{21}^{\boldsymbol{N}}\cdot\hat{\mathcal{P}}_{23}^{\boldsymbol{N}}\,\bar{\eta}_3\cdot\Gamma(\bar{\bar{\eta}}_2\cdot\Gamma\,\hat{\mathcal{P}}_{24}^{\boldsymbol{N}}(\A_{34})^{n_v}\hat{\mathcal{P}}_{24}^{\boldsymbol{N}}(\A_{34})^{n_v}\,\bar{\bar{\eta}}_4\cdot\Gamma)_{cs_3})_{cs_4}\\
&=(x_3^{2\xi}\bar{\eta}_2\cdot\Gamma\,\hat{\mathcal{P}}_{21}^{\boldsymbol{N}}\cdot\hat{\mathcal{P}}_{23}^{\boldsymbol{N}}\,\bar{\eta}_3\cdot\Gamma(\bar{\bar{\eta}}_2\cdot\Gamma\,\hat{\mathcal{P}}_{24}^{\boldsymbol{N}}(\A_{34})^{n_v}\,\bar{\bar{\eta}}_4\cdot\Gamma)_{cs_3})_{cs_4}\\
&=(x_3^{2\xi}\bar{\eta}_2\cdot\Gamma\,\hat{\mathcal{P}}_{21}^{\boldsymbol{N}}\cdot\hat{\mathcal{P}}_{23}^{\boldsymbol{N}}\,\bar{\eta}_3\cdot\Gamma(\bar{\bar{\eta}}_2\cdot\Gamma\,(\A_{24})^{n_v}\hat{\mathcal{P}}_{23}^{\boldsymbol{N}}(\A_{243})^{n_v}\,\bar{\bar{\eta}}_4\cdot\Gamma)_{cs_3})_{cs_4}\\
&=(x_3^{2\xi}\bar{\eta}_2\cdot\Gamma\,\hat{\mathcal{P}}_{21}^{\boldsymbol{N}}\cdot\hat{\mathcal{P}}_{23}^{\boldsymbol{N}}\,\bar{\eta}_3\cdot\Gamma(\bar{\bar{\eta}}_2\cdot\Gamma\,(\A_{243})^{n_v}\,\bar{\bar{\eta}}_4\cdot\Gamma)_{cs_3})_{cs_4}\\
&=(x_3^{2\xi}\bar{\eta}_2\cdot\Gamma\,(\A_{21})^{n_v}\hat{\mathcal{P}}_{23}^{\boldsymbol{N}}(\A_{21})^{n_v}\hat{\mathcal{P}}_{23}^{\boldsymbol{N}}\,\bar{\eta}_3\cdot\Gamma(\bar{\bar{\eta}}_2\cdot\Gamma\,(\A_{243})^{n_v}\,\bar{\bar{\eta}}_4\cdot\Gamma)_{cs_3})_{cs_4}\\
&=(x_3^{2\xi}\bar{\eta}_2\cdot\Gamma\,(\A_{21})^{n_v}\hat{\mathcal{P}}_{23}^{\boldsymbol{N}}\,\bar{\eta}_3\cdot\Gamma(\bar{\bar{\eta}}_2\cdot\Gamma\,(\A_{243})^{n_v}\,\bar{\bar{\eta}}_4\cdot\Gamma)_{cs_3})_{cs_4}\\
&=(x_3^{2\xi}\bar{\eta}_2\cdot\Gamma\,(\A_{123})^{n_v}\bar{\eta}_3\cdot\Gamma\,\hat{\mathcal{P}}_{13}^{\boldsymbol{N}}(\A_{23})^{n_v}(\bar{\bar{\eta}}_2\cdot\Gamma\,(\A_{243})^{n_v}\,\bar{\bar{\eta}}_4\cdot\Gamma)_{cs_3})_{cs_4}.
}[EqJ4]
In the first line above, the last projection operator was replaced using \eqref{EqetaP}.  Then, half of the generated metrics and one of the last two projection operators were annihilated.  Subsequently, in the third and fourth lines, the two previous steps were repeated with the last projection operator. Further, in the fourth and fifth lines, the same two steps were then performed on the first projection operator.  Finally, in the last equality, the sole remaining projection operator was moved to the middle, between the three- and four-point conformal substitutions, and a set of metrics was introduced using \eqref{EqetaP}.

It is interesting to note here that contrary to \eqref{EqJ3}, it is impossible to remove the last projection operator in \eqref{EqJ4}.  This is expected, since in the case of four-point correlation functions, there are no half-projectors for the exchanged quasi-primary operators, unlike for three-point correlation functions.  A projection operator is therefore necessary to ensure that the conformal blocks are in the appropriate irreducible representation.  Given this result as well as \eqref{Eq3ptCB}, the four-point conformal blocks assume the form
\eqna{
\mathscr{G}_{(a|b]}^{ij|m|kl}&=\sum_{b'=1}^{N_{klm}}(-1)^{2\xi_m}\lambda_{\boldsymbol{N}_m}\tOPE{a}{i}{j}{m}{1}{2}(R_{klm})_{bb'}\\
&\phantom{=}\qquad\cdot(x_3^{2\xi_m}\bar{\eta}_2\cdot\Gamma\,(\A_{123})^{n_v^m}\,\bar{\eta}_3\cdot\Gamma\,\hat{\mathcal{P}}_{13}^{\boldsymbol{N}_m}(\A_{23})^{n_v}(\bar{\bar{\eta}}_2\cdot\Gamma\,(\A_{243})^{n_v^m}\,\bar{\bar{\eta}}_4\cdot\Gamma)_{cs_3})_{cs_4}\cdot\tCF{b'}{k}{l}{m}{3}{4}\\
&=\tOPE{a}{i}{j}{m}{1}{2}\cdot((-x_3)^{2\xi_m}\bar{\eta}_2\cdot\Gamma\,(\A_{123})^{n_v^m}\,\bar{\eta}_3\cdot\Gamma\,\hat{\mathcal{P}}_{13}^{\boldsymbol{N}_m}(\A_{23})^{n_v^m}\,\bar{\bar{\eta}}_2\cdot\Gamma\,\FCF{b}{k}{l}{m}{3}{4}(\A_{34},\Gamma_{34},\epsilon_{34};\A_{34}\cdot\bar{\bar{\eta}}_2))_{cs_4}\\
&=\tOPE{a}{i}{j}{m}{1}{2}\cdot\left((-x_3)^{2\xi_m}(\A_{123})^{n_v^m}\bar{\eta}_2\cdot\Gamma\,\bar{\eta}_3\cdot\Gamma\,\hat{\mathcal{P}}_{13}^{\boldsymbol{N}_m}\,\bar{\bar{\eta}}_2\cdot\Gamma\,\FCF{b}{k}{l}{m}{3}{4}(\A_{34},\Gamma_{34},\epsilon_{34};\A_{34}\cdot\bar{\bar{\eta}}_2)\right)_{cs_4},
}[Eq4ptCB]
with the proper parameters for the exchanged quasi-primary operator $h_1=h_{ijm}$, $n_1=n_a$, $h_2=h_{klm}$, $n_2=n_b$, $\Delta=\Delta_m$ and $\boldsymbol{N}=\boldsymbol{N}_m$ in the four-point conformal substitution \eqref{Eqcs4}.

In \cite{Fortin:2019gck}, we remarked that the conformal blocks feature the simplest form in the mixed basis.  However, for the implementation of the conformal bootstrap, it is more convenient to work in a pure tensor structure basis, \textit{e.g.}\ the three-point basis.  Following the discussion in the previous subsections [see \eqref{EqRbases} and \eqref{EqRcoeff}], the conformal blocks in the pure three-point basis can be determined from their mixed counterparts by acting with the rotation matrices as in
\eqn{\mathscr{G}_{[a|b]}^{ij|m|kl}=\sum_{a'=1}^{N_{ijm}}(R_{ijm})_{aa'}\mathscr{G}_{(a'|b]}^{ij|m|kl}.}
Clearly, in the interest of setting up the bootstrap, it is therefore necessary to compute not only the conformal blocks in the mixed basis but also the corresponding rotation matrices.


\subsubsection{Four-Point Tensorial Function}

In \cite{Fortin:2019fvx,Fortin:2019dnq}, the four-point tensorial function $\bar{I}_{12;34}^{(d,h,n;p_3,p_4)}$ was found to be given by
\eqn{\bar{I}_{12;34}^{(d,h,n;p_3,p_4)}=\sum_{\substack{q_0,q_1,q_2,q_3,q_4\geq0\\\bar{q}=2q_0+q_1+q_2+q_3+q_4=n}}S_{(\boldsymbol{q})}\rho^{(d,h;p_3+p_4)}x_3^{p_3+p_4+h+q_0+q_2+q_3+q_4}K_{12;34;3}^{(d,h;p_3,p_4;q_0,q_1,q_2,q_3,q_4)}(x_3;y_4),}[EqIb4]
with the totally symmetric object $S_{(\boldsymbol{q})}$ defined by
\eqn{S_{(\boldsymbol{q})}^{A_1\cdots A_{\bar{q}}}=g^{(A_1A_2}\cdots g^{A_{2q_0-1}A_{2q_0}}\bar{\eta}_1^{A_{2q_0+1}}\cdots\bar{\eta}_1^{A_{2q_0+q_1}}\cdots\bar{\eta}_4^{A_{\bar{q}-q_4+1}}\cdots\bar{\eta}_4^{A_{\bar{q}})},}[EqS4]
where $\bar{q}=2q_0+q_1+q_2+q_3+q_4$ and $y_4=1-x_3/x_4$.  The function \eqref{EqS4} is the natural extension of \eqref{EqK3} to four points.

The $K$-function appearing in the four-point $\bar{I}$-function is given by 
\eqna{
K_{12;34;3}^{(d,h;\boldsymbol{p};\boldsymbol{q})}(x_3;y_4)&=\frac{(-1)^{q_0+q_3+q_4}(-2)^{\bar{q}-q_0}\bar{q}!}{q_0!q_1!q_2!q_3!q_4!}\frac{(-h-\bar{q})_{\bar{q}-q_0-q_2}(p_3)_{q_3}(p_3+p_4+h)_{\bar{q}-q_0-q_1}}{(p_3+p_4)_{q_3+q_4}(p_3+p_4+1-d/2)_{-q_0-q_1-q_2}}(p_4)_{q_4}\\
&\phantom{=}\qquad\times K_{12;34;3}^{(d+2\bar{q}-2q_0,h+q_0+q_2;p_3+q_3,p_4+q_4)}(x_3;y_4),
}[EqK4]
where
\eqna{
K_{12;34;3}^{(d,h;p_3,p_4)}(x_3;y_4)&=\sum_{n_4,n_{34}\geq0}\frac{(-h)_{n_{34}}(p_3)_{n_{34}}(p_3+p_4+h)_{n_4}}{(p_3+p_4)_{n_4+n_{34}}(p_3+p_4+1-d/2)_{n_{34}}}\frac{(p_4)_{n_4}}{n_{34}!(n_4-n_{34})!}y_4^{n_4}\left(\frac{x_3}{y_4}\right)^{n_{34}}\\
&=G(p_4,p_3+p_4+h,p_3+p_4+1-d/2,p_3+p_4;u/v,1-1/v),
}[EqK0]
is the usual Exton $G$-function $G(\alpha,\beta,\gamma,\delta;x,y)$ with appropriately shifted parameters \cite{Exton_1995}.  Hence, the tensorial $\bar{I}$-function is constructed from linear combinations of the Exton $G$-function.  Alternatively, we may regard it as built from the conformal block for scalar exchange in the scalar four-point correlator.

Like the three-point tensorial function, the four-point $\bar{I}$-function \eqref{EqIb4} satisfies a set of contiguous relations \cite{Fortin:2019fvx,Fortin:2019dnq},
\eqna{
g\cdot\bar{I}_{12;34}^{(d,h,n;p)}&=0,\\
\bar{\eta}_1\cdot\bar{I}_{12;34}^{(d,h,n;p_3,p_4)}&=\bar{I}_{12;34}^{(d,h+1,n-1;p_3,p_4)},\\
\bar{\eta}_2\cdot\bar{I}_{12;34}^{(d,h,n;p_3,p_4)}&=\rho^{(d,1;-h-n)}\bar{I}_{12;34}^{(d,h,n-1;p_3,p_4)},\\
\bar{\eta}_3\cdot\bar{I}_{12;34}^{(d,h,n;p_3,p_4)}&=\bar{I}_{12;34}^{(d,h+1,n-1;p_3-1,p_4)},\\
\bar{\eta}_4\cdot\bar{I}_{12;34}^{(d,h,n;p_3,p_4)}&=\bar{I}_{12;34}^{(d,h+1,n-1;p_3,p_4-1)},
}[EqCont4]
which enable easy contractions.


\subsection{Summary of Three- and Four-Point Functions}

To recapitulate the above review of the embedding space OPE formalism, we summarize here the key quantities of interest, namely the three- and four-point correlation functions.  Although we ultimately seek to compute four-point conformal blocks, the determination of the three-point correlation functions is necessary for the extraction of the rotation matrices.  As discussed above, the latter are instrumental in allowing us to translate between the mixed basis, where the four-point conformal blocks are simplest in form, to a pure tensor structure basis, which is more convenient for bootstrap purposes.

Reintroducing the dummy indices in the correlation functions \eqref{Eq3pt}
\eqna{
\Vev{\Op{i}{1}\Op{j}{2}\Op{m}{3}}&=\frac{(\mathcal{T}_{12}^{\boldsymbol{N}_i}\Gamma)^{\{Aa\}}(\mathcal{T}_{21}^{\boldsymbol{N}_j}\Gamma)^{\{Bb\}}(\mathcal{T}_{31}^{\boldsymbol{N}_m}\Gamma)^{\{Ee\}}}{\ee{1}{2}{\frac{1}{2}(\tau_i+\tau_j-\chi_m)}\ee{1}{3}{\frac{1}{2}(\chi_i-\chi_j+\tau_m)}\ee{2}{3}{\frac{1}{2}(-\chi_i+\chi_j+\chi_m)}}\\
&\phantom{=}\qquad\cdot\sum_{a=1}^{N_{ijm}}\cCF{a}{i}{j}{m}(\mathscr{G}_{(a|}^{ij|m})_{\{aA\}\{bB\}\{eE\}},
}[Eq3ptind]
and \eqref{Eq4pt}
\eqna{
&\Vev{\Op{i}{1}\Op{j}{2}\Op{k}{3}\Op{l}{4}}\\
&\qquad=\frac{(\mathcal{T}_{12}^{\boldsymbol{N}_i}\Gamma)^{\{Aa\}}(\mathcal{T}_{21}^{\boldsymbol{N}_j}\Gamma)^{\{Bb\}}(\mathcal{T}_{34}^{\boldsymbol{N}_k}\Gamma)^{\{Cc\}}(\mathcal{T}_{43}^{\boldsymbol{N}_l}\Gamma)^{\{Dd\}}}{\ee{1}{2}{\frac{1}{2}(\tau_i-\chi_i+\tau_j+\chi_j)}\ee{1}{3}{\frac{1}{2}(\chi_i-\chi_j+\chi_k-\chi_l)}\ee{1}{4}{\frac{1}{2}(\chi_i-\chi_j-\chi_k+\chi_l)}\ee{3}{4}{\frac{1}{2}(-\chi_i+\chi_j+\tau_k+\tau_l)}}\\
&\qquad\phantom{=}\qquad\cdot\sum_m\sum_{a=1}^{N_{ijm}}\sum_{b=1}^{N_{klm}}\cOPE{a}{i}{j}{m}\aCF{b}{k}{l}{m}(\mathscr{G}_{(a|b]}^{ij|m|kl})_{\{aA\}\{bB\}\{cC\}\{dD\}},
}[Eq4ptind]
leads to the conformal blocks \eqref{Eq3ptCB}
\eqn{(\mathscr{G}_{(a|}^{ij|m})_{\{aA\}\{bB\}\{eE\}}=\lambda_{\boldsymbol{N}_m}\left((\A_{321E}^{\phantom{321E}E'})^{n_v^m}(\bar{\eta}_3\cdot\Gamma\,\bar{\eta}_2\cdot\Gamma)_e^{\phantom{e}e'}\right)_{cs_3}(\tCF{a}{i}{j}{m}{1}{2})_{\{aA\}\{bB\}\{e'E'\}\{F\}},}[Eq3ptCBind]
and \eqref{Eq4ptCB}
\eqna{
(\mathscr{G}_{(a|b]}^{ij|m|kl})_{\{aA\}\{bB\}\{cC\}\{dD\}}&=(\tOPE{a}{i}{j}{m}{1}{2})_{\{aA\}\{bB\}}^{\phantom{\{aA\}\{bB\}}\{Ee\}\{F\}}\left((-x_3)^{2\xi_m}(\A_{123E}^{\phantom{123E}E'})^{n_v^m}(\bar{\bar{\eta}}_2\cdot\Gamma)_e^{\phantom{e}e'}\right.\\
&\phantom{=}\qquad\left.\times(\bar{\eta}_3\cdot\Gamma\,\hat{\mathcal{P}}_{13}^{\boldsymbol{N}_m}\,\bar{\eta}_2\cdot\Gamma)_{e'E'}^{\phantom{e'E'}E''e''}(\FCF{b}{k}{l}{m}{3}{4})_{\{cC\}\{dD\}\{e''E''\}}\right)_{cs_4}.
}[Eq4ptCBind]
Here, the explicit $F$-indices on the tensor structures are contracted through the OPE differential operator \eqref{EqOPE} with the implicit $F$-indices of the conformal substitutions $cs_3$ and $cs_4$, respectively.

For the remainder of this paper, we will focus on developing a set of simple and efficient rules for the determination of the rotation matrices and conformal blocks in the mixed basis for quasi-primary operators in arbitrary irreducible representations, given some input data, namely the projection operators and tensor structures.  We now turn to a discussion of this input group theoretic data.


\section{Input Data}\label{SecInput}

It is apparent from \eqref{Eq3ptCB} and \eqref{Eq4ptCB} that the input data consists of the projection operators and the tensor structures.  Although the tensor structures are obtained once the projection operators are determined, it is simpler to discuss the tensor structures first.  In this section, we will introduce a simple basis of three-point tensor structures, which is made out of products of a small set of allowed constituents.  The projection operators are built from their corresponding irreducible representations.  In this work, we will primarily focus on the projection operators for the exchanged quasi-primary operators, which are more intricate, due to the existence of infinite towers of exchanged quasi-primary operators in $\boldsymbol{N}_m+\ell\boldsymbol{e}_1$, resulting in $\ell$-dependent projectors.


\subsection{Bases of Tensor Structures}

The simplest available basis of tensor structures is the three-point basis \eqref{Eq3ptTS}, where tensor structures are simply constructed from products of allowed constituents.  There is also its analog for the OPE basis.  However, these two bases are not related straightforwardly \cite{Fortin:2019pep}.  Indeed, a change of basis is necessary, which calls for a computation of the rotation matrices mentioned above.

Before proceeding, let us first consider the tensor structures for $\boldsymbol{N}_m\to\boldsymbol{N}_m+\ell\boldsymbol{e}_1$, with $\boldsymbol{N}_m$ chosen to have $N_1=0$, \textit{i.e.}\ a vanishing first Dynkin index.  This observation will allow us to compute conformal blocks for infinite towers of exchanged quasi-primary operators.  Indeed, if it is possible (impossible) to exchange a quasi-primary operator in the irreducible representation $\boldsymbol{N}_m+\ell\boldsymbol{e}_1$ for some fixed $\ell\geq\ell_{\text{min}}$ ($\ell<\ell_{\text{min}}$), then it is straightforward to conclude that all quasi-primary operators in irreducible representations $\boldsymbol{N}_m+\ell\boldsymbol{e}_1$ with $\ell\geq\ell_{\text{min}}$ can also be exchanged, leading to an infinite tower of exchanged quasi-primary operators with the same seed irreducible representation $\boldsymbol{N}_m+\ell_{\text{min}}\boldsymbol{e}_1$.  We remark here that both $\boldsymbol{N}_m$ and $\ell_{\text{min}}$ depend on the irreducible representations of the quasi-primary operators of interest.

For exchanged quasi-primary operators in the $\boldsymbol{N}_m+\ell\boldsymbol{e}_1$ irreducible representation, the three-point basis can therefore be separated as follows:
\eqn{
\begin{gathered}
\FCF{b}{k}{l}{,m+\ell}{3}{4}=\FCF{b}{k}{l}{,m+i_b}{3}{4}(\A_{34}\cdot\bar{\bar{\eta}}_2)^{\ell-i_b},\\
\FCF{a}{i}{j}{,m+\ell}{1}{2}=\FCF{a}{i}{j}{,m+i_a}{1}{2}(\A_{12}\cdot\bar{\eta}_3)^{\ell-i_a}\to\tCF{a}{i}{j}{,m+\ell}{1}{2}=\tCF{a}{i}{j}{,m+i_a}{1}{2}(\A_{12})^{\ell-i_a},
\end{gathered}
}[EqTS]
where the $(\A_{34}\cdot\bar{\bar{\eta}}_2)_{E''_{i_b+1}}\cdots(\A_{34}\cdot\bar{\bar{\eta}}_2)_{E''_\ell}$ and $(\A_{12}\cdot\bar{\eta}_3)_{E_{i_a+1}}\cdots(\A_{12}\cdot\bar{\eta}_3)_{E_\ell}$ are the symmetrized $\ell$-dependent parts of the respective tensor structures.\footnote{Therefore, $i_a$ and $i_b$ are $\ell$-independent nonnegative integers, \textit{i.e.}\ $i_a$ and $i_b$ are fixed even when $\ell\to\infty$.}  We observe that in the second line of \eqref{EqTS}, the OPE basis is obtained from the three-point basis by simply transforming all $\A_{12}\cdot\bar{\eta}_3\to\A_{12}$ with the extra index contracting with the OPE differential operator, for example
\eqn{(\A_{12}\cdot\bar{\eta}_3)_{E_{i_a+1}}\cdots(\A_{12}\cdot\bar{\eta}_3)_{E_\ell}\to\A_{12E'_{i_a+1}F_{i_a+1}}\cdots\A_{12E'_\ell F_\ell}.}[Eq3pTStoOPETS]
It follows that the OPE basis used here does not satisfy the projection property \eqref{EqPTS} of the most natural tensor structures from the OPE point of view \cite{Fortin:2019dnq,Fortin:2019pep}.  However, its simple form will be of great advantage when we determine the three- and four-point conformal blocks.  We note also that the number of $(\A_{34}\cdot\bar{\bar{\eta}}_2)$'s in $\FCF{b}{k}{l}{,m+\ell}{3}{4}$ is given by $n_b$, while that of $(\A_{12}\cdot\bar{\eta}_3)$'s in $\FCF{a}{i}{j}{,m+\ell}{1}{2}$ by $n_a$.

In \eqref{EqTS}, the undetermined parts of the tensor structures, \textit{i.e.}\ $\tCF{a}{i}{j}{,m+i_a}{1}{2}$ and $\FCF{b}{k}{l}{,m+i_b}{3}{4}$, are fixed by the knowledge of the specific irreducible representations of the quasi-primary operators under consideration.  In the following, we dub them the ``special" parts of the tensor structures and specify them only for particular examples with known quasi-primary operators.


\subsection{Projection Operators}

Since there are no half-projectors for the exchanged quasi-primary operators in the four-point correlation function \eqref{Eq4ptind}, the projection operator to the exchanged representation necessarily appears explicitly in the four-point conformal blocks \eqref{Eq4ptCBind}.  As for the tensor structures, we work here with exchanged quasi-primary operators in the $\boldsymbol{N}_m+\ell\boldsymbol{e}_1$ infinite tower of irreducible representations.  To determine the four-point blocks, it is simpler to expand the projection operators as
\eqn{\hat{\mathcal{P}}_{13}^{\boldsymbol{N}_m+\ell\boldsymbol{e}_1}=\sum_t\mathscr{A}_t(d,\ell)\hat{\mathcal{Q}}_{13|t}^{\boldsymbol{N}_m+\ell_t\boldsymbol{e}_1}\hat{\mathcal{P}}_{13|d+d_t}^{(\ell-\ell_t)\boldsymbol{e}_1},}[EqPExp]
where $\mathscr{A}_t(d,\ell)$ are constants dependent on $d$ and $\ell$.  The sum is finite and $\ell$-independent here, and the number of terms depends on the irreducible representation $\boldsymbol{N}_m$.  Moreover, the tensor quantities $\hat{\mathcal{Q}}_{13|t}^{\boldsymbol{N}_m+\ell_t\boldsymbol{e}_1}$ encode information about the special parts of the irreducible representation $\boldsymbol{N}_m+\ell_t\boldsymbol{e}_1$, while the remaining indices are carried by shifted projection operators for some $d'$ and $\ell'$, denoted by
\eqna{
(\hat{\mathcal{P}}_{13|d'}^{\ell'\boldsymbol{e}_1})_{E'_\ell\cdots E'_1}^{\phantom{E'_\ell\cdots E'_1}E''_1\cdots E''_\ell}&=\sum_{i=0}^{\lfloor\ell'/2\rfloor}\frac{(-\ell')_{2i}}{2^{2i}i!(-\ell'+2-d'/2)_i}\A_{13(E'_1E'_2}\A_{13}^{(E''_1E''_2}\cdots\A_{13E'_{2i-1}E'_{2i}}\A_{13}^{E''_{2i-1}E''_{2i}}\\
&\phantom{=}\qquad\times\A_{13E'_{2i+1}}^{\phantom{13E'_{2i+1}}E''_{2i+1}}\cdots\A_{13E'_\ell)}^{\phantom{13E'_{\ell'})}E''_{\ell'})}.
}[EqPShift]
It is important to notice here that the $\ell$ $E'$-indices in \eqref{EqPExp}, distributed among $\hat{\mathcal{Q}}_{13|t}^{\boldsymbol{N}_m+\ell_t\boldsymbol{e}_1}$ and $\hat{\mathcal{P}}_{13|d+d_t}^{(\ell-\ell_t)\boldsymbol{e}_1}$, are symmetrized.  The same is true of the $\ell$ $E''$-indices.  Furthermore, we point out that the shifted projection operators $\hat{\mathcal{P}}_{13|d+d_t}^{(\ell-\ell_t)\boldsymbol{e}_1}$ \eqref{EqPShift} are not traceless when $d_t\neq0$.

For future convenience, we include some properties of the shifted projection operators under extraction of indices.  Indeed, in the computation of four-point blocks, it is often necessary to extract $n'_E$ $E'$-indices and $n''_E$ $E''$-indices from the shifted projection operators.  These are special indices that ultimately contract with the special parts of the tensor structures.  The general form for this extraction is
\eqna{
(\hat{\mathcal{P}}_{13|d}^{\ell\boldsymbol{e}_1})_{\{E'\}}^{\phantom{\{E'\}}\{E''\}}&=\sum_{\substack{r,\boldsymbol{r}',\boldsymbol{r}''\geq0\\r+2r'_0+r'_1+r'_2=n'_E\\r+2r''_0+r''_1+r''_2=n''_E\\r'_0+r'_1+r'_3=r''_0+r''_1+r''_3}}\mathscr{C}_{n'_E,n''_E}^{(d,\ell)}(r,\boldsymbol{r}',\boldsymbol{r}'')(\A_{13E'_s}^{\phantom{13E'_s}E''_s})^r\\
&\phantom{=}\qquad\times(\A_{13E'_sE'_s})^{r'_0}(\A_{13E'_sE'})^{r'_1}(\A_{13E'_s}^{\phantom{13E'_s}E''})^{r'_2}(\A_{13E'E'})^{r'_3}\\
&\phantom{=}\qquad\times(\A_{13}^{E''_sE''_s})^{r''_0}(\A_{13}^{E''_sE''})^{r''_1}(\A_{13E'}^{\phantom{13E'}E''_s})^{r''_2}(\A_{13}^{E''E''})^{r''_3}\\
&\phantom{=}\qquad\times(\hat{\mathcal{P}}_{13|d+2(r+r'_2+r''_2)+r'_0+r'_1+r'_3+r''_0+r''_1+r''_3}^{[\ell-(r+r'_0+r'_1+r'_2+r'_3+r''_0+r''_1+r''_2+r''_3)]\boldsymbol{e}_1})_{\{E'\}}^{\phantom{\{E'\}}\{E''\}},
}[EqPExtract]
where it is understood that the sets of special indices $\{E'_s\}$ and $\{E''_s\}$ and the remaining sets of indices $\{E'\}$ and $\{E''\}$ are all symmetrized independently.  From the identity
\eqna{
(\hat{\mathcal{P}}_{13|d}^{\ell\boldsymbol{e}_1})_{\{E'\}}^{\phantom{\{E'\}}\{E''\}}&=\A_{13E'_s}^{\phantom{13E'_s}(E''}(\hat{\mathcal{P}}_{13|d+2}^{(\ell-1)\boldsymbol{e}_1})_{\{E'\}}^{\phantom{\{E'\}}\{E''\})}\\
&\phantom{=}\qquad+\frac{\ell-1}{2(-\ell+2-d/2)}\A_{13E'_s(E'}\A_{13}^{(E''E''}(\hat{\mathcal{P}}_{13|d+2}^{(\ell-2)\boldsymbol{e}_1})_{\{E'\})}^{\phantom{\{E'\})}\{E''\})},
}[EqPExtract10]
which corresponds to \eqref{EqPExtract} with $n'_E=1$ and $n''_E=0$, as can be seen directly from \eqref{EqPShift}, it is easy to obtain the recurrence relation
\eqna{
&\mathscr{C}_{n'_E+1,n''_E}^{(d,\ell)}(r,\boldsymbol{r}',\boldsymbol{r}'')\\
&\qquad=\frac{r'_1+1}{\ell-n'_E}\mathscr{C}_{n'_E,n''_E}^{(d,\ell)}(r,\boldsymbol{r}'-\boldsymbol{e}_0+\boldsymbol{e}_1,\boldsymbol{r}'')+\frac{2(r'_3+1)}{\ell-n'_E}\mathscr{C}_{n'_E,n''_E}^{(d,\ell)}(r,\boldsymbol{r}'-\boldsymbol{e}_1+\boldsymbol{e}_3,\boldsymbol{r}'')\\
&\qquad\phantom{=}+\frac{r''_2+1}{\ell-n'_E}\mathscr{C}_{n'_E,n''_E}^{(d,\ell)}(r-1,\boldsymbol{r}',\boldsymbol{r}''+\boldsymbol{e}_2)\\
&\qquad\phantom{=}-\frac{-\ell+r+r'_0+r'_1+r'_2+r'_3+r''_0+r''_1+r''_2+r''_3-1}{\ell-n'_E}\mathscr{C}_{n'_E,n''_E}^{(d,\ell)}(r,\boldsymbol{r}'-\boldsymbol{e}_2,\boldsymbol{r}'')\\
&\qquad\phantom{=}+\frac{(-\ell+r+r'_0+r'_1+r'_2+r'_3+r''_0+r''_1+r''_2+r''_3-2)_2}{2(\ell-n'_E)[-\ell+(r'_0+r'_1+r'_3+r''_0+r''_1+r''_3-2)/2+2-d/2]}\mathscr{C}_{n'_E,n''_E}^{(d,\ell)}(r,\boldsymbol{r}'-\boldsymbol{e}_1,\boldsymbol{r}''-\boldsymbol{e}_3),
}[EqRR]
as well as the analog recurrence relation for double-primed quantities.  With the unique nonvanishing boundary condition $\mathscr{C}_{0,0}^{(d,\ell)}(0,\boldsymbol{0},\boldsymbol{0})=1$, the solution to \eqref{EqRR} and its double-primed analog is given by
\eqna{
\mathscr{C}_{n'_E,n''_E}^{(d,\ell)}(r,\boldsymbol{r}',\boldsymbol{r}'')&=\frac{(-1)^{r+r'_2+r'_3+n'_E+r''_2+r''_3+n''_E}n'_E!n''_E!}{2^{r'_0+r'_3+r''_0+r''_3}r!r'_0!r'_1!r'_2!r'_3!r''_0!r''_1!r''_2!r''_3!}\frac{[(r'_0+r'_1-r'_3+r''_0+r''_1-r''_3)/2]!}{(-\ell)_{n'_E}(-\ell)_{n''_E}}\\
&\phantom{=}\qquad\times(-r'_0-r'_1)_{r''_3}(-r''_0-r''_1)_{r'_3}\frac{(-\ell)_{r+r'_0+r'_1+r'_2+r'_3+r''_0+r''_1+r''_2+r''_3}}{(-\ell+2-d/2)_{(r'_0+r'_1+r'_3+r''_0+r''_1+r''_3)/2}}.
}[EqCC]
This property under extraction of indices will greatly simplify the computation of four-point conformal blocks.

Finally, we remark that the undetermined parts of the projection operators in \eqref{EqPExp}, denoted by $\mathscr{A}_t(d,\ell)$ and $\hat{\mathcal{Q}}_{13|t}^{\boldsymbol{N}_m+\ell_t\boldsymbol{e}_1}$, are fixed by the knowledge of the specific irreducible representation of the exchanged quasi-primary operator under consideration.  This mirrors the analysis of the tensor structures.  By analogy, we also refer to them as the special parts of the projection operators and fix them once we consider specific examples of four-point conformal blocks.  It turns out that we can define a convenient diagrammatic notation that would allow us to easily enumerate the various terms arising from the index separation.  We discuss this next.


\subsubsection{Diagrammatic Notation}

The extraction of indices \eqref{EqPExtract} leads to specific partitions of $n'_E$ and $n''_E$, given by
\eqn{n'_E=r+2r'_0+r'_1+r'_2\qquad\text{and}\qquad n''_E=r+2r''_0+r''_1+r''_2,}
respectively.  There is also the extra condition $r'_0+r'_1+r'_3=r''_0+r''_1+r''_3$, where the maximum values for $r'_3$ and $r''_3$ are $r'_3\leq r''_0+r''_1$ and $r''_3\leq r'_0+r'_1$.  We introduce here a bookkeeping technique to easily generate the appropriate partitions of $n'_E$ and $n''_E$ that appear in the computation of four-point conformal blocks.

To proceed, let us symbolize the shifted projection operator \eqref{EqPShift} by the vertex
\eqn{(\hat{\mathcal{P}}_{13|d}^{\ell\boldsymbol{e}_1})_{\{E'\}}^{\phantom{\{E'\}}\{E''\}}=\Diag{1.5}{0}{0}{0}{0}{0}{0}{0}.}[EqPDiag]
Here the solid line represents the metrics of the form $\A_{13E'E'}$; the dotted line represents the metrics of the form $\A_{13}^{E''E''}$; and the dashed line represents the metrics of the form $\A_{13E'}^{\phantom{13E'}E''}$.  The chosen convention sets the $E'E'$-line as a solid line, the $E''E''$-line as a dotted line, and the hybrid $E'E''$-line as a hybrid dashed line.

We are now interested in extracting $n'_E$ $E'$-indices and $n''_E$ $E''$-indices, which are all denoted by the subscript $s$ in \eqref{EqPExtract}.  These are the special indices that do not contract with the $\ell$-dependent parts of the tensor structures \eqref{EqTS}, in contrast to the non-special indices in the shifted projection operator \eqref{EqPExtract}, which are denoted by $E'$ and $E''$ and are contracted with the known $\ell$-dependent parts of the tensor structures \eqref{EqTS}.

With these conventions, the extraction of indices can proceed as follows.  On the one hand, any special $E'_s$-index can be extracted from the $E'E'$-line or the $E'E''$-line, resulting in metrics of the form $\A_{13E'_sE'}$ and $\A_{13E'_s}^{\phantom{13E'_s}E''}$, respectively.  On the other hand, any special $E''_s$-index can be extracted from the $E''E''$-line or the $E'E''$-line, leading to metrics of the form $\A_{13}^{E''_sE''}$ and $\A_{13E'}^{\phantom{13E'}E''_s}$, respectively.  The extraction of an $E'_s$-index ($E''_s$-index) is denoted by an extra external solid (dotted) line emerging from the appropriate line of the original vertex.  Moreover, to account for metrics with two special indices, loops are also allowed on each vertex line.  Thus, metrics of the form $\A_{13E'_sE'_s}$ are represented by solid loops on the $E'E'$-line, metrics of the form $\A_{13}^{E''_sE''_s}$ by dotted loops on the $E''E''$-line, and lastly, metrics of the form $\A_{13E'_s}^{\phantom{13E'_s}E''_s}$ by dashed loops on the $E'E''$-line.  The partitions of $n'_E$ and $n''_E$ are therefore constructed by dressing the three lines of the original vertex \eqref{EqPDiag} with external lines such that the number of external solid (dotted) lines add up to $n'_E$ ($n''_E$).  Here, the external loops count for two lines, with the dashed loops counting for one solid and one dotted line each.  In addition, due to the extra condition, one diagram can represent several extended partitions (extended partitions include also $r'_3$ and $r''_3$, respectively).  From the extra condition, the number of extended partitions per diagram is given by $\text{min}\{r'_0+r'_1,r''_0+r''_1\}+1$, \textit{i.e.}\ the minimum value between the number of solid external lines and loops on the vertex solid line and the number of dotted external lines and loops on the vertex dotted line, plus one.  The number of extended partitions is encoded in prefactors in front of each diagram.

Hence, the partitions of interest for the identity \eqref{EqPExtract10} are easily obtained diagrammatically, with the diagrammatic equation for the identity \eqref{EqPExtract10} written as
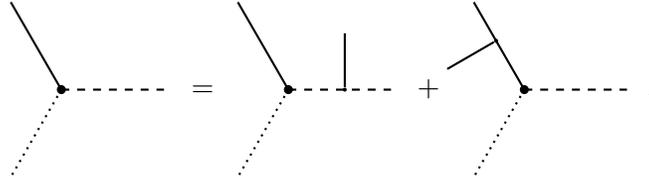
\eqn{\Diag{1.5}{0}{0}{0}{0}{0}{0}{0}=\Diag{1.5}{0}{0}{0}{1}{0}{0}{0}+\Diag{1.5}{0}{0}{1}{0}{0}{0}{0}.}
The first diagram on the RHS represents the partitions $n'_E=0+2\times0+0+1=1$ and $n''_E=0+2\times0+0+0=0$ and corresponds to the first term on the RHS of \eqref{EqPExtract10} while the second diagram on the RHS represents the partitions $n'_E=0+2\times0+1+0=1$ and $n''_E=0+2\times0+0+0=0$ and corresponds to the second term on the RHS of \eqref{EqPExtract10}.  The allowed values of $r'_3$ and $r''_3$ are obtained from the extra condition, which can be computed from the diagrams by counting the number of external lines and loops on the original solid and dotted lines, respectively, showing that both diagrams correspond to only one term each, as in \eqref{EqPExtract10}.  Finally, the associated coefficients are computed directly with \eqref{EqCC}.

As a more complicated example, the diagrammatic equation for $n'_E=2$ and $n''_E=1$ is given by
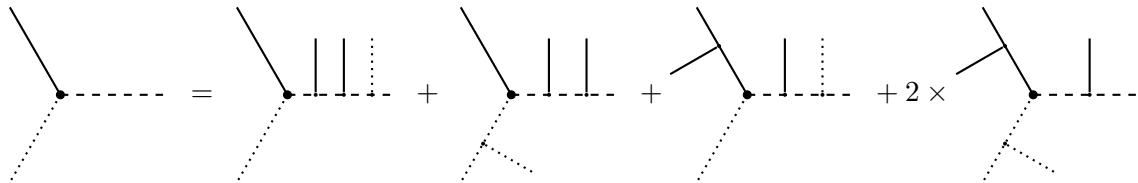
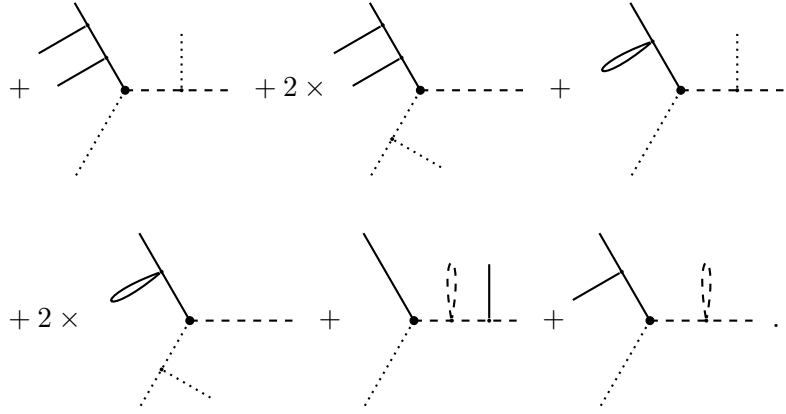
\eqna{
\Diag{1.5}{0}{0}{0}{0}{0}{0}{0}&=\Diag{1.5}{0}{0}{0}{2}{0}{0}{1}+\Diag{1.5}{0}{0}{0}{2}{0}{1}{0}+\Diag{1.5}{0}{0}{1}{1}{0}{0}{1}+2\times\Diag{1.5}{0}{0}{1}{1}{0}{1}{0}\\
&\phantom{=}\qquad+\Diag{1.5}{0}{0}{2}{0}{0}{0}{1}+2\times\Diag{1.5}{0}{0}{2}{0}{0}{1}{0}+\Diag{1.5}{0}{1}{0}{0}{0}{0}{1}\\
&\phantom{=}\qquad+2\times\Diag{1.5}{0}{1}{0}{0}{0}{1}{0}+\Diag{1.5}{1}{0}{0}{1}{0}{0}{0}+\Diag{1.5}{1}{0}{1}{0}{0}{0}{0}.
}
For each diagram, counting the total number of external lines and loops on the original solid line as well as the original dotted line shows that all the diagrams correspond to one term in \eqref{EqPExtract}, apart from
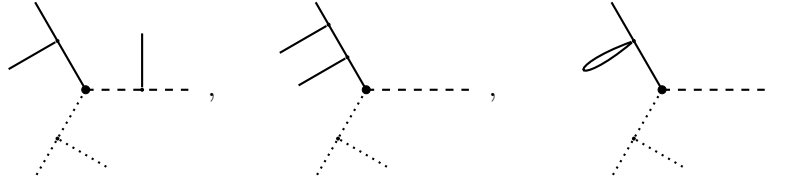
\eqn{\Diag{1.5}{0}{0}{1}{1}{0}{1}{0},\qquad\Diag{1.5}{0}{0}{2}{0}{0}{1}{0},\qquad\Diag{1.5}{0}{1}{0}{0}{0}{1}{0},}
which correspond to two terms each, with $(r'_3,r''_3)\in\{(0,0),(1,1)\}$, $(r'_3,r''_3)\in\{(0,1),(1,2)\}$, and $(r'_3,r''_3)\in\{(0,0),(1,1)\}$, respectively.  Again, \eqref{EqCC} gives the appropriate coefficients for each term.  Having established some convenient notation for the index separation, we next turn to the determination of the rotation matrices.


\section{Three-Point Functions and Rotation Matrices}\label{SecRM}

In this section, we use the tensor structure basis introduced above to compute rotation matrices.  Initial results imply $\ell$-dependent sums that must eventually be re-summed, and we show how this can be done in all generality with the help of an identity for hypergeometric functions.


\subsection{Three-Point Conformal Blocks}

From the tensor structures \eqref{EqTS}, the three-point conformal blocks \eqref{Eq3ptCBind} can be expressed as
\eqna{
(\mathscr{G}_{(a|}^{ij|m+\ell})_{\{aA\}\{bB\}\{eE\}}&=\lambda_{\boldsymbol{N}_{m+\ell}}\left((\A_{321E}^{\phantom{321E}E'})^{n_v^m+\ell}(\bar{\eta}_3\cdot\Gamma\,\bar{\eta}_2\cdot\Gamma)_e^{\phantom{e}e'}\right)_{cs_3}(\tCF{a}{i}{j}{,m+\ell}{1}{2})_{\{aA\}\{bB\}\{e'E'\}\{F\}}\\
&=\lambda_{\boldsymbol{N}_{m+\ell}}(\tCF{a}{i}{j}{,m+i_a}{1}{2})_{\{aA\}\{bB\}\{e'E'\}\{F\}}\\
&\phantom{=}\qquad\times\left((\A_{321E}^{\phantom{321E}E'})^{n_v^m+i_a}(\A_{321E}^{\phantom{321E}E'})^{\ell-i_a}(\bar{\eta}_3\cdot\Gamma\,\bar{\eta}_2\cdot\Gamma)_e^{\phantom{e}e'}\right)_{cs_3}(\A_{12E'F})^{\ell-i_a}\\
&=\lambda_{\boldsymbol{N}_{m+\ell}}(\tCF{a}{i}{j}{,m+i_a}{1}{2})_{\{aA\}\{bB\}\{e'E'\}\{F\}}\\
&\phantom{=}\qquad\times\left((\A_{321E}^{\phantom{321E}E'})^{n_v^m+i_a}(\bar{\eta}_3\cdot\Gamma\,\bar{\eta}_2\cdot\Gamma)_e^{\phantom{e}e'}(\A_{321E}^{\phantom{321E}E'})^{\ell-i_a}\right)_{cs_3}(-\bar{\eta}_{2E'}\bar{\eta}_{1F})^{\ell-i_a},
}
since
\eqn{\A_{12E'F}=g_{E'F}-\bar{\eta}_{1E'}\bar{\eta}_{2F}-\bar{\eta}_{2E'}\bar{\eta}_{1F}\to-\bar{\eta}_{2E'}\bar{\eta}_{1F}.}[EqA12EpF]
It is straightforward to see that this simplification is true.  Indeed, the contraction of the term $\bar{\eta}_{1E'}\bar{\eta}_{2F}$ vanishes straightforwardly, while the contraction of $g_{E'F}$ vanishes from the definition of the conformal substitution cast in terms of the OPE differential operator as $\D_{12}^{(d,h,n)F^n}=\ee{1}{2}{-\frac{n}{2}}\D_{12}^{2(h+n)}(\eta_2^F)^n$ \cite{Fortin:2019dnq}.

Now, owing to the contraction with the half-projector $(\mathcal{T}_{31}^{\boldsymbol{N}_m}\Gamma)$ in \eqref{Eq3ptind}, it is self-evident that the $\A_{321}$-metrics can be simplified to
\eqn{\A_{321E}^{\phantom{321E}E'}=g_E^{\phantom{E}E'}-\bar{\eta}_{1E}\bar{\eta}_2^{E'}-\bar{\eta}_{2E}\bar{\eta}_3^{E'}+\bar{\eta}_{2E}\bar{\eta}_2^{E'}\to g_E^{\phantom{E}E'}-\bar{\eta}_{2E}\bar{\eta}_3^{E'}+\bar{\eta}_{2E}\bar{\eta}_2^{E'}.}
Therefore, inside the three-point conformal blocks we can expand these as
\eqna{
(\A_{321E}^{\phantom{321E}E'})^{n_v^m+i_a}&=\sum_\sigma\sum_{r_0,r_3\geq0}\binom{n_v^m+i_a}{r_0+r_3}\binom{r_0+r_3}{r_3}\frac{(-1)^{r_3}}{(n_v^m+i_a)!}g_{E_{\sigma(1)}}^{\phantom{E_{\sigma(1)}}{E'_{\sigma(1)}}}\cdots g_{E_{\sigma(r_0)}}^{\phantom{E_{\sigma(r_0)}}{E'_{\sigma(r_0)}}}\\
&\phantom{=}\qquad\times\bar{\eta}_{2E_{\sigma(r_0+1)}}\bar{\eta}_3^{E'_{\sigma(r_0+1)}}\cdots\bar{\eta}_{2E_{\sigma(r_0+r_3)}}\bar{\eta}_3^{E'_{\sigma(r_0+r_3)}}\\
&\phantom{=}\qquad\times\bar{\eta}_{2E_{\sigma(r_0+r_3+1)}}\bar{\eta}_2^{E'_{\sigma(r_0+r_3+1)}}\cdots\bar{\eta}_{2E_{\sigma(n_v^m+i_a)}}\bar{\eta}_2^{E'_{\sigma(n_v^m+i_a)}},\\
(\A_{321})^{\ell-i_a}&=\sum_{t_0,t_3\geq0}\binom{\ell-i_a}{t_0+t_3}\binom{t_0+t_3}{t_3}(-1)^{t_3}(g_E^{\phantom{E}E'})^{t_0}(\bar{\eta}_{2E}\bar{\eta}_3^{E'})^{t_3}(\bar{\eta}_{2E}\bar{\eta}_2^{E'})^{\ell-i_a-t_0-t_3}.
}
In the first group above, the indices are not necessarily symmetrized by the special part of the tensor structure.  Hence, the expansion must take into account the different indices, which forces a sum over all the permutations $\sigma$ of the $n_v^m+i_a$ pairs of indices.  In the second group, the indices are all already symmetrized from their contraction with the half-projector on one side and the known $\ell$-dependent part of the tensor structure $(-\bar{\eta}_{2E'}\bar{\eta}_{1F})^{\ell-i_a}$ on the other, thus simplifying the expansion by allowing all the indices to be treated on an equal footing.

Proceeding with the conformal substitution \eqref{EqJ3} by simply counting the appropriate powers, the three-point conformal blocks become
\eqna{
&(\mathscr{G}_{(a|}^{ij|m+\ell})_{\{aA\}\{bB\}\{eE\}}\\
&\qquad=\lambda_{\boldsymbol{N}_{m+\ell}}(\tCF{a}{i}{j}{,m+i_a}{1}{2})_{\{aA\}\{bB\}\{e'E'\}\{F\}}(\bar{\eta}_3\cdot\Gamma\Gamma_F)_e^{\phantom{e}e'}(-\bar{\eta}_{2E'}\bar{\eta}_{1F})^{\ell_a}\\
&\qquad\phantom{=}\times\sum_\sigma\sum_{r_0,r_3,t_0,t_3\geq0}\binom{n_v^{m+i_a}}{r_0+r_3}\binom{r_0+r_3}{r_3}\frac{(-1)^{r_3}}{n_v^{m+i_a}!}\binom{\ell_a}{t_0+t_3}\binom{t_0+t_3}{t_3}(-1)^{t_3}\\
&\qquad\phantom{=}\times g_{E_{\sigma(1)}}^{\phantom{E_{\sigma(1)}}{E'_{\sigma(1)}}}\cdots g_{E_{\sigma(r_0)}}^{\phantom{E_{\sigma(r_0)}}{E'_{\sigma(r_0)}}}\bar{\eta}_3^{E'_{\sigma(r_0+1)}}\cdots\bar{\eta}_3^{E'_{\sigma(r_0+r_3)}}(g_E^{\phantom{E}E'})^{t_0}(\bar{\eta}_3^{E'})^{t_3}\\
&\qquad\phantom{=}\times\bar{I}_{12}^{(d,h'-\ell_a+2t_0+t_3,n'+2\ell_a-2t_0-t_3;p'-t_0)}{}_{E_{\sigma(r_0+1)}\cdots E_{\sigma(n_v^{m+i_a})}E^{\ell-i_a-t_0}}^{E'_{\sigma(r_0+r_3+1)}\cdots E'_{\sigma(n_v^{m+i_a})}(E')^{\ell_a-t_0-t_3}F^{n_a+2\xi_m}},
}
with
\eqn{
\begin{gathered}
h=h_{ij,m+\ell}-2n_v^m-2\xi_m-\ell-n_a/2-i_a,\\
n=2n_v^m+2\xi_m+n_a+2i_a,\\
p=\Delta_{m+\ell}+n_v^m+\ell,\\
n_v^{m+i_a}=n_v^m+i_a,\\
\ell_a=\ell-i_a,
\end{gathered}
}[EqParam3]
and
\eqn{h'=h+2r_0+r_3,\qquad\qquad n'=n-2r_0-r_3,\qquad\qquad p'=p-r_0.}
Contracting the $\ell$-dependent part of the tensor structure by using the contiguous relations \eqref{EqCont3} gives
\eqna{
&(\mathscr{G}_{(a|}^{ij|m+\ell})_{\{aA\}\{bB\}\{eE\}}\\
&\qquad=\lambda_{\boldsymbol{N}_{m+\ell}}(\tCF{a}{i}{j}{,m+i_a}{1}{2})_{\{aA\}\{bB\}\{e'E'\}\{F\}}(\bar{\eta}_3\cdot\Gamma\Gamma_F)_e^{\phantom{e}e'}(-1)^{\ell_a}\\
&\qquad\phantom{=}\times\sum_\sigma\sum_{r_0,r_3,t_0,t_3\geq0}\binom{n_v^{m+i_a}}{r_0+r_3}\binom{r_0+r_3}{r_3}\frac{(-1)^{r_3}}{n_v^{m+i_a}!}\binom{\ell_a}{t_0+t_3}\binom{t_0+t_3}{t_3}(-1)^{t_3}\\
&\qquad\phantom{=}\times\rho^{(d,\ell_a-t_0-t_3;-h-n-\ell_a)}g_{E_{\sigma(1)}}^{\phantom{E_{\sigma(1)}}{E'_{\sigma(1)}}}\cdots g_{E_{\sigma(r_0)}}^{\phantom{E_{\sigma(r_0)}}{E'_{\sigma(r_0)}}}\bar{\eta}_3^{E'_{\sigma(r_0+1)}}\cdots\bar{\eta}_3^{E'_{\sigma(r_0+r_3)}}\\
&\qquad\phantom{=}\times\bar{I}_{12}^{(d,h'+2t_0+t_3,n'-t_0;p'-t_0)}{}_{E_{\sigma(r_0+1)}\cdots E_{\sigma(n_v^{m+i_a})}E^{\ell_a-t_0}}^{E'_{\sigma(r_0+r_3+1)}\cdots E'_{\sigma(n_v^{m+i_a})}F^{n_a-\ell_a+2\xi_m}}(\bar{\eta}_{2E})^{t_0}.
}
We next consider the $\bar{I}$-function.  Denoting all of its (symmetrized) $E'_\sigma$- and $F$-indices by $Z$-indices, we find with the help of \eqref{EqIb3}
\eqn{\bar{I}_{12}^{(d,h'+2t_0+t_3,n'-t_0;p'-t_0)}{}_{E_\sigma^{k_2}E^{\ell_a-t_0}}^{Z^{k_1}}=\sum_{\substack{q_0,q_1,q_2,q_3\geq0\\\bar{q}=n'-t_0}}\widetilde{K}^{(d,h'+2t_0+t_3;p'-t_0;q_0,q_1,q_2,q_3)}S_{(q_0,q_1,q_2,q_3)}{}_{E_\sigma^{k_2}E^{\ell_a-t_0}}^{Z^{k_1}},}
where
\eqn{
\begin{gathered}
k_1=n-n_v^{m+i_a}-\ell_a-r_0-r_3,\\
k_2=n_v^{m+i_a}-r_0.
\end{gathered}
}
We now aim to extract the $E^{\ell_a-t_0}$ indices from $S_{(\boldsymbol{q})}$ to eventually re-sum over $t_0$ and $t_3$, which are both $\ell$-dependent sums, \textit{i.e.}\ both sums grow as $\ell$ grows.  This will simplify the computation of rotation matrices by replacing $\ell$-dependent summations by $\ell$-independent ones.

To begin with, we note that all the $E$-indices (including the $E_\sigma$-indices) in $S_{(\boldsymbol{q})}$ must be carried by either metrics or $\bar{\eta}_2$'s, due to their contraction with the half-projector in \eqref{Eq3ptind}.  Moreover, since there cannot be $g_{EE}$'s in $S_{(\boldsymbol{q})}$ due to the tracelessness condition of the same half-projector, there is a minimum number of $\bar{\eta}_{2E}$'s in $S_{(\boldsymbol{q})}$ which is given by the absolute value of the number of $E$-indices minus the number of $Z$-indices, \textit{i.e.}\ $|\ell_a-t_0+k_2-k_1|$.  Moreover, from the following identities (see Section 4 of \cite{Fortin:2019dnq})
\eqna{
S_{(\boldsymbol{q})}^{A_1\cdots A_{\bar{q}}}&=\frac{2q_0}{\bar{q}}g^{A_{\bar{q}}(A_1}S_{(\boldsymbol{q}-\boldsymbol{e}_0)}^{A_2\cdots A_{\bar{q}-1})}+\sum_{r\neq0}\frac{q_r}{\bar{q}}\bar{\eta}_r^{A_{\bar{q}}}S_{(\boldsymbol{q}-\boldsymbol{e}_r)}^{A_1\cdots A_{\bar{q}-1}}\\
&=\frac{2q_0}{\bar{q}(\bar{q}-1)}g^{A_{\bar{q}}A_{\bar{q}-1}}S_{(\boldsymbol{q}-\boldsymbol{e}_0)}^{A_1\cdots A_{\bar{q}-2}}+\frac{4q_0(q_0-1)}{\bar{q}(\bar{q}-1)}g^{A_{\bar{q}}(A_1}g^{|A_{\bar{q}-1}|A_2}S_{(\boldsymbol{q}-2\boldsymbol{e}_0)}^{A_3\cdots A_{\bar{q}-2})}\\
&\phantom{=}\qquad+\sum_{r\neq0}\frac{4q_0q_r}{\bar{q}(\bar{q}-1)}\bar{\eta}_r^{(A_{\bar{q}}}g^{A_{\bar{q}-1})(A_1}S_{(\boldsymbol{q}-\boldsymbol{e}_0-\boldsymbol{e}_r)}^{A_2\cdots A_{\bar{q}-2})}+\sum_{r,s\neq0}\frac{q_r(q_s-\delta_{rs})}{\bar{q}(\bar{q}-1)}\bar{\eta}_r^{A_{\bar{q}}}\bar{\eta}_s^{A_{\bar{q}-1}}S_{(\boldsymbol{q}-\boldsymbol{e}_r-\boldsymbol{e}_s)}^{A_1\cdots A_{\bar{q}-2}},
}
it is easy by recurrence to extract all the symmetrized $E^{\ell_a-t_0}$ indices, such as
\eqna{
S_{(q_0,q_1,q_2,q_3)}{}_{E_\sigma^{k_2}E^{\ell_a-t_0}}^{Z^{k_1}}&=\sum_{s_0\geq0}\binom{\ell_a-t_0}{s_0}\frac{(-2)^{s_0}(-q_0)_{s_0}(-q_2)_{\ell_a-s_0-t_0}}{(k_1+1)_{\ell_a+k_2-t_0}}(k_1-s_0+1)_{k_2}\\
&\phantom{=}\qquad\times(g_E^{(Z})^{s_0}S_{(q_0-s_0,q_1,q_2-\ell_a+s_0+t_0,q_3)E_\sigma^{k_2}}^{Z^{k_1-s_0})}(-\bar{\eta}_{2E})^{\ell_a-s_0-t_0}.
}
Next, shifting $q_0\to q_0+s_0$ and $q_2\to q_2+\ell_a-s_0-t_0$ leads to
\eqna{
&\bar{I}_{12}^{(d,h'+2t_0+t_3,n'-t_0;p'-t_0)}{}_{E_\sigma^{k_2}E^{\ell_a-t_0}}^{Z^{k_1}}\\
&\qquad=\sum_{s_0\geq0}\sum_{\substack{q_0,q_1,q_2,q_3\geq0\\\bar{q}=n'-\ell_a-s_0}}\binom{\ell_a-t_0}{s_0}\frac{(-2)^{s_0}(-q_0-s_0)_{s_0}(-q_2-\ell_a+s_0+t_0)_{\ell_a-s_0-t_0}}{(k_1+1)_{\ell_a+k_2-t_0}}(k_1-s_0+1)_{k_2}\\
&\qquad\phantom{=}\times\widetilde{K}_{\bar{q}=n'-t_0}^{(d,h'+2t_0+t_3;p'-t_0;q_0+s_0,q_1,q_2+\ell_a-s_0-t_0,q_3)}(g_E^{(Z})^{s_0}S_{(q_0,q_1,q_2,q_3)E_\sigma^{k_2}}^{Z^{k_1-s_0})}(-\bar{\eta}_{2E})^{\ell_a-s_0-t_0},
}
where all the dependence on $t_0$ and $t_3$ in the tensorial components has been removed, and the subscript on the $\widetilde{K}$-function is to remind us that its value of $\bar{q}$ is not the same as the one for $S_{(\boldsymbol{q})}$ anymore.  With this identity, the three-point conformal blocks then assume the form
\eqna{
&(\mathscr{G}_{(a|}^{ij|m+\ell})_{\{aA\}\{bB\}\{eE\}}\\
&\qquad=\lambda_{\boldsymbol{N}_{m+\ell}}(\tCF{a}{i}{j}{,m+i_a}{1}{2})_{\{aA\}\{bB\}\{e'E'\}\{F\}}(\bar{\eta}_3\cdot\Gamma\Gamma_F)_e^{\phantom{e}e'}\\
&\qquad\phantom{=}\times\sum_\sigma\sum_{r_0,r_3,s_0,t_0,t_3\geq0}\frac{(-1)^{\ell_a+r_3+t_0+t_3}}{n_v^{m+i_a}!}\binom{n_v^{m+i_a}}{r_0+r_3}\binom{r_0+r_3}{r_3}\binom{\ell_a}{t_0+t_3}\binom{t_0+t_3}{t_3}\\
&\qquad\phantom{=}\times\sum_{\substack{q_0,q_1,q_2,q_3\geq0\\\bar{q}=n'-\ell_a-s_0}}\binom{\ell_a-t_0}{s_0}\frac{(-2)^{s_0}(-q_0-s_0)_{s_0}(-q_2-\ell_a+s_0+t_0)_{\ell_a-s_0-t_0}}{(k_1+1)_{\ell_a+k_2-t_0}}(k_1-s_0+1)_{k_2}\\
&\qquad\phantom{=}\times\rho^{(d,\ell_a-t_0-t_3;-h-n-\ell_a)}\widetilde{K}_{\bar{q}=n'-t_0}^{(d,h'+2t_0+t_3;p'-t_0;q_0+s_0,q_1,q_2+\ell_a-s_0-t_0,q_3)}\\
&\qquad\phantom{=}\times g_{E_{\sigma(1)}}^{\phantom{E_{\sigma(1)}}{E'_{\sigma(1)}}}\cdots g_{E_{\sigma(r_0)}}^{\phantom{E_{\sigma(r_0)}}{E'_{\sigma(r_0)}}}\bar{\eta}_3^{E'_{\sigma(r_0+1)}}\cdots\bar{\eta}_3^{E'_{\sigma(r_0+r_3)}}(g_E^{(Z})^{s_0}S_{(q_0,q_1,q_2,q_3)E_\sigma^{k_2}}^{Z^{k_1-s_0})}(-\bar{\eta}_{2E})^{\ell_a-s_0}\\
&\qquad=\lambda_{\boldsymbol{N}_{m+\ell}}(\tCF{a}{i}{j}{,m+i_a}{1}{2})_{\{aA\}\{bB\}\{e'E'\}\{F\}}(\bar{\eta}_3\cdot\Gamma\Gamma_F)_e^{\phantom{e}e'}\\
&\qquad\phantom{=}\times\sum_\sigma\sum_{r_0,r_3,s_0\geq0}\frac{(-1)^{\ell_a+r_3}(-2)^{\ell_a+2s_0}}{n_v^{m+i_a}!s_0!}\binom{n_v^{m+i_a}}{r_0+r_3}\binom{r_0+r_3}{r_3}\frac{(-n')_{\ell_a+s_0}(-h-n)_{\ell_a+s_0}}{(k_1+1)_{\ell_a+k_2}}(k_1-s_0+1)_{k_2}\\
&\qquad\phantom{=}\times\sum_{\substack{q_0,q_1,q_2,q_3\geq0\\\bar{q}=n'-\ell_a-s_0}}\frac{(-q_0-s_0)_{s_0}(-q_2-\ell_a+s_0)_{\ell_a-s_0}}{(p'+h+n-s_0-q_0-q_1)_{-\ell_a-s_0}}\rho^{(d,\ell_a;-h-n-\ell_a)}\widetilde{K}_{\bar{q}=n'-\ell_a-s_0}^{(d,h';p';q_0+s_0,q_1,q_2+\ell_a-s_0,q_3)}\Sigma_t\\
&\qquad\phantom{=}\times g_{E_{\sigma(1)}}^{\phantom{E_{\sigma(1)}}{E'_{\sigma(1)}}}\cdots g_{E_{\sigma(r_0)}}^{\phantom{E_{\sigma(r_0)}}{E'_{\sigma(r_0)}}}\bar{\eta}_3^{E'_{\sigma(r_0+1)}}\cdots\bar{\eta}_3^{E'_{\sigma(r_0+r_3)}}(g_E^{(Z})^{s_0}S_{(q_0,q_1,q_2,q_3)E_\sigma^{k_2}}^{Z^{k_1-s_0})}(-\bar{\eta}_{2E})^{\ell_a-s_0},
}
where we have used the definitions \eqref{EqK3}.  Here the sums over $t_0$ and $t_3$ are included in $\Sigma_t$, which is given explicitly by
\eqna{
\Sigma_t&=\sum_{t_0,t_3\geq0}(-1)^{t_0+t_3}s_0!\binom{\ell_a}{t_0+t_3}\binom{t_0+t_3}{t_3}\binom{\ell_a-t_0}{s_0}\\
&\phantom{=}\qquad\times\frac{(-p'+1)_{t_0}(p'+h+n-s_0-q_0-q_1)_{t_3}(p'+h'+1-d/2)_{t_0+t_3}}{(h'+\ell_a+q_0+q_2+1)_{t_0+t_3}(h+n+d/2)_{t_0+t_3}}.
}
Equipped with this result, we can finally transform the $\ell$-dependent sums over $t_0$ and $t_3$ in $\Sigma_t$ into $\ell$-independent sums.\footnote{Clearly, from the binomial coefficient $\binom{\ell_a}{t_0+t_3}$, the sums over $t_0$ and $t_3$ stop at $\ell_a$, which grows like $\ell$ for large $\ell$.}

First of all, we shift $t_0\to t_0-t_3$ and rewrite the sum over $t_3$ as a hypergeometric function, which leads to
\eqna{
\Sigma_t&=\sum_{t_0\geq0}\binom{\ell_a}{t_0}\frac{(\ell_a-s_0-t_0+1)_{s_0}(p'-t_0)_{t_0}(p'+h'+1-d/2)_{t_0}}{(h'+\ell_a+q_0+q_2+1)_{t_0}(h+n+d/2)_{t_0}}\\
&\phantom{=}\qquad\times{}_3F_2\left[\begin{array}{c}p'+h+n-s_0-q_0-q_1,\ell_a-t_0+1,-t_0\\p'-t_0,\ell_a-s_0-t_0+1\end{array};1\right].
}
We then use the well known identity\footnote{For $n$ and $m$ nonnegative integers and $\alpha$, $\beta$ and $\gamma$ arbitrary complex numbers.}
\eqn{{}_3F_2\left[\begin{array}{c}\alpha,\beta,-n\\\gamma,\beta-m\end{array};1\right]=\frac{(\gamma-\alpha)_n}{(\gamma)_n}{}_3F_2\left[\begin{array}{c}\alpha,-m,-n\\\alpha-\gamma-n+1,\beta-m\end{array};1\right],}[Eq3F2]
to transform the first $\ell$-dependent sum,
\eqna{
\Sigma_t&=\sum_{t_0,t\geq0}\binom{\ell_a}{t_0}\frac{(\ell_a-s_0-t_0+1)_{s_0}(-h-n+s_0-t_0+q_0+q_1)_{t_0}(p'+h'+1-d/2)_{t_0}}{(h'+\ell_a+q_0+q_2+1)_{t_0}(h+n+d/2)_{t_0}}\\
&\phantom{=}\qquad\times\frac{(p'+h+n-s_0-q_0-q_1)_t(-s_0)_t(-t_0)_t}{(h+n-s_0-q_0-q_1+1)_t(\ell_a-s_0-t_0+1)_tt!},
}
where the index of summation of the new hypergeometric function is $t$.  To further transform the sum over $t_0$, we now shift $t_0\to t_0+t$ and re-express the sum as a hypergeometric function such that
\eqna{
\Sigma_t&=(-1)^{s_0}(-\ell_a)_{s_0}\sum_{t\geq0}\frac{(-s_0)_t(p'+h+n-s_0-q_0-q_1)_t(p'+h'+1-d/2)_t}{(h'+\ell_a+q_0+q_2+1)_t(h+n+d/2)_tt!}\\
&\phantom{=}\qquad\times{}_3F_2\left[\begin{array}{c}p'+h'+t+1-d/2,h+n-s_0+t-q_0-q_1+1,-\ell_a+s_0\\h+n+t+d/2,h'+\ell_a+t+q_0+q_2+1\end{array};1\right].
}
We next apply the identity \eqref{Eq3F2} once again.  This leads to
\eqna{
\Sigma_t&=(-1)^{s_0}(-\ell_a)_{s_0}\sum_{s_3,t\geq0}\frac{(-s_0)_t(p'+h+n-s_0-q_0-q_1)_t(p'+h'+1-d/2)_t}{(h'+\ell_a+q_0+q_2+1)_t(h+n+d/2)_tt!}\\
&\phantom{=}\qquad\times\frac{(-p'+n'-1+d)_{\ell_a-s_0}}{(h+n+t+d/2)_{\ell_a-s_0}}\frac{(p'+h'+t+1-d/2)_{s_3}(-q_3)_{s_3}(-\ell_a+s_0)_{s_3}}{(p'-n'-\ell_a+s_0+2-d)_{s_3}(h'+\ell_a+t+q_0+q_2+1)_{s_3}s_3!},
}[EqSigma]
where the index of summation of the new hypergeometric function was chosen to be $s_3$.  At this point, the two $\ell$-dependent sums have been transformed into two $\ell$-independent sums, and we can return to the three-point conformal blocks.

Now, inserting \eqref{EqSigma} in the three-point conformal blocks, shifting $q_3\to q_3+s_3$ using the fact that
\eqn{S_{(q_0,q_1,q_2,q_3+s_3)}{}_{E_\sigma^{k_2}}^{Z^{k_1-s_0}}=\frac{(k_1-s_0-s_3+1)_{k_2}}{(k_1-s_0+1)_{k_2}}(\bar{\eta}_3^{(Z})^{s_3}S_{(q_0,q_1,q_2,q_3)}{}_{E_\sigma^{k_2}}^{Z^{k_1-s_0-s_3})},}
since all $\bar{\eta}_3$'s must have $Z$-indices only, and finally re-summing the $\boldsymbol{q}$'s into an $\bar{I}$-function \eqref{EqIb3}, our final result becomes
\eqna{
&(\mathscr{G}_{(a|}^{ij|m+\ell})_{\{aA\}\{bB\}\{eE\}}\\
&\qquad=\lambda_{\boldsymbol{N}_{m+\ell}}(\tCF{a}{i}{j}{,m+i_a}{1}{2})_{\{aA\}\{bB\}\{e'E'\}\{F\}}(\bar{\eta}_3\cdot\Gamma\Gamma_F)_e^{\phantom{e}e'}\\
&\qquad\phantom{=}\times\sum_{r_0,r_3,s_0,s_3,t\geq0}\frac{(-1)^{r_0+s_0}(-2)^{\ell_a+s_0-t}}{n_v^{m+i_a}!r_0!r_3!s_0!s_3!t!}(-n_v^{m+i_a})_{r_0+r_3}(-\ell_a)_{s_0+s_3}(-s_0)_t\\
&\qquad\phantom{=}\times(-n+n_v^{m+i_a}+\ell_a+r_0+r_3)_{s_0+s_3}(-h-n-\ell_a)_{\ell_a+s_0-t}\\
&\qquad\phantom{=}\times(-h-n-\ell_a+1-d/2)_{s_0-t}(p-n-\ell_a+r_0+r_3+s_0+s_3+2-d)_{\ell_a-s_0-s_3}\\
&\qquad\phantom{=}\times\sum_\sigma g_{E_{\sigma(1)}}^{\phantom{E_{\sigma(1)}}{E'_{\sigma(1)}}}\cdots g_{E_{\sigma(r_0)}}^{\phantom{E_{\sigma(r_0)}}{E'_{\sigma(r_0)}}}\bar{\eta}_3^{E'_{\sigma(r_0+1)}}\cdots\bar{\eta}_3^{E'_{\sigma(r_0+r_3)}}(g_E^{(Z})^{s_0}(\bar{\eta}_3^Z)^{s_3}\\
&\qquad\phantom{=}\times\bar{I}_{12}^{(d+2\ell_a,h+\ell_a+2r_0+r_3+s_3+t,n-\ell_a-2r_0-r_3-s_0-s_3;p-r_0)}{}_{E_\sigma^{n_v^{m+i_a}-r_0}}^{Z^{n-n_v^{m+i_a}-\ell_a-r_0-r_3-s_0-s_3})}(-\bar{\eta}_{2E})^{\ell_a-s_0},
}[EqCB3]
where the $Z$-indices belong to
\eqn{Z\in\{E'_{\sigma(r_0+r_3+1)},\ldots,E'_{\sigma(n_v^{m+i_a})},F^{n_a-\ell_a+2\xi_m}\},}
and the different parameters are defined in \eqref{EqParam3}.  Here, the notation was chosen such that $r_0$ and $r_3$ represent the number of non-symmetrized free metrics and $\bar{\eta}_3$'s, respectively, while $s_0$ and $s_3$ represent the number of free symmetrized (\textit{i.e.}\ with $Z$-indices) metrics and $\bar{\eta}_3$'s, respectively.

Clearly, from \eqref{EqCB3}, we have the following bounds on the different indices of summation:
\eqn{0\leq r_0+r_3\leq n_v^{m+i_a},\qquad\qquad0\leq s_0+s_3\leq n_v^{m+i_a}+2\xi_m+n_a-\ell_a-r_0-r_3,\qquad\qquad0\leq t\leq s_0,}
which are all $\ell$-independent, as desired.


\subsection{Rotation Matrix}

It is now straightforward to find the rotation matrix from \eqref{EqCB3}.  First, we contract the remaining $\A_{12E'F}\to-\bar{\eta}_{2E'}\bar{\eta}_{1F}$ [see \eqref{EqA12EpF}] from the special part of the tensor structure $\tCF{a}{i}{j}{,m+i_a}{1}{2}$, using the contiguous relations \eqref{EqCont3} when appropriate.  Then, we simply expand the $\bar{I}$-function and contract the remaining factors from the special part of the tensor structure.  Finally, we replace all free $\bar{\eta}$'s by $\A_{12}\cdot\bar{\eta}_3$ with the appropriate sign [as in \eqref{Eq3pTStoOPETS}].  Equipped with this result, we can determine the rotation matrix from the relation
\eqna{
\mathscr{G}_{(a|}^{ij|m+\ell}&=\sum_{a'=1}^{N_{ij,m+\ell}}(R_{ij,m+\ell}^{-1})_{aa'}\,\bar{\eta}_3\cdot\Gamma\,\FCF{a'}{i}{j}{,m+\ell}{1}{2}(\A_{12},\Gamma_{12},\epsilon_{12};\A_{12}\cdot\bar{\eta}_3)\\
&=\sum_{a'=1}^{N_{ij,m+\ell}}(R_{ij,m+\ell}^{-1})_{aa'}\,\bar{\eta}_3\cdot\Gamma\,\FCF{a'}{i}{j}{,m+i_{a'}}{1}{2}(\A_{12},\Gamma_{12},\epsilon_{12};\A_{12}\cdot\bar{\eta}_3)(\A_{12}\cdot\bar{\eta}_3)^{\ell-i_{a'}},
}[EqRM]
using the symmetry properties of the irreducible representations of the three quasi-primary operators under consideration to match with the three-point tensor structure basis.

It is also possible to first expand \eqref{EqCB3} and then contract with the tensor structure.  With the definition
\eqna{
&{}_a\kappa_{(\boldsymbol{q},r_0,r_3,s_0,s_3,t)}^{ij|m+\ell}\\
&\qquad=\lambda_{\boldsymbol{N}_m+\ell\boldsymbol{e}_1}(-1)^{2\xi_m+n_a-\ell+i_a-r_0-r_3-s_3-q_0-q_1-q_2}(-2)^{h_{ij,m+\ell}+n_a/2-q_0}\\
&\qquad\phantom{=}\times\frac{(2n_v^m+2\xi_m+n_a-\ell+3i_a-2r_0-r_3-s_0-s_3)!}{(n_v^m+i_a)!q_0!q_1!q_2!q_3!r_0!r_3!s_0!s_3!t!}(-\ell+i_a)_{s_0+s_3}\\
&\qquad\phantom{=}\times(-n_v^m-i_a)_{r_0+r_3}(-n_v^m-2\xi_m-n_a+\ell-2i_a+r_0+r_3)_{s_0+s_3}(-s_0)_t\\
&\qquad\phantom{=}\times(-h_{ij,m+\ell}-n_a/2+1-d/2)_{s_0-t}(-h_{ij,m+\ell}-n_a/2)_{\ell-i_a+s_0-t+q_0+q_1+q_3}\\
&\qquad\phantom{=}\times(\Delta_{m+\ell}-n_v^m-2\xi_m-n_a-i_a+r_0+r_3+s_0+s_3+2-d)_{\ell-i_a-s_0-s_3}\\
&\qquad\phantom{=}\times(\Delta_{m+\ell}+n_v^m+\ell-r_0)_{h_{ij,m+\ell}+n_a/2-\ell+i_a-s_0+t-q_0-q_1}\\
&\qquad\phantom{=}\times(\Delta_{m+\ell}+n_v^m+i_a-r_0-q_0-q_1-q_2+1-d/2)_{h_{ij,m+\ell}+n_a/2-\ell+i_a-s_0+t-q_0-q_3},
}[Eqkappa]
the three-point conformal blocks \eqref{EqCB3} take on the form
\eqna{
(\mathscr{G}_{(a|}^{ij|m+\ell})_{\{aA\}\{bB\}\{eE\}}&=(\tCF{a}{i}{j}{,m+i_a}{1}{2})_{\{aA\}\{bB\}\{e'E'\}\{F\}}(\bar{\eta}_3\cdot\Gamma\Gamma_F)_e^{\phantom{e}e'}\\
&\phantom{=}\qquad\times\sum_{r_0,r_3,s_0,s_3,t\geq0}\sum_{q_0,q_1,q_2,q_3\geq0}{}_a\kappa_{(\boldsymbol{q},r_0,r_3,s_0,s_3,t)}^{ij|m+\ell}\\
&\phantom{=}\qquad\times\sum_\sigma g_{E_{\sigma(1)}}^{\phantom{E_{\sigma(1)}}{E'_{\sigma(1)}}}\cdots g_{E_{\sigma(r_0)}}^{\phantom{E_{\sigma(r_0)}}{E'_{\sigma(r_0)}}}\bar{\eta}_3^{E'_{\sigma(r_0+1)}}\cdots\bar{\eta}_3^{E'_{\sigma(r_0+r_3)}}(g_E^{(Z})^{s_0}(\bar{\eta}_3^Z)^{s_3}\\
&\phantom{=}\qquad\times S_{(q_0,q_1,q_2,q_3)}{}_{E_\sigma^{n_v^{m+i_a}-r_0}}^{Z^{n-n_v^{m+i_a}-\ell_a-r_0-r_3-s_0-s_3})}(-\bar{\eta}_{2E})^{\ell_a-s_0},
}
where
\eqn{2q_0+q_1+q_2+q_3=2n_v^m+2\xi_m+n_a-\ell+3i_a-2r_0-r_3-s_0-s_3.}
Thus, each element of the rotation matrix under consideration may be conveniently expressed as a sum of $\kappa$'s \eqref{Eqkappa} with suitable shifts.

We observe that in order to obtain the four-point conformal blocks in the three-point basis, it is also necessary to invert the rotation matrix computed from \eqref{EqRM}.  Despite the fact that the size of the rotation matrices is $\ell$-dependent, it turns out that one can invert them without worrying about their $\ell$-dependent size.

Surprisingly, the determination of the rotation matrix is the most contrived calculation in our quest for the four-point conformal blocks.  We next consider the blocks themselves.


\section{Four-Point Functions and Conformal Blocks}\label{SecCB}

In this section, we use the tensor structures and the projection operators in their general form to compute the most general four-point conformal blocks in terms of the special case-dependent parts.  Just like for the rotation matrices, our goal is to complete the $\ell$-dependent computations and express the final result purely in terms of the special parts of the tensor structures and the projection operators.  With such a result, the determination of the infinite towers of conformal blocks would be reduced to simple manipulations of the $\ell$-independent special parts under consideration.


\subsection{Four-Point Conformal Blocks}

Using the tensor structures \eqref{EqTS} and the projection operators \eqref{EqPExp} in \eqref{Eq4ptCBind} leads to the conformal blocks
\eqna{
&(\mathscr{G}_{(a|b]}^{ij|m+\ell|kl})_{\{aA\}\{bB\}\{cC\}\{dD\}}\\
&\qquad=\sum_t\mathscr{A}_t(d,\ell)(\tOPE{a}{i}{j}{,m+i_a}{1}{2})_{\{aA\}\{bB\}}^{\phantom{\{aA\}\{bB\}}\{Ee\}\{F\}}(-\bar{\eta}_2^E\bar{\eta}_1^F)^{\ell-i_a}\left((-x_3)^{2\xi_m}(\A_{123E}^{\phantom{123E}E'})^{n_v^m+\ell}(\bar{\bar{\eta}}_2\cdot\Gamma)_e^{\phantom{e}e'}\right.\\
&\qquad\phantom{=}\left.\times(\bar{\eta}_3\cdot\Gamma\,\hat{\mathcal{Q}}_{13|t}^{\boldsymbol{N}_m+\ell_t\boldsymbol{e}_1}\hat{\mathcal{P}}_{13|d+d_t}^{(\ell-\ell_t)\boldsymbol{e}_1}\,\bar{\eta}_2\cdot\Gamma)_{e'E'}^{\phantom{e'E'}E''e''}(\FCF{b}{k}{l}{,m+i_b}{3}{4})_{\{cC\}\{dD\}\{e''E''\}}[(\A_{34}\cdot\bar{\bar{\eta}}_2)_{E''}]^{\ell-i_b}\right)_{cs_4}.
}
Here the property \eqref{EqA12EpF}, which is also true for $\A_{12}^{EF}$ by the same logical argument, allowed us to substitute $\A_{12}^{EF}\to-\bar{\eta}_2^E\bar{\eta}_1^F$ in the previous equation.  Clearly, the metrics $\A_{12}^{EF}$ from the special part of the tensor structure can also be simplified with the help of \eqref{EqA12EpF}, \textit{i.e.}\ $\A_{12}^{EF}\to-\bar{\eta}_2^E\bar{\eta}_1^F$.

Our goal now is to manipulate the projection operator such that the symmetrizations on the $\ell$ $E'$- and $E''$-indices may be removed.  Clearly, since the $\ell$-dependent parts of the tensor structures have $\ell-i_a$ $E$-indices and $\ell-i_b$ $E''$-indices symmetrized, respectively, it is only necessary to extract $i_a$ $E'$-indices and $i_b$ $E''$-indices from the symmetrizations.  This is accomplished by a simple double expansion, leading to
\eqna{
&(\bar{\eta}_3\cdot\Gamma\,\hat{\mathcal{Q}}_{13|t}^{\boldsymbol{N}_m+\ell_t\boldsymbol{e}_1}\,\bar{\eta}_2\cdot\Gamma)_{e'{E'}^{n_v^m}({E'}^{\ell_t}}^{\phantom{e'{E'}^{n_v^m}({E'}^{\ell_t}}({E''}^{\ell_t}|{E''}^{n_v^m}e''|}(\hat{\mathcal{P}}_{13|d+d_t}^{(\ell-\ell_t)\boldsymbol{e}_1})_{{E'}^{\ell-\ell_t})}^{\phantom{{E'}^{\ell-\ell_t})}{E''}^{\ell-\ell_t})}\\
&\qquad=\sum_{j_a,j_b\geq0}\binom{i_a}{j_a}\binom{i_b}{j_b}\frac{(-\ell_t)_{i_a-j_a}(-\ell_t)_{i_b-j_b}(-\ell+\ell_t)_{j_a}(-\ell+\ell_t)_{j_b}}{(-\ell)_{i_a}(-\ell)_{i_b}}\\
&\qquad\phantom{=}\times(\bar{\eta}_3\cdot\Gamma\,\hat{\mathcal{Q}}_{13|t}^{\boldsymbol{N}_m+\ell_t\boldsymbol{e}_1}\,\bar{\eta}_2\cdot\Gamma)_{e'{E'}^{n_v^m}({E'}^{\ell_t-i_a+j_a}({E'_s}^{i_a-j_a}}^{\phantom{e'{E'}^{n_v^m}({E'}^{\ell_t-i_a+j_a}({E'_s}^{i_a-j_a}}({E''_s}^{i_b-j_b}({E''}^{\ell_t-i_b+j_b}|{E''}^{n_v^m}e''|}\\
&\qquad\phantom{=}\times\sum_{\substack{r,\boldsymbol{r}',\boldsymbol{r}''\geq0\\r+2r'_0+r'_1+r'_2=j_a\\r+2r''_0+r''_1+r''_2=j_b\\r'_0+r'_1+r'_3=r''_0+r''_1+r''_3}}\mathscr{C}_{j_a,j_b}^{(d+d_t,\ell-\ell_t)}(r,\boldsymbol{r}',\boldsymbol{r}'')(\A_{13E'_s}^{\phantom{13E'_s}E''_s})^r\\
&\qquad\phantom{=}\times(\A_{13E'_sE'_s})^{r'_0}(\A_{13E'_sE'})^{r'_1}(\A_{13E'_s)}^{\phantom{13E'_s)}E''})^{r'_2}(\A_{13E'E'})^{r'_3}\\
&\qquad\phantom{=}\times(\A_{13}^{E''_sE''_s})^{r''_0}(\A_{13}^{E''_sE''})^{r''_1}(\A_{13E'}^{\phantom{13E'}E''_s)})^{r''_2}(\A_{13}^{E''E''})^{r''_3}\\
&\qquad\phantom{=}\times(\hat{\mathcal{P}}_{13|d+d_t+2(r+r'_2+r''_2)+r'_0+r'_1+r'_3+r''_0+r''_1+r''_3}^{[\ell-\ell_t-(r+r'_0+r'_1+r'_2+r'_3+r''_0+r''_1+r''_2+r''_3)]\boldsymbol{e}_1})_{\{E'\})}^{\phantom{\{E'\})}\{E''\})},
}
where we used \eqref{EqPExtract}.  We note here that apart from the indices for $\boldsymbol{N}_m$ appearing in the special parts of the projection operator, all the $i_a$ special $E'_s$-indices are symmetrized together.  The same is true for the $i_b$ special $E''_s$-indices, the $\ell-i_a$ $E'$-indices, and the $\ell-i_b$ $E''$-indices.

Upon substituting this result into the four-point block, we may remove the two explicit symmetrizations on the $\ell-i_a$ $E'$- and $\ell-i_b$ $E''$-indices and contract the $\ell$-dependent part of the three-point tensor structure straightforwardly.  This gives
\eqna{
&(\mathscr{G}_{(a|b]}^{ij|m+\ell|kl})_{\{aA\}\{bB\}\{cC\}\{dD\}}\\
&\qquad=\sum_t\mathscr{A}_t(d,\ell)\sum_{j_a,j_b\geq0}\binom{i_a}{j_a}\binom{i_b}{j_b}\frac{(-\ell_t)_{i_a-j_a}(-\ell_t)_{i_b-j_b}(-\ell+\ell_t)_{j_a}(-\ell+\ell_t)_{j_b}}{(-\ell)_{i_a}(-\ell)_{i_b}}\\
&\qquad\phantom{=}\times\sum_{\substack{r,\boldsymbol{r}',\boldsymbol{r}''\geq0\\r+2r'_0+r'_1+r'_2=j_a\\r+2r''_0+r''_1+r''_2=j_b\\r'_0+r'_1+r'_3=r''_0+r''_1+r''_3}}(-2)^{r''_3}\mathscr{C}_{j_a,j_b}^{(d+d_t,\ell-\ell_t)}(r,\boldsymbol{r}',\boldsymbol{r}'')(\tOPE{a}{i}{j}{,m+i_a}{1}{2})_{\{aA\}\{bB\}}^{\phantom{\{aA\}\{bB\}}\{Ee\}\{F\}}(-\bar{\eta}_2^E\bar{\eta}_1^F)^{\ell-i_a}\\
&\qquad\phantom{=}\times\left((-x_3)^{2\xi_m}(\A_{123(E_s}^{\phantom{123E_s}E''_s})^r(\A_{12E_sE_s})^{r'_0}(\A_{12EE_s})^{r'_1}(-\A_{123}\cdot\bar{\bar{\eta}}_{4E_s})^{r'_2}(\A_{12EE})^{r'_3}\right.\\
&\qquad\phantom{=}\times(\bar{\bar{\eta}}_2\cdot\Gamma\,\bar{\eta}_3\cdot\Gamma\,(\A_{123})^{n_v^m+\ell_t}\hat{\mathcal{Q}}_{13|t}^{\boldsymbol{N}_m+\ell_t\boldsymbol{e}_1}\,\bar{\eta}_2\cdot\Gamma)_{|eE^{n_v^m}E^{\ell_t-i_a+j_a}|E_s^{i_a-j_a})}^{\phantom{|eE^{n_v^m}E^{\ell_t-i_a+j_a}|E_s^{i_a-j_a})}({E''_s}^{i_b-j_b}|{E''}^{\ell_t-i_b+j_b}{E''}^{n_v^m}e''|}\\
&\qquad\phantom{=}\times(\A_{13}^{E''_sE''_s})^{r''_0}(\bar{\bar{\eta}}_2^{E''_s})^{r''_1}(\A_{123E}^{\phantom{123E}E''_s)})^{r''_2}(\FCF{b}{k}{l}{,m+i_b}{3}{4})_{\{cC\}\{dD\}\{e''E''\}}[(\A_{34}\cdot\bar{\bar{\eta}}_2)_{E''}]^{\ell_t-i_b+j_b}\\
&\qquad\phantom{=}\left.\times((\A_{123})^{\ell'}\hat{\mathcal{P}}_{13|d'}^{\ell'\boldsymbol{e}_1}(\A_{34}\cdot\bar{\bar{\eta}}_2)^{\ell'})_{\{E\}}\right)_{cs_4},
}
where
\eqn{
\begin{gathered}
d'=d+d_t+2(r+r'_2+r''_2)+r'_0+r'_1+r'_3+r''_0+r''_1+r''_3,\\
\ell'=\ell-\ell_t-(r+r'_0+r'_1+r'_2+r'_3+r''_0+r''_1+r''_2+r''_3).
\end{gathered}
}[Eqdplp]

Since none of the special $E_s$-indices are contracted with $-\bar{\eta}_2^E\bar{\eta}_1^F$ and none of the special $E''_s$-indices are contracted with $\A_{34}\cdot\bar{\bar{\eta}}_{2E''}$,\footnote{As a reminder, both sets of indices originate from the $\ell\boldsymbol{e}_1$ part of the projection operator for the exchanged quasi-primary operator.  They are considered special because they were not symmetrized with the remaining $\ell\boldsymbol{e}_1$ indices in the tensor structures.  Hence, they cannot possibly contract with the symmetrized part of their respective tensor structures.} we can make the replacements
\eqn{\A_{12E_sE_s}\to g_{E_sE_s},\qquad\qquad\A_{13}^{E''_sE''_s}\to g^{E''_sE''_s}.}
Moreover, all the non-special $\ell$-dependent $E$-indices, \textit{i.e.}\ all $E$-indices (except for the $E^{n_v^m}$ indices on the special part of the projection operator), must contract with $-\bar{\eta}_2^E\bar{\eta}_1^F$.  Hence, we may replace
\eqn{\A_{12EE_s}\to-\bar{\eta}_{1E}\bar{\eta}_{2E_s},\qquad\qquad\A_{12EE}\to-2\bar{\eta}_{1E}\bar{\eta}_{2E}.}
With these simplifications, the four-point conformal blocks assume the form
\eqna{
&(\mathscr{G}_{(a|b]}^{ij|m+\ell|kl})_{\{aA\}\{bB\}\{cC\}\{dD\}}\\
&\qquad=\sum_t\mathscr{A}_t(d,\ell)\sum_{j_a,j_b\geq0}\binom{i_a}{j_a}\binom{i_b}{j_b}\frac{(-\ell_t)_{i_a-j_a}(-\ell_t)_{i_b-j_b}(-\ell+\ell_t)_{j_a}(-\ell+\ell_t)_{j_b}}{(-\ell)_{i_a}(-\ell)_{i_b}}\\
&\qquad\phantom{=}\times\sum_{\substack{r,\boldsymbol{r}',\boldsymbol{r}''\geq0\\r+2r'_0+r'_1+r'_2=j_a\\r+2r''_0+r''_1+r''_2=j_b\\r'_0+r'_1+r'_3=r''_0+r''_1+r''_3}}(-1)^{\ell+2\xi_m-i_a+r'_1+r'_2}(-2)^{r'_3+r''_3}\mathscr{C}_{j_a,j_b}^{(d+d_t,\ell-\ell_t)}(r,\boldsymbol{r}',\boldsymbol{r}'')\\
&\qquad\phantom{=}\times(\tOPE{a}{i}{j}{,m+i_a}{1}{2})_{\{aA\}\{bB\}}^{\phantom{\{aA\}\{bB\}}\{Ee\}\{F\}}(\bar{\eta}_2^E)^{\ell-i_a-r'_1-r'_3}(\bar{\eta}_1^F)^{\ell-i_a}\left((x_3)^{2\xi_m}(x_3x_4)^{\xi_m}\right.\\
&\qquad\phantom{=}\times(\bar{\eta}_2\cdot\Gamma\,\bar{\eta}_3\cdot\Gamma\,(\A_{123})^{n_v^m+\ell_t}\hat{\mathcal{Q}}_{13|t}^{\boldsymbol{N}_m+\ell_t\boldsymbol{e}_1}\,\bar{\eta}_2\cdot\Gamma)_{eE^{n_v^m}E^{\ell_t-i_a+j_a}(E_s^{i_a-j_a}}^{\phantom{eE^{n_v^m}E^{\ell_t-i_a+j_a}(E_s^{i_a-j_a}}({E''_s}^{i_b-j_b}|{E''}^{\ell_t-i_b+j_b}{E''}^{n_v^m}e''|}\\
&\qquad\phantom{=}\times(\A_{123E_s}^{\phantom{123E_s}E''_s})^r(g_{E_sE_s})^{r'_0}(\bar{\eta}_{2E_s})^{r'_1}(\A_{123}\cdot\bar{\bar{\eta}}_{4E_s)})^{r'_2}(\bar{\eta}_{2E})^{r'_3}\\
&\qquad\phantom{=}\times(g^{E''_sE''_s})^{r''_0}(\bar{\bar{\eta}}_2^{E''_s})^{r''_1}(\A_{123E}^{\phantom{123E}E''_s)})^{r''_2}(\FCF{b}{k}{l}{,m+i_b}{3}{4})_{\{cC\}\{dD\}\{e''E''\}}[(\A_{34}\cdot\bar{\bar{\eta}}_2)_{E''}]^{\ell_t-i_b+j_b}\\
&\qquad\phantom{=}\left.\times((\A_{123})^{\ell'}\hat{\mathcal{P}}_{13|d'}^{\ell'\boldsymbol{e}_1}(\A_{34}\cdot\bar{\bar{\eta}}_2)^{\ell'})_{\{E\}}\right)_{cs_4}.
}
We now extract all the allowed factors from the conformal substitution to rewrite the result as
\eqna{
&(\mathscr{G}_{(a|b]}^{ij|m+\ell|kl})_{\{aA\}\{bB\}\{cC\}\{dD\}}\\
&\qquad=\sum_t\mathscr{A}_t(d,\ell)\sum_{j_a,j_b\geq0}\binom{i_a}{j_a}\binom{i_b}{j_b}\frac{(-\ell_t)_{i_a-j_a}(-\ell_t)_{i_b-j_b}(-\ell+\ell_t)_{j_a}(-\ell+\ell_t)_{j_b}}{(-\ell)_{i_a}(-\ell)_{i_b}}\\
&\qquad\phantom{=}\times\sum_{\substack{r,\boldsymbol{r}',\boldsymbol{r}''\geq0\\r+2r'_0+r'_1+r'_2=j_a\\r+2r''_0+r''_1+r''_2=j_b\\r'_0+r'_1+r'_3=r''_0+r''_1+r''_3}}(-1)^{\ell+2\xi_m-i_a+r'_1+r'_2}(-2)^{\ell-\ell_t-j_a-j_b+r+r'_0+r''_0}\mathscr{C}_{j_a,j_b}^{(d+d_t,\ell-\ell_t)}(r,\boldsymbol{r}',\boldsymbol{r}'')\\
&\qquad\phantom{=}\times(\tOPE{a}{i}{j}{,m+i_a}{1}{2})_{\{aA\}\{bB\}}^{\phantom{\{aA\}\{bB\}}\{Ee\}\{F\}}(g_{E_sE_s})^{r'_0}(\bar{\eta}_2^E)^{\ell-i_a-r'_1-r'_3}(\bar{\eta}_1^F)^{\ell-i_a}\\
&\qquad\phantom{=}\times\left((x_3)^{-r'_2+2\xi_m}(x_3x_4)^{(r'_2+r''_1+\ell_t+j_b+n_b-\ell)/2+\xi_m}\right.\\
&\qquad\phantom{=}\times(\bar{\eta}_2^{F''})^{4\xi_m}((\A_{123})^{n_v^m+\ell_t})_{E^{n_v^m}E^{\ell_t-i_a+j_a}E_s^{i_a-j_a}}^{\phantom{E^{n_v^m}E^{\ell_t-i_a+j_a}E_s^{i_a-j_a}}{E'_s}^{i_a-j_a}{E'}^{\ell_t-i_a+j_a}{E'}^{n_v^m}}\\
&\qquad\phantom{=}\times(\A_{123E_s}^{\phantom{123E_s}E''_s})^r(\bar{\eta}_{2E_s})^{r'_1}(\A_{123E_s}^{\phantom{123E_s}E''})^{r'_2}(\bar{\eta}_{2E})^{r'_3}(\bar{\eta}_2^{E''_s})^{r''_1}(\A_{123E}^{\phantom{123E}E''_s})^{r''_2}(\bar{\eta}_2^{F''})^{n_b-\ell+i_b}(\bar{\eta}_2^{F''})^{\ell_t-i_b+j_b}\\
&\qquad\phantom{=}\left.\times((\A_{123})^{\ell'}\hat{\mathcal{P}}_{13|d'}^{\ell'\boldsymbol{e}_1}((-2)^{-1}\A_{34}\cdot\bar{\bar{\eta}}_2)^{\ell'})_{\{E\}}\right)_{cs_4}\\
&\qquad\phantom{=}\times(\Gamma_{F''}\,\bar{\eta}_3\cdot\Gamma\,\hat{\mathcal{Q}}_{13|t}^{\boldsymbol{N}_m+\ell_t\boldsymbol{e}_1}\,\Gamma_{F''})_{e{E'}^{n_v^m}{E'}^{\ell_t-i_a+j_a}{E'}_s^{i_a-j_a}}^{\phantom{e{E'}^{n_v^m}{E'}^{\ell_t-i_a+j_a}{E'}_s^{i_a-j_a}}{E''_s}^{i_b-j_b}{E''}^{\ell_t-i_b+j_b}{E''}^{n_v^m}e''}\\
&\qquad\phantom{=}\times(g^{E''_sE''_s})^{r''_0}(\tCF{b}{k}{l}{,m+i_b}{3}{4})_{\{cC\}\{dD\}\{e''E''\}\{F''\}}(\A_{34E''F''})^{\ell_t-i_b+j_b}(\bar{\eta}_{4E''})^{r'_2},
}
where $n_b-\ell+i_b$ is the number of $(\A_{34}\cdot\bar{\bar{\eta}}_2)$'s in the three-point tensor structure $\FCF{b}{k}{l}{,m+i_b}{3}{4}$.  Also, note that the latter was changed to the associated OPE tensor structure $\tCF{b}{k}{l}{,m+i_b}{3}{4}$ to allow its extraction outside the conformal substitution $cs_4$.  Moreover, we stop explicitly symmetrizing over the special indices to avoid cluttering the computation too much.

Equipped with this form, we now proceed with the conformal substitution
\eqna{
CS'_4&=(\bar{\eta}_2^E)^{\ell-i_a-r'_1-r'_3}(\bar{\eta}_1^F)^{\ell-i_a}\left((x_3)^{-r'_2+2\xi_m}(x_3x_4)^{(r'_2+r''_1+\ell_t+j_b+n_b-\ell)/2+\xi_m}\right.\\
&\phantom{=}\qquad\times(\bar{\eta}_2^{F''})^{4\xi_m}((\A_{123})^{n_v^m+\ell_t})_{E^{n_v^m}E^{\ell_t-i_a+j_a}E_s^{i_a-j_a}}^{\phantom{E^{n_v^m}E^{\ell_t-i_a+j_a}E_s^{i_a-j_a}}{E'_s}^{i_a-j_a}{E'}^{\ell_t-i_a+j_a}{E'}^{n_v^m}}\\
&\phantom{=}\qquad\times(\A_{123E_s}^{\phantom{123E_s}E''_s})^r(\bar{\eta}_{2E_s})^{r'_1}(\A_{123E_s}^{\phantom{123E_s}E''})^{r'_2}(\bar{\eta}_{2E})^{r'_3}(\bar{\eta}_2^{E''_s})^{r''_1}(\A_{123E}^{\phantom{123E}E''_s})^{r''_2}(\bar{\eta}_2^{F''})^{n_b-\ell+i_b}(\bar{\eta}_2^{F''})^{\ell_t-i_b+j_b}\\
&\phantom{=}\qquad\left.\times((\A_{123})^{\ell'}\hat{\mathcal{P}}_{13|d'}^{\ell'\boldsymbol{e}_1}((-2)^{-1}\A_{34}\cdot\bar{\bar{\eta}}_2)^{\ell'})_{\{E\}}\right)_{cs_4}.
}
We begin by analyzing the shifted projection operator.  We find that for fixed $d'$ and $\ell'$, the contracted shifted projection operator behaves as
\eqna{
&((\A_{123})^{\ell'}\hat{\mathcal{P}}_{13|d'}^{\ell'\boldsymbol{e}_1}((-2)^{-1}\A_{34}\cdot\bar{\bar{\eta}}_2)^{\ell'})_{\{E\}}\\
&\qquad=\sum_{i=0}^{\lfloor\ell'/2\rfloor}\frac{(-\ell')_{2i}}{2^{2i}i!(-\ell'+2-d'/2)_i}(\bar{\eta}_{1E}\bar{\eta}_{2E})^i\left(\frac{[(\bar{\eta}_4-\bar{\eta}_2)x_4-(\bar{\eta}_3-\bar{\eta}_2)x_3]_E}{2\sqrt{x_3x_4}}\right)^{\ell'-2i},
}
as all the $E$-indices are ultimately contracted with $(\bar{\eta}_2^E)^{\ell'}$ outside the conformal substitution.  Assuming for a moment that the remaining factors in the conformal substitution are set to one, we expect the conformal substitution to give
\eqna{
CS''_4&=(\bar{\eta}_2^E)^{\ell'}(\bar{\eta}_1^F)^{\ell-i_a}\left(((\A_{123})^{\ell'}\hat{\mathcal{P}}_{13|d'}^{\ell'\boldsymbol{e}_1}((-2)^{-1}\A_{34}\cdot\bar{\bar{\eta}}_2)^{\ell'})_{\{E\}}\right)_{cs_4}\\
&=(\bar{\eta}_2^E)^{\ell'}(\bar{\eta}_1^F)^{\ell-i_a}\\
&\phantom{=}\qquad\times\left(\sum_{i=0}^{\lfloor\ell'/2\rfloor}\frac{(-\ell')_{2i}}{2^{2i}i!(-\ell'+2-d'/2)_i}(\bar{\eta}_{1E}\bar{\eta}_{2E})^i\left(\frac{[(\bar{\eta}_4-\bar{\eta}_2)x_4-(\bar{\eta}_3-\bar{\eta}_2)x_3]_E}{2\sqrt{x_3x_4}}\right)^{\ell'-2i}\right)_{cs_4}\\
&=\frac{\ell'!}{2^{\ell'}(d'/2-1)_{\ell'}}\left(C_{\ell'}^{(d'/2-1)}(X)\right)_{s''}.
}

We encounter here the Gegenbauer polynomials $C_n^{(\lambda)}(X)$ in the variable
\eqn{X=\frac{(\alpha_4-\alpha_2)x_4-(\alpha_3-\alpha_2)x_3}{2},}[EqX]
with the substitution
\eqn{s'':\alpha_2^{s_2}\alpha_3^{s_3}\alpha_4^{s_4}x_3^{r_3}x_4^{r_4}\to G_{(\ell'-\ell,n_a-\ell,n_a+\ell'-2\ell+2i_a,0,0)F^{n_a-\ell+i_a}}^{ij|m+\ell|kl}.}[EqSpp]
The variable \eqref{EqX} and substitution \eqref{EqSpp} were introduced to allow the re-summation into the Gegenbauer polynomials, with
\eqn{\sum_{i=0}^{\lfloor\ell'/2\rfloor}\frac{(-\ell')_{2i}}{2^{2i}i!(-\ell'+2-d'/2)_i}X^{\ell'-2i}=\frac{\ell'!}{2^{\ell'}(d'/2-1)_{\ell'}}C_{\ell'}^{(d'/2-1)}(X).}
Indeed, since in the expansion of $([(\bar{\eta}_4-\bar{\eta}_2)x_4-(\bar{\eta}_3-\bar{\eta}_2)x_3]_E)^{\ell'-2i}$ we have $s_2+s_3+s_4=r_3+r_4=\ell'-2i$, we can replace $i$ by either one of
\eqn{i=\frac{\ell'-(s_2+s_3+s_4)}{2},\qquad\qquad i=\frac{\ell'-(r_3+r_4)}{2}.}
Hence, in the conformal substition $cs_4$ there are
\eqna{
\text{number of $\bar{\eta}_2$'s}&=i+s_2=\frac{\ell'+s_2-s_3-s_4}{2},\\
\text{number of $x_3$'s}&=-\frac{\ell'}{2}+i+r_3=\frac{r_3-r_4}{2},\\
\text{number of $x_4$'s}&=-\frac{\ell'}{2}+i+r_4=-\frac{r_3-r_4}{2},
}
which straightforwardly lead to the substitution \eqref{EqSpp} for the variable $X$ \eqref{EqX}, with the quantity
\eqna{
&G_{(n_1,n_2,n_3,n_4,n_5)A_1\cdots A_n}^{ij|m+\ell|kl}\\
&\qquad=\rho^{(d,(\ell+s_2-s_3-s_4+n_1)/2;-h_{ijm}-(\ell+n_2)/2)}x_3^{-s_3}x_4^{-s_4}\\
&\qquad\phantom{=}\times\bar{I}_{12;34}^{(d,h_{ijm}-(s_2-s_3-s_4+n_3)/2,n;-h_{klm}+(r_3-r_4+n_4)/2,\chi_m+h_{klm}-(r_3-r_4+n_5)/2)}{}_{A_1\cdots A_n}.
}[EqG]
Here, $G$ is the quantity that naturally appears in the conformal substitutions \eqref{Eq4ptCBind} \cite{Fortin:2019gck}.  It turns out to have some interesting properties, as we will discuss shortly.

As for the remaining factors inside the conformal substitution, we find that these can be easily taken into account by simply noting that the conformal substitution depends only on the powers of $\bar{\eta}_2$, $x_3$ and $x_4$.  Moreover, since powers add up under multiplication of factors and these powers appear directly inside \eqref{EqG}, a direct consequence is the fundamental property
\eqn{G_{(n_1,n_2,n_3,n_4,n_5)A^n}^{ij|m+\ell|kl}G_{(m_1,m_2,m_3,m_4,m_5)B^m}^{ij|m+\ell|kl}=G_{(n_1+m_1,n_2+m_2,n_3+m_3,n_4+m_4,n_5+m_5)A^nB^m}^{ij|m+\ell|kl},}[EqGG]
which is understood as long as the definition of $G$ in terms of the $\bar{I}$-function \eqref{EqG} is not used until there is only one $G$ per term.

Hence, the initial conformal substitution may be rewritten as
\eqn{CS'_4=\frac{\ell'!}{2^{\ell'}(d'/2-1)_{\ell'}}(\bar{\eta}_2^E)^{\ell-i_a-r'_1-r'_3-\ell'}\left(C_{\ell'}^{(d'/2-1)}(X)\right)_{s'},}
with the new substitution
\eqna{
s':\alpha_2^{s_2}\alpha_3^{s_3}\alpha_4^{s_4}x_3^{r_3}x_4^{r_4}&\to(\mathcal{S}^{n_v^m+\ell_t})_{E^{n_v^m}E^{\ell_t-i_a+j_a}(E_s^{i_a-j_a}}^{\phantom{E^{n_v^m}E^{\ell_t-i_a+j_a}E_s^{i_a-j_a}}{E'_s}^{i_a-j_a}{E'}^{\ell_t-i_a+j_a}{E'}^{n_v^m}}(\mathcal{S}_{E_s}^{\phantom{E_s}E''_s})^r(\mathcal{S}_{E_s}^{\phantom{E_s}E''})^{r'_2}(\mathcal{S}_{E}^{\phantom{E}E''_s})^{r''_2}\\
&\phantom{\to}\qquad\times\left(G_{(\ell'-\ell,n_a-\ell,n'_3,n'_4,n'_5)}^{ij|m+\ell|kl}\right)_{F^{n_a-\ell+i_a}E^{r'_3}E_s^{r'_1}}^{{E''_s}^{r''_1}{F''}^{4\xi_m}{F''}^{n_b-\ell+i_b}{F''}^{\ell_t-i_b+j_b}},
}[EqSp]
with
\eqn{
\begin{gathered}
n'_3=n_a+2n_b-3\ell+2i_a-j_a+2j_b+\ell_t+r'_1+2r''_1-r''_2+8\xi_m,\\
n'_4=n_b-\ell+j_b+\ell_t-r'_2+r''_1+6\xi_m,\\
n'_5=\ell-n_b-j_b-\ell_t-r'_2-r''_1-2\xi_m.
\end{gathered}
}[EqParam4]
Owing to the fundamental property \eqref{EqGG}, the new quantity
\eqn{\mathcal{S}_A^{\phantom{A}B}=g_A^{\phantom{A}B}G_{(0,0,0,0,0)}^{ij|m+\ell|kl}-G_{(0,0,2,0,0)A}^{ij|m+\ell|kl}\bar{\eta}_1^B-\bar{\eta}_{3A}(G_{(0,0,2,2,0)}^{ij|m+\ell|kl})^B+(G_{(0,0,4,2,0)}^{ij|m+\ell|kl})_A^{\phantom{A}B},}[EqSS]
originates directly from the free
\eqn{\A_{123A}^{\phantom{123A}B}=g_A^{\phantom{A}B}-\bar{\eta}_{2A}\bar{\eta}_1^B-x_3\bar{\eta}_{3A}\bar{\eta}_2^B+x_3\bar{\eta}_{2A}\bar{\eta}_2^B,}
inside the conformal substitution.

Upon combining the remaining contractions outside the conformal substitution, we arrive at the following form for the four-point conformal blocks:
\eqna{
&(\mathscr{G}_{(a|b]}^{ij|m+\ell|kl})_{\{aA\}\{bB\}\{cC\}\{dD\}}\\
&\qquad=\sum_t\mathscr{A}_t(d,\ell)\sum_{j_a,j_b\geq0}\binom{i_a}{j_a}\binom{i_b}{j_b}\frac{(-\ell_t)_{i_a-j_a}(-\ell_t)_{i_b-j_b}(-\ell+\ell_t)_{j_a}(-\ell+\ell_t)_{j_b}}{(-\ell)_{i_a}(-\ell)_{i_b}}\\
&\qquad\phantom{=}\times\sum_{\substack{r,\boldsymbol{r}',\boldsymbol{r}''\geq0\\r+2r'_0+r'_1+r'_2=j_a\\r+2r''_0+r''_1+r''_2=j_b\\r'_0+r'_1+r'_3=r''_0+r''_1+r''_3}}(-1)^{\ell-\ell'-i_a+r'_1+r'_2}\frac{(-2)^{r'_3+r''_3}\ell'!}{(d'/2-1)_{\ell'}}\mathscr{C}_{j_a,j_b}^{(d+d_t,\ell-\ell_t)}(r,\boldsymbol{r}',\boldsymbol{r}'')\\
&\qquad\phantom{=}\times\left(C_{\ell'}^{(d'/2-1)}(X)\right)_{s_{(a|b)}^{ij|m+\ell|kl}(t,j_a,j_b,r,\boldsymbol{r}',\boldsymbol{r}'')},
}[EqCB4]
with the substitutions
\eqna{
&s_{(a|b)}^{ij|m+\ell|kl}(t,j_a,j_b,r,\boldsymbol{r}',\boldsymbol{r}''):\alpha_2^{s_2}\alpha_3^{s_3}\alpha_4^{s_4}x_3^{r_3}x_4^{r_4}\\
&\qquad\to(-1)^{2\xi_m}(\tOPE{a}{i}{j}{,m+i_a}{1}{2})_{\{aA\}\{bB\}}^{\phantom{\{aA\}\{bB\}}\{Ee\}\{F\}}(g_{E_sE_s})^{r'_0}(\mathcal{S}_{E_s}^{\phantom{E_s}E''_s})^r[(\mathcal{S}\cdot\bar{\eta}_4)_{E_s}]^{r'_2}\\
&\qquad\phantom{\to}\times\left(G_{(\ell'-\ell+2r'_3,n_a-\ell,n'_3,n'_4,n'_5)}^{ij|m+\ell|kl}\right)_{F^{n_a-\ell+i_a}E_s^{r'_1}}^{{E''_s}^{r''_1}{F''}^{4\xi_m}{F''}^{n_b-\ell+i_b}{F''}^{\ell_t-i_b+j_b}}(\bar{\eta}_2^E)^{\ell_t-i_a+j_a}\\
&\qquad\phantom{\to}\times(\Gamma_{F''}\,\bar{\eta}_3\cdot\Gamma\,\mathcal{S}^{n_v^m+\ell_t}\hat{\mathcal{Q}}_{13|t}^{\boldsymbol{N}_m+\ell_t\boldsymbol{e}_1}\,\Gamma_{F''})_{eE^{n_v^m}E^{\ell_t-i_a+j_a}E_s^{i_a-j_a}}^{\phantom{eE^{n_v^m}E^{\ell_t-i_a+j_a}E_s^{i_a-j_a}}{E''_s}^{i_b-j_b}{E''}^{\ell_t-i_b+j_b}{E''}^{n_v^m}e''}\\
&\qquad\phantom{\to}\times[(\bar{\eta}_2\cdot\mathcal{S})^{E''_s}]^{r''_2}(g^{E''_sE''_s})^{r''_0}(\tCF{b}{k}{l}{,m+i_b}{3}{4})_{\{cC\}\{dD\}\{e''E''\}\{F''\}}(\A_{34E''F''})^{\ell_t-i_b+j_b},
}[EqS]
with the quantity $\mathcal{S}_A{}^B$ defined in \eqref{EqSS}, and the various parameters given in \eqref{EqParam4} and \eqref{Eqdplp}.  We remind the reader that the $E_s$-indices and the $E''_s$-indices are symmetrized separately in \eqref{EqS}.

The form \eqref{EqCB4} for the four-point conformal blocks in the mixed basis features linear combinations of Gegenbauer polynomials in the variable $X$ \eqref{EqX}, each with a specific substitution given by \eqref{EqS}.  Like for the rotation matrices, all the $\ell$ dependence has been taken into account, and we simply need to work with the $\ell$-independent part of the projection operators and tensor structures to generate the complete infinite tower of associated conformal blocks.  Once the substitutions are implemented, the blocks are expressed in terms of the $\bar{I}$-functions \eqref{EqIb4} which are tensorial generalizations of the Exton $G$-function, as evident from \eqref{EqK4} and \eqref{EqK0}.

For each value of $t$ appearing in the decomposition \eqref{EqPExp} and every $j_a$ and $j_b$ arising in the double expansion of the four-point conformal blocks \eqref{EqCB4}, there are associated partitions which correspond to the allowed Gegenbauer polynomials.  Using the diagrammatic notation introduced above, we may represent these partitions by diagrams, which are distinguished by their values of $r'_3$ and $r''_3$.  It follows that the four-point conformal blocks can be represented by a set of (nonperturbative Feynman-like) diagrams, where each diagram corresponds to a set of Gegenbauer polynomials, with their associated substitutions \eqref{EqS}.

Finally, we remark that the special parts of the tensor structures may sometimes be simplified.  For example, except for the simplification $\A_{12}^{EF}\to-\bar{\eta}_2^E\bar{\eta}_1^F$ that we mentioned above, we can make the substitutions
\eqn{
\begin{gathered}
\A_{12A}^{\phantom{12A}E}\to g_A^{\phantom{A}E},\qquad\qquad\A_{12B}^{\phantom{12B}E}\to g_B^{\phantom{B}E},\qquad\qquad\A_{12A}^{\phantom{12A}F}\to g_A^{\phantom{A}F},\qquad\qquad\A_{12B}^{\phantom{12B}F}\to g_B^{\phantom{B}F},\\
\A_{34CE}\to g_{CE},\qquad\qquad\A_{34DE}\to g_{DE},\qquad\qquad\A_{34CF}\to g_{CF},\qquad\qquad\A_{34DF}\to g_{DF},
\end{gathered}
}
due to the contractions of these metrics with the external half-projectors in \eqref{Eq4ptind}.


\subsection{Properties of \texorpdfstring{$G$}{G}}

The final substitutions \eqref{EqS} necessitate the multiplication of several $G$'s together, according to \eqref{EqGG}, which we repeat here
\eqn{G_{(n_1,n_2,n_3,n_4,n_5)A^n}^{ij|m+\ell|kl}G_{(m_1,m_2,m_3,m_4,m_5)B^m}^{ij|m+\ell|kl}=G_{(n_1+m_1,n_2+m_2,n_3+m_3,n_4+m_4,n_5+m_5)A^nB^m}^{ij|m+\ell|kl},}
in addition to contractions with known tensorial objects.  In fact, the contiguous relations \eqref{EqCont4} translate directly to the $G$'s as
\eqna{
g\cdot G_{(n_1,n_2,n_3,n_4,n_5)}^{ij|m+\ell|kl}&=0,\\
\bar{\eta}_1\cdot G_{(n_1,n_2,n_3,n_4,n_5)}^{ij|m+\ell|kl}&=G_{(n_1,n_2,n_3-2,n_4,n_5)}^{ij|m+\ell|kl},\\
\bar{\eta}_2\cdot G_{(n_1,n_2,n_3,n_4,n_5)}^{ij|m+\ell|kl}&=G_{(n_1+2,n_2,n_3,n_4,n_5)}^{ij|m+\ell|kl},\\
\bar{\eta}_3\cdot G_{(n_1,n_2,n_3,n_4,n_5)}^{ij|m+\ell|kl}&=G_{(n_1,n_2,n_3-2,n_4-2,n_5)}^{ij|m+\ell|kl},\\
\bar{\eta}_4\cdot G_{(n_1,n_2,n_3,n_4,n_5)}^{ij|m+\ell|kl}&=G_{(n_1,n_2,n_3-2,n_4,n_5+2)}^{ij|m+\ell|kl}.
}[EqContG]
Moreover, the quantity \eqref{EqSS} often appears contracted in some specific ways that we expound here for completeness.  In particular,
\eqn{
\begin{gathered}
(\bar{\eta}_2\cdot\mathcal{S})^B=\bar{\eta}_2^BG_{(0,0,0,0,0)}^{ij|m+\ell|kl}-G_{(2,0,2,0,0)}^{ij|m+\ell|kl}\bar{\eta}_1^B-x_3^{-1}(G_{(0,0,2,2,0)}^{ij|m+\ell|kl})^B+(G_{(2,0,4,2,0)}^{ij|m+\ell|kl})^B,\\
(\mathcal{S}\cdot\bar{\eta}_4)_A=\bar{\eta}_{4A}G_{(0,0,0,0,0)}^{ij|m+\ell|kl}-G_{(0,0,2,0,0)A}^{ij|m+\ell|kl}-\bar{\eta}_{3A}G_{(0,0,0,2,2)}^{ij|m+\ell|kl}+G_{(0,0,2,2,2)A}^{ij|m+\ell|kl},\\
(\mathcal{S}\cdot\A_{13}\cdot\mathcal{S}^T)_{AB}=g_{AB}G_{(0,0,0,0,0)}^{ij|m+\ell|kl}-\bar{\eta}_{1A}G_{(0,0,2,0,0)B}^{ij|m+\ell|kl}-G_{(0,0,2,0,0)A}^{ij|m+\ell|kl}\bar{\eta}_{1B}.
\end{gathered}
}
We note that the last identity can be directly shown to hold from the original $\A_{123}$ that gives rise to $\mathcal{S}$.  Indeed, one has
\eqn{\A_{123}\cdot\A_{13}\cdot\A_{123}^T=\A_{12},}
which matches the above result after substitution.


\section{Summary of Results}\label{SecSum}

In this section, we summarize the main results for the three- and four-point conformal blocks derived in the previous two sections.  The solutions are given for infinite towers of exchanged quasi-primary operators in irreducible representations $\boldsymbol{N}_m+\ell\boldsymbol{e}_1$, with the universal $\ell$-dependent part already processed.

The tensor structures \eqref{EqTS} that enter the results are decomposed into a universal $\ell$-dependent part and a special part.  The universal $\ell$-dependent parts have $\ell-i_a$ or $\ell-i_b$ $\boldsymbol{e}_1$ indices on the exchanged quasi-primary operators contracted with some (or all) of the $n_a$ or $n_b$ free indices on the OPE differential operators.  These have already been accounted for in the results.  Meanwhile, the special parts, which arise from $\boldsymbol{N}_m+i_a\boldsymbol{e}_1$ and $\boldsymbol{N}_m+i_b\boldsymbol{e}_1$, respectively, appear directly in the results and must be contracted properly for any given case under consideration.


\subsection{Three-Point Conformal Blocks}

The derivation in Section \ref{SecRM} leads to the following form for the three-point conformal blocks [see \eqref{EqCB3}]
\vspace{10pt}

\begin{center}
\noindent\fbox{
\parbox{0.9\textwidth}{
\eqna{
&(\mathscr{G}_{(a|}^{ij|m+\ell})_{\{aA\}\{bB\}\{eE\}}\\
&\qquad=\lambda_{\boldsymbol{N}_{m+\ell}}(\tCF{a}{i}{j}{,m+i_a}{1}{2})_{\{aA\}\{bB\}\{e'E'\}\{F\}}(\bar{\eta}_3\cdot\Gamma\Gamma_F)_e^{\phantom{e}e'}\\
&\qquad\phantom{=}\times\sum_{r_0,r_3,s_0,s_3,t\geq0}\frac{(-1)^{r_0+s_0}(-2)^{\ell_a+s_0-t}}{n_v^{m+i_a}!r_0!r_3!s_0!s_3!t!}(-n_v^{m+i_a})_{r_0+r_3}(-\ell_a)_{s_0+s_3}(-s_0)_t\\
&\qquad\phantom{=}\times(-n+n_v^{m+i_a}+\ell_a+r_0+r_3)_{s_0+s_3}(-h-n-\ell_a)_{\ell_a+s_0-t}\\
&\qquad\phantom{=}\times(-h-n-\ell_a+1-d/2)_{s_0-t}(p-n-\ell_a+r_0+r_3+s_0+s_3+2-d)_{\ell_a-s_0-s_3}\\
&\qquad\phantom{=}\times\sum_\sigma g_{E_{\sigma(1)}}^{\phantom{E_{\sigma(1)}}{E'_{\sigma(1)}}}\cdots g_{E_{\sigma(r_0)}}^{\phantom{E_{\sigma(r_0)}}{E'_{\sigma(r_0)}}}\bar{\eta}_3^{E'_{\sigma(r_0+1)}}\cdots\bar{\eta}_3^{E'_{\sigma(r_0+r_3)}}(g_E^{(Z})^{s_0}(\bar{\eta}_3^Z)^{s_3}\\
&\qquad\phantom{=}\times\bar{I}_{12}^{(d+2\ell_a,h+\ell_a+2r_0+r_3+s_3+t,n-\ell_a-2r_0-r_3-s_0-s_3;p-r_0)}{}_{E_\sigma^{n_v^{m+i_a}-r_0}}^{Z^{n-n_v^{m+i_a}-\ell_a-r_0-r_3-s_0-s_3})}(-\bar{\eta}_{2E})^{\ell_a-s_0}
}
}
}
\end{center}
\vspace{10pt}

\noindent Here the symmetrized $Z$-indices belong to $\{E'_{\sigma(r_0+r_3+1)},\ldots,E'_{\sigma(n_v^{m+i_a})},F^{n_a-\ell_a+2\xi_m}\}$, and the remaining parameters are found in \eqref{EqParam3}.  From this result, it is relatively straightforward to determine the rotation matrices.


\subsection{Four-Point Conformal Blocks}

From the decomposition \eqref{EqPExp} and the proof laid out in Section \ref{SecCB}, the four-point conformal blocks are given by [see \eqref{EqCB4} and \eqref{EqS}]
\vspace{10pt}

\begin{center}
\noindent\fbox{
\parbox{0.9\textwidth}{
\eqna{
&(\mathscr{G}_{(a|b]}^{ij|m+\ell|kl})_{\{aA\}\{bB\}\{cC\}\{dD\}}\\
&\qquad=\sum_t\mathscr{A}_t(d,\ell)\sum_{j_a,j_b\geq0}\binom{i_a}{j_a}\binom{i_b}{j_b}\frac{(-\ell_t)_{i_a-j_a}(-\ell_t)_{i_b-j_b}(-\ell+\ell_t)_{j_a}(-\ell+\ell_t)_{j_b}}{(-\ell)_{i_a}(-\ell)_{i_b}}\\
&\qquad\phantom{=}\times\sum_{\substack{r,\boldsymbol{r}',\boldsymbol{r}''\geq0\\r+2r'_0+r'_1+r'_2=j_a\\r+2r''_0+r''_1+r''_2=j_b\\r'_0+r'_1+r'_3=r''_0+r''_1+r''_3}}(-1)^{\ell-\ell'-i_a+r'_1+r'_2}\frac{(-2)^{r'_3+r''_3}\ell'!}{(d'/2-1)_{\ell'}}\mathscr{C}_{j_a,j_b}^{(d+d_t,\ell-\ell_t)}(r,\boldsymbol{r}',\boldsymbol{r}'')\\
&\qquad\phantom{=}\times\left(C_{\ell'}^{(d'/2-1)}(X)\right)_{s_{(a|b)}^{ij|m+\ell|kl}(t,j_a,j_b,r,\boldsymbol{r}',\boldsymbol{r}'')}
}
\eqna{
&s_{(a|b)}^{ij|m+\ell|kl}(t,j_a,j_b,r,\boldsymbol{r}',\boldsymbol{r}''):\alpha_2^{s_2}\alpha_3^{s_3}\alpha_4^{s_4}x_3^{r_3}x_4^{r_4}\\
&\qquad\to(-1)^{2\xi_m}(\tOPE{a}{i}{j}{,m+i_a}{1}{2})_{\{aA\}\{bB\}}^{\phantom{\{aA\}\{bB\}}\{Ee\}\{F\}}(g_{E_sE_s})^{r'_0}(\mathcal{S}_{E_s}^{\phantom{E_s}E''_s})^r[(\mathcal{S}\cdot\bar{\eta}_4)_{E_s}]^{r'_2}\\
&\qquad\phantom{\to}\times\left(G_{(\ell'-\ell+2r'_3,n_a-\ell,n'_3,n'_4,n'_5)}^{ij|m+\ell|kl}\right)_{F^{n_a-\ell+i_a}E_s^{r'_1}}^{{E''_s}^{r''_1}{F''}^{4\xi_m}{F''}^{n_b-\ell+i_b}{F''}^{\ell_t-i_b+j_b}}(\bar{\eta}_2^E)^{\ell_t-i_a+j_a}\\
&\qquad\phantom{\to}\times(\Gamma_{F''}\,\bar{\eta}_3\cdot\Gamma\,\mathcal{S}^{n_v^m+\ell_t}\hat{\mathcal{Q}}_{13|t}^{\boldsymbol{N}_m+\ell_t\boldsymbol{e}_1}\,\Gamma_{F''})_{eE^{n_v^m}E^{\ell_t-i_a+j_a}E_s^{i_a-j_a}}^{\phantom{eE^{n_v^m}E^{\ell_t-i_a+j_a}E_s^{i_a-j_a}}{E''_s}^{i_b-j_b}{E''}^{\ell_t-i_b+j_b}{E''}^{n_v^m}e''}\\
&\qquad\phantom{\to}\times[(\bar{\eta}_2\cdot\mathcal{S})^{E''_s}]^{r''_2}(g^{E''_sE''_s})^{r''_0}(\tCF{b}{k}{l}{,m+i_b}{3}{4})_{\{cC\}\{dD\}\{e''E''\}\{F''\}}(\A_{34E''F''})^{\ell_t-i_b+j_b}
}
}
}
\end{center}
\vspace{10pt}

\noindent The conformal blocks are thus represented by linear combinations of the Gegenbauer polynomials in the variable $X$ \eqref{EqX}, coupled with associated substitutions.  Here $j_a$ and $j_b$ are the numbers of extracted indices from the shifted projection operators appearing in the decomposition \eqref{EqPExp}, with the remaining extracted indices appearing in the special part of the projection operator.  Moreover, the indices of summation $r$, $\boldsymbol{r}'$, and $\boldsymbol{r}''$ determine how the special indices are extracted from the shifted projection operators, as in \eqref{EqPExtract}.  Finally, the quantity $\mathcal{S}$ \eqref{EqSS} is built out of the quantity $G$ \eqref{EqG}, which encodes the action of the OPE differential operator.  The latter satisfies some interesting properties, listed in \eqref{EqGG} and \eqref{EqContG}.  We again stress here that the special $E'_s$- and $E''_s$-indices are symmetrized independently.

Having established the essential results necessary for computing arbitrary four-point conformal blocks, we next apply these results to a series of examples.


\section{Examples}\label{SecEx}

This section makes use of the results for the three- and four-point conformal blocks in the context of simple examples, with external quasi-primary operators in scalar, vector, and fermion irreducible representations.  Known blocks are compared with previously computed blocks obtained from the embedding space OPE formalism in \cite{Fortin:2019gck}, and new results are compared with the literature when possible.  Although quite straightforward, most steps in the computations are done explicitly for all examples to elucidate the methods developed in the previous sections.


\subsection{\texorpdfstring{$\vev{SSSS}$}{SSSS}}

For the four-point correlation function of four scalars, the exchanged quasi-primary operators are in the $\ell\boldsymbol{e}_1$ irreducible representation, with the projection operator simply given by $\hat{\mathcal{P}}_{13}^{\ell\boldsymbol{e}_1}=\hat{\mathcal{P}}_{13|d}^{\ell\boldsymbol{e}_1}$ [see \eqref{EqPle1}].  Hence, $\boldsymbol{N}_m=\boldsymbol{0}$ with $n_v^m=0$, and the decomposition \eqref{EqPExp} is straightforward with only one term with $(d_1,\ell_{t=1})=(0,0)$, $\mathscr{A}_1(d,\ell)=1$, and $\hat{\mathcal{Q}}_{13|1}^{\boldsymbol{0}}=1$.  Moreover, there is just a single tensor structure for both OPEs and, following \eqref{EqTS} and \eqref{Eq3pTStoOPETS}, the forms are
\eqn{
\begin{gathered}
(\FCF{b=1}{k}{l}{,m+\ell}{3}{4})_{\{cC\}\{dD\}\{e''E''\}}=[(\A_{34}\cdot\bar{\bar{\eta}}_2)_{E''}]^\ell\to(\tCF{b=1}{k}{l}{,m+\ell}{3}{4})_{\{cC\}\{dD\}\{e''E''\}\{F''\}}=(\A_{34E''F''})^\ell,\\
(\FCF{a=1}{i}{j}{,m+\ell}{1}{2})_{\{aA\}\{bB\}\{eE\}}=[(\A_{12}\cdot\bar{\eta}_3)_E]^\ell\to(\tOPE{a=1}{i}{j}{,m+\ell}{1}{2})_{\{aA\}\{bB\}}^{\phantom{\{aA\}\{bB\}}\{Ee\}\{F\}}=(\A_{12}^{EF})^\ell,
\end{gathered}
}[EqSSSS-TS]
implying that $n_{a=1}=n_{b=1}=\ell$, $i_{a=1}=i_{b=1}=0$ and $\tOPE{a=1}{i}{j}{m}{1}{2}=\tCF{b=1}{k}{l}{m}{3}{4}=1$.  Hence, the special parts of the tensor structures are all trivial, and no extraction of special indices is necessary.

Therefore, the sums over $t$, $j_{a=1}$, $j_{b=1}$, $r$, $\boldsymbol{r}'$ and $\boldsymbol{r}''$ all collapse to a single term so that the conformal blocks \eqref{EqCB4} are
\eqn{\mathscr{G}_{(1|1]}^{ij|m+\ell|kl}=\left(\Diag{1}{0}{0}{0}{0}{0}{0}{0}\right)_{s_{(1|1)}^1}=\frac{\ell!}{(d/2-1)_\ell}\left(C_\ell^{(d/2-1)}(X)\right)_{s_{(1|1)}^1},}
with the substitution \eqref{EqS}
\eqn{s_{(1|1)}^1\equiv s_{(1|1)}^{ij|m+\ell|kl}(1,0,0,0,\boldsymbol{0},\boldsymbol{0}):\alpha_2^{s_2}\alpha_3^{s_3}\alpha_4^{s_4}x_3^{r_3}x_4^{r_4}\to G_{(0,0,0,0,0)}^{ij|m+\ell|kl}.}
As expected, this is the desired result \cite{Fortin:2019gck}.

In the same manner, the rotation matrix can be computed straightforwardly from the three-point conformal blocks \eqref{EqCB3}.  With $n_v^m=0$, $n_{a=1}=\ell$ and $i_{a=1}=0$, all the sums over $r_0$, $r_3$, $s_0$, $s_3$ and $t$ in \eqref{EqCB3} also collapse to only one term, leading straightforwardly to
\eqn{\mathscr{G}_{(1|}^{ij|m+\ell}=\lambda_{\ell\boldsymbol{e}_1}(-2)^\ell(-h_{ij,m+\ell}-\ell/2)_\ell(\Delta_{m+\ell}-\ell+2-d)_\ell\rho^{(d+2\ell,h_{ij,m+\ell}-\ell/2;\Delta_{m+\ell}+\ell)}(-\bar{\eta}_{2E})^\ell.}
Hence, from \eqref{EqRM} and using the notation \eqref{Eqkappa} the rotation matrix is
\eqn{(R_{ij,m+\ell}^{-1})_{1,1}={}_{a=1}\kappa_{(0,0,0,0,0,0,0,0,0)}^{ij|m+\ell},}[EqSSSS-RM]
as already obtained in \cite{Fortin:2019gck}.

It is evident that the basic rules introduced here are quite efficient, generating the rotation matrix and the conformal blocks effortlessly.  As we will see in subsequent examples, their potential is in full display for four-point correlation functions with external quasi-primary operators in nontrivial irreducible representations of the Lorentz group.


\subsection{\texorpdfstring{$\vev{SSSR}$}{SSSR}}

With two scalars $SS$, the only possible exchanged irreducible representations (in the $s$-channel) are the $\ell\boldsymbol{e}_1$ representations with projection operators \eqref{EqPle1}.  Hence, in this case, we need to deal with the same decomposition \eqref{EqPExp} as in the previous example with $\vev{SSSS}$, and also the same tensor structure $\tCF{a=1}{i}{j}{,m+\ell}{1}{2}$ \eqref{EqSSSS-TS}.  Consequently, the rotation matrix transforming the blocks to the pure three-point basis is the same as for $\vev{SSSS}$ and is given by \eqref{EqSSSS-RM}.  We may therefore restrict attention to the $SR$ part of the correlation function.

Representation theory implies that the only possible irreducible representations $R$ among the remaining defining representations (\textit{i.e.}\ antisymmetric tensors and fermions) for the last external quasi-primary operator must either be $\boldsymbol{e}_1$ or $\boldsymbol{e}_2$.  Since they do not have the same number of conformal blocks, we treat them separately below.


\subsubsection{\texorpdfstring{$\vev{SSSV}$}{SSSV}}

In the case of the four-point correlation function of three scalars and one vector, there are two $34$-tensor structures in $d>3$, given by
\eqna{
b=1:\qquad&(\FCF{b}{k}{l}{,m+\ell}{3}{4})_{\{cC\}\{dD\}\{e''E''\}}=(\A_{34}\cdot\bar{\bar{\eta}}_2)_D[(\A_{34}\cdot\bar{\bar{\eta}}_2)_{E''}]^\ell\\
&\qquad\to(\tCF{b}{k}{l}{,m+\ell}{3}{4})_{\{cC\}\{dD\}\{e''E''\}\{F''\}}=\A_{34DF''}(\A_{34E''F''})^\ell,\\
b=2:\qquad&(\FCF{b}{k}{l}{,m+\ell}{3}{4})_{\{cC\}\{dD\}\{e''E''\}}=\A_{34DE''_1}[(\A_{34}\cdot\bar{\bar{\eta}}_2)_{E''}]^{\ell-1}\\
&\qquad\to(\tCF{b}{k}{l}{,m+\ell}{3}{4})_{\{cC\}\{dD\}\{e''E''\}\{F''\}}=\A_{34DE''_1}(\A_{34E''F''})^{\ell-1},
}
which imply that
\eqna{
b=1:\qquad&n_b=\ell+1,\qquad i_b=0,\qquad(\tCF{b}{k}{l}{m}{3}{4})_{DF''}=\A_{34DF''},\\
b=2:\qquad&n_b=\ell-1,\qquad i_b=1,\qquad(\tCF{b}{k}{l}{,m+1}{3}{4})_{DE''_1}=\A_{34DE''_1}.
}
From these forms, it is apparent that no indices need to be extracted for the first tensor structure, while for the second tensor structure, one $E''$-index must be extracted.  These correspond diagrammatically to
\eqna{
b=1:\qquad&\hat{\mathcal{P}}_{13|d}^{\ell\boldsymbol{e}_1}=\Diag{1}{0}{0}{0}{0}{0}{0}{0},\\
b=2:\qquad&\hat{\mathcal{P}}_{13|d}^{\ell\boldsymbol{e}_1}=\Diag{1}{0}{0}{0}{0}{0}{0}{1}+\Diag{1}{0}{0}{0}{0}{0}{1}{0}.
}
Through their associated partitions, these diagrams in turn directly give the conformal blocks in terms of Gegenbauer polynomials,
\eqna{
&\mathscr{G}_{(1|1]}^{ij|m+\ell|kl}=\frac{\ell!}{(d/2-1)_\ell}\left(C_\ell^{(d/2-1)}(X)\right)_{s_{(1|1)}^1},\\
&\mathscr{G}_{(1|2]}^{ij|m+\ell|kl}=-\frac{(\ell-1)!}{(d/2)_{\ell-1}}\left(C_{\ell-1}^{(d/2)}(X)\right)_{s_{(1|2)}^1}+\frac{(\ell-1)!}{(d/2)_{\ell-1}}\left(C_{\ell-2}^{(d/2)}(X)\right)_{s_{(1|2)}^2},
}
as in \eqref{EqCB4}.  Moreover, contracting the special parts of the tensor structures with the special part of the projection operator as in \eqref{EqS} with the help of the partitions obtained through the diagrams, the associated substitutions are
\eqna{
s_{(1|1)}^1:\alpha_2^{s_2}\alpha_3^{s_3}\alpha_4^{s_4}x_3^{r_3}x_4^{r_4}&\to G_{(0,0,2,1,-1)D}^{ij|m+\ell|kl},\\
s_{(1|2)}^1:\alpha_2^{s_2}\alpha_3^{s_3}\alpha_4^{s_4}x_3^{r_3}x_4^{r_4}&\to(\bar{\eta}_2\cdot\mathcal{S})_DG_{(-1,0,-1,0,0)}^{ij|m+\ell|kl},\\
s_{(1|2)}^2:\alpha_2^{s_2}\alpha_3^{s_3}\alpha_4^{s_4}x_3^{r_3}x_4^{r_4}&\to G_{(0,0,2,1,-1)D}^{ij|m+\ell|kl}.
}
We note here that we have replaced $\A_{34DF''}\to g_{DF''}$ and $\A_{34DE''_1}\to g_{DE''_1}$, due to their contractions with the half-projectors in \eqref{Eq4ptind}.  Moreover, the substitutions $s_{(1|1)}^1$ and $s_{(1|2)}^2$ are exactly the same, although they do not originate from the same contractions.


\subsubsection{\texorpdfstring{$\vev{SSS\boldsymbol{e}_2}$}{SSSe2}}

The tensor structure for a scalar, a two-index antisymmetric tensor, and exchanged $\ell\boldsymbol{e}_1$ is
\eqna{
b=1:\qquad&(\FCF{b}{k}{l}{,m+\ell}{3}{4})_{\{cC\}\{dD\}\{e''E''\}}=\A_{34D_1E''_1}(\A_{34}\cdot\bar{\bar{\eta}}_2)_{D_2}[(\A_{34}\cdot\bar{\bar{\eta}}_2)_{E''}]^{\ell-1}\\
&\qquad\to(\tCF{b}{k}{l}{,m+\ell}{3}{4})_{\{cC\}\{dD\}\{e''E''\}\{F''\}}=\A_{34D_1E''_1}\A_{34D_2F''}(\A_{34E''F''})^{\ell-1},
}
where the $D$-indices are antisymmetrized once contracted with the $\boldsymbol{e}_2$ half-projector.  Clearly, we have
\eqn{b=1:\qquad n_b=\ell,\qquad i_b=1,\qquad(\tCF{b}{k}{l}{,m+1}{3}{4})_{D_2D_1E''_1F''}=\A_{34D_1E''_1}\A_{34D_2F''},}
and we must extract one $E''$-index.  Hence, the partitions are labeled diagrammatically by
\eqn{b=1:\qquad\hat{\mathcal{P}}_{13|d}^{\ell\boldsymbol{e}_1}=\Diag{1}{0}{0}{0}{0}{0}{0}{1}+\Diag{1}{0}{0}{0}{0}{0}{1}{0},}
which leads to the conformal blocks \eqref{EqCB4} in terms of Gegenbauer polynomials given by
\eqn{\mathscr{G}_{(1|1]}^{ij|m+\ell|kl}=-\frac{(\ell-1)!}{(d/2)_{\ell-1}}\left(C_{\ell-1}^{(d/2)}(X)\right)_{s_{(1|1)}^1}+\frac{(\ell-1)!}{(d/2)_{\ell-1}}\left(C_{\ell-2}^{(d/2)}(X)\right)_{s_{(1|1)}^2}.}
Using the partitions associated with each diagram and contracting the special parts together according to \eqref{EqS}, we obtain the associated substitutions
\eqna{
s_{(1|1)}^1:\alpha_2^{s_2}\alpha_3^{s_3}\alpha_4^{s_4}x_3^{r_3}x_4^{r_4}&\to(\bar{\eta}_2\cdot\mathcal{S})_{D_1}G_{(-1,0,1,1,-1)D_2}^{ij|m+\ell|kl},\\
s_{(1|1)}^2:\alpha_2^{s_2}\alpha_3^{s_3}\alpha_4^{s_4}x_3^{r_3}x_4^{r_4}&\to G_{(0,0,4,2,-2)D_1D_2}^{ij|m+\ell|kl}.
}
Due to the antisymmetry of the $D$-indices (from their contraction with the half-projector) and the fact that $G$ is totally symmetric in its indices, the second term in the conformal blocks vanishes, and we easily get a result that matches the one in\cite{Fortin:2019gck}.


\subsection{\texorpdfstring{$\vev{SRSR}$}{SRSR} and \texorpdfstring{$\vev{SSRR}$}{SSRR}}

With the conformal bootstrap in mind, we need to determine not only the conformal blocks of $\vev{SRSR}$ four-point correlation functions, but also the blocks of $\vev{SSRR}$ (the $s$- and $t$-channels).  For $R$ a defining representation which is not a scalar, there are several different cases to consider.  Here we proceed in some detail for the cases $R=V$ and $R=F$, and leave the blocks for $R$ in antisymmetric tensor representations to a forthcoming work.

Once again, in the $\vev{SSRR}$ case, the exchanged quasi-primary operators are in the $\ell\boldsymbol{e}_1$ irreducible representation, implying a decomposition \eqref{EqPExp} as in the $\vev{SSSS}$ example, and their $12$-tensor structures are given by \eqref{EqSSSS-TS}.  Hence, their rotation matrices are given by \eqref{EqSSSS-RM}, and we can focus only on the $RR$ side of the computation with $\ell\boldsymbol{e}_1$ exchange.


\subsubsection{\texorpdfstring{$\vev{SVSV}$}{SVSV} and \texorpdfstring{$\vev{SSVV}$}{SSVV}}

For $\vev{SVSV}$, there are two possible infinite towers of exchanged quasi-primary operators, $\ell\boldsymbol{e}_1$ and $\boldsymbol{e}_2+\ell\boldsymbol{e}_1$.  We first consider $\ell\boldsymbol{e}_1$ exchange.  In this case, the tensor structures are explicitly given by
\eqna{
b=1:\qquad&(\FCF{b}{k}{l}{,m+\ell}{3}{4})_{\{cC\}\{dD\}\{e''E''\}}=(\A_{34}\cdot\bar{\bar{\eta}}_2)_D[(\A_{34}\cdot\bar{\bar{\eta}}_2)_{E''}]^\ell\\
&\qquad\to(\tCF{b}{k}{l}{,m+\ell}{3}{4})_{\{cC\}\{dD\}\{e''E''\}\{F''\}}=\A_{34DF''}(\A_{34E''F''})^\ell,\\
b=2:\qquad&(\FCF{b}{k}{l}{,m+\ell}{3}{4})_{\{cC\}\{dD\}\{e''E''\}}=\A_{34DE''_1}[(\A_{34}\cdot\bar{\bar{\eta}}_2)_{E''}]^{\ell-1}\\
&\qquad\to(\tCF{b}{k}{l}{,m+\ell}{3}{4})_{\{cC\}\{dD\}\{e''E''\}\{F''\}}=\A_{34DE''_1}(\A_{34E''F''})^{\ell-1},\\
a=1:\qquad&(\FCF{a}{i}{j}{,m+\ell}{1}{2})_{\{aA\}\{bB\}\{eE\}}=(\A_{12}\cdot\bar{\eta}_3)_B[(\A_{12}\cdot\bar{\eta}_3)_E]^\ell\\
&\qquad\to(\tOPE{a}{i}{j}{,m+\ell}{1}{2})_{\{aA\}\{bB\}}^{\phantom{\{aA\}\{bB\}}\{Ee\}\{F\}}=\A_{12B}^{\phantom{12B}F}(\A_{12}^{EF})^\ell,\\
a=2:\qquad&(\FCF{a}{i}{j}{,m+\ell}{1}{2})_{\{aA\}\{bB\}\{eE\}}=\A_{12BE_1}[(\A_{12}\cdot\bar{\eta}_3)_E]^{\ell-1}\\
&\qquad\to(\tOPE{a}{i}{j}{,m+\ell}{1}{2})_{\{aA\}\{bB\}}^{\phantom{\{aA\}\{bB\}}\{Ee\}\{F\}}=\A_{12B}^{\phantom{12B}E_1}(\A_{12}^{EF})^{\ell-1},
}
so that we have
\eqna{
b=1:\qquad&n_b=\ell+1,\qquad i_b=0,\qquad(\tCF{b}{k}{l}{m}{3}{4})_{DF''}=\A_{34DF''},\\
b=2:\qquad&n_b=\ell-1,\qquad i_b=1,\qquad(\tCF{b}{k}{l}{,m+1}{3}{4})_{DE''_1}=\A_{34DE''_1},\\
a=1:\qquad&n_a=\ell+1,\qquad i_a=0,\qquad(\tCF{a}{i}{j}{m}{1}{2})_B^{\phantom{B}F}=\A_{12B}^{\phantom{12B}F},\\
a=2:\qquad&n_a=\ell-1,\qquad i_a=1,\qquad(\tCF{a}{i}{j}{,m+1}{1}{2})_B^{\phantom{B}E_1}=\A_{12B}^{\phantom{12B}E_1}.
}
With the aid of the diagrams, we can easily extract indices to find the contributions
\eqna{
a=1,b=1:\qquad&\hat{\mathcal{P}}_{13|d}^{\ell\boldsymbol{e}_1}=\Diag{1}{0}{0}{0}{0}{0}{0}{0},\\
a=1,b=2:\qquad&\hat{\mathcal{P}}_{13|d}^{\ell\boldsymbol{e}_1}=\Diag{1}{0}{0}{0}{0}{0}{0}{1}+\Diag{1}{0}{0}{0}{0}{0}{1}{0},\\
a=2,b=1:\qquad&\hat{\mathcal{P}}_{13|d}^{\ell\boldsymbol{e}_1}=\Diag{1}{0}{0}{0}{1}{0}{0}{0}+\Diag{1}{0}{0}{1}{0}{0}{0}{0},\\
a=2,b=2:\qquad&\hat{\mathcal{P}}_{13|d}^{\ell\boldsymbol{e}_1}=\Diag{1}{0}{0}{0}{1}{0}{0}{1}+\Diag{1}{0}{0}{0}{1}{0}{1}{0}+\Diag{1}{0}{0}{1}{0}{0}{0}{1}+2\times\Diag{1}{0}{0}{1}{0}{0}{1}{0}+\Diag{1}{1}{0}{0}{0}{0}{0}{0},
}
which lead directly to four conformal blocks expressed in terms of Gegenbauer polynomials \eqref{EqCB4}, namely
\eqna{
\mathscr{G}_{(1|1]}^{ij|m+\ell|kl}&=\frac{\ell!}{(d/2-1)_\ell}\left(C_\ell^{(d/2-1)}(X)\right)_{s_{(1|1)}^1},\\
\mathscr{G}_{(1|2]}^{ij|m+\ell|kl}&=-\frac{(\ell-1)!}{(d/2)_{\ell-1}}\left(C_{\ell-1}^{(d/2)}(X)\right)_{s_{(1|2)}^1}+\frac{(\ell-1)!}{(d/2)_{\ell-1}}\left(C_{\ell-2}^{(d/2)}(X)\right)_{s_{(1|2)}^2},\\
\mathscr{G}_{(2|1]}^{ij|m+\ell|kl}&=-\frac{(\ell-1)!}{(d/2)_{\ell-1}}\left(C_{\ell-1}^{(d/2)}(X)\right)_{s_{(2|1)}^1}+\frac{(\ell-1)!}{(d/2)_{\ell-1}}\left(C_{\ell-2}^{(d/2)}(X)\right)_{s_{(2|1)}^2},\\
\mathscr{G}_{(2|2]}^{ij|m+\ell|kl}&=\frac{(\ell-1)!}{\ell(d/2+1)_{\ell-2}}\left(C_{\ell-2}^{(d/2+1)}(X)\right)_{s_{(2|2)}^1}-\frac{(\ell-1)!}{\ell(d/2+1)_{\ell-2}}\left(C_{\ell-3}^{(d/2+1)}(X)\right)_{s_{(2|2)}^2}\\
&\phantom{=}\qquad-\frac{(\ell-1)!}{\ell(d/2+1)_{\ell-2}}\left(C_{\ell-3}^{(d/2+1)}(X)\right)_{s_{(2|2)}^3}-\frac{(\ell-1)!}{\ell(d/2)_{\ell-1}}\left(C_{\ell-2}^{(d/2)}(X)\right)_{s_{(2|2)}^4}\\
&\phantom{=}\qquad+\frac{(\ell-1)!}{\ell(d/2+1)_{\ell-2}}\left(C_{\ell-4}^{(d/2+1)}(X)\right)_{s_{(2|2)}^5}+\frac{(\ell-1)!}{\ell(d/2)_{\ell-1}}\left(C_{\ell-1}^{(d/2)}(X)\right)_{s_{(2|2)}^6}.
}
Using the tensor structures and the extended partitions in \eqref{EqS}, we find the substitutions
\eqna{
s_{(1|1)}^1:\alpha_2^{s_2}\alpha_3^{s_3}\alpha_4^{s_4}x_3^{r_3}x_4^{r_4}&\to G_{(0,1,3,1,-1)BD}^{ij|m+\ell|kl},\\
s_{(1|2)}^1:\alpha_2^{s_2}\alpha_3^{s_3}\alpha_4^{s_4}x_3^{r_3}x_4^{r_4}&\to(\bar{\eta}_2\cdot\mathcal{S})_DG_{(-1,1,0,0,0)B}^{ij|m+\ell|kl},\\
s_{(1|2)}^2:\alpha_2^{s_2}\alpha_3^{s_3}\alpha_4^{s_4}x_3^{r_3}x_4^{r_4}&\to G_{(0,1,3,1,-1)BD}^{ij|m+\ell|kl},\\
s_{(2|1)}^1:\alpha_2^{s_2}\alpha_3^{s_3}\alpha_4^{s_4}x_3^{r_3}x_4^{r_4}&\to(\mathcal{S}\cdot\bar{\eta}_4)_BG_{(-1,-1,2,0,-2)D}^{ij|m+\ell|kl},\\
s_{(2|1)}^2:\alpha_2^{s_2}\alpha_3^{s_3}\alpha_4^{s_4}x_3^{r_3}x_4^{r_4}&\to G_{(-2,-1,3,1,-1)BD}^{ij|m+\ell|kl},\\
s_{(2|2)}^1:\alpha_2^{s_2}\alpha_3^{s_3}\alpha_4^{s_4}x_3^{r_3}x_4^{r_4}&\to(\mathcal{S}\cdot\bar{\eta}_4)_B(\bar{\eta}_2\cdot\mathcal{S})_DG_{(-2,-1,-1,-1,-1)}^{ij|m+\ell|kl},\\
s_{(2|2)}^2:\alpha_2^{s_2}\alpha_3^{s_3}\alpha_4^{s_4}x_3^{r_3}x_4^{r_4}&\to(\mathcal{S}\cdot\bar{\eta}_4)_BG_{(-1,-1,2,0,-2)D}^{ij|m+\ell|kl},\\
s_{(2|2)}^3:\alpha_2^{s_2}\alpha_3^{s_3}\alpha_4^{s_4}x_3^{r_3}x_4^{r_4}&\to(\bar{\eta}_2\cdot\mathcal{S})_DG_{(-3,-1,0,0,0)B}^{ij|m+\ell|kl},\\
s_{(2|2)}^4:\alpha_2^{s_2}\alpha_3^{s_3}\alpha_4^{s_4}x_3^{r_3}x_4^{r_4}&\to G_{(-2,-1,3,1,-1)BD}^{ij|m+\ell|kl},\\
s_{(2|2)}^5:\alpha_2^{s_2}\alpha_3^{s_3}\alpha_4^{s_4}x_3^{r_3}x_4^{r_4}&\to G_{(-2,-1,3,1,-1)BD}^{ij|m+\ell|kl},\\
s_{(2|2)}^6:\alpha_2^{s_2}\alpha_3^{s_3}\alpha_4^{s_4}x_3^{r_3}x_4^{r_4}&\to\mathcal{S}_{BD}G_{(-1,-1,0,0,0)}^{ij|m+\ell|kl},
}
by straightforward contraction.

Before proceeding to consider the remaining infinite tower of exchanged quasi-primary operators, we determine the rotation matrix $\ell\boldsymbol{e}_1$.  Applying \eqref{EqCB3}, it is straightforward to get
\eqna{
(R_{ij,m+\ell}^{-1})_{1,1}&={}_1\kappa_{(0,0,0,0,0,0,0,1,0)}^{ij|m+\ell}+{}_1\kappa_{(0,0,0,1,0,0,0,0,0)}^{ij|m+\ell},\\
(R_{ij,m+\ell}^{-1})_{1,2}&={}_1\kappa_{(0,0,0,0,0,0,1,0,0)}^{ij|m+\ell}+{}_1\kappa_{(0,0,0,0,0,0,1,0,1)}^{ij|m+\ell},\\
(R_{ij,m+\ell}^{-1})_{2,1}&=-{}_2\kappa_{(0,0,1,0,0,0,0,1,0)}^{ij|m+\ell}-{}_2\kappa_{(0,0,1,0,0,1,0,0,0)}^{ij|m+\ell}-\frac{1}{2}{}_2\kappa_{(0,0,1,1,0,0,0,0,0)}^{ij|m+\ell},\\
(R_{ij,m+\ell}^{-1})_{2,2}&={}_2\kappa_{(0,0,0,0,1,0,0,0,0)}^{ij|m+\ell}-{}_2\kappa_{(0,0,1,0,0,0,1,0,0)}^{ij|m+\ell}-{}_2\kappa_{(0,0,1,0,0,0,1,0,1)}^{ij|m+\ell}+{}_2\kappa_{(1,0,0,0,0,0,0,0,0)}^{ij|m+\ell},
}
which agrees with \cite{Fortin:2019gck} when $\ell=1$.

The remaining infinite tower of exchanged quasi-primary operators corresponds to the irreducible representations $\boldsymbol{e}_2+\ell\boldsymbol{e}_1$, with the projection operators given by \eqref{EqPemple1} with $m=2$.  With $E_1$ and $E_2$ denoting the antisymmetric indices on the projection operator, the tensor structures are
\eqna{
b=1:\qquad&(\FCF{b}{k}{l}{,m+\ell}{3}{4})_{\{cC\}\{dD\}\{e''E''\}}=\A_{34DE''_1}(\A_{34}\cdot\bar{\bar{\eta}}_2)_{E''_2}[(\A_{34}\cdot\bar{\bar{\eta}}_2)_{E''}]^\ell\\
&\qquad\to(\tCF{b}{k}{l}{,m+\ell}{3}{4})_{\{cC\}\{dD\}\{e''E''\}\{F''\}}=\A_{34DE''_1}\A_{34E''_2F''}(\A_{34E''F''})^\ell,\\
a=1:\qquad&(\FCF{a}{i}{j}{,m+\ell}{1}{2})_{\{aA\}\{bB\}\{eE\}}=\A_{12BE_1}(\A_{12}\cdot\bar{\eta}_3)_{E_2}[(\A_{12}\cdot\bar{\eta}_3)_E]^\ell\\
&\qquad\to(\tOPE{a}{i}{j}{,m+\ell}{1}{2})_{\{aA\}\{bB\}}^{\phantom{\{aA\}\{bB\}}\{Ee\}\{F\}}=\A_{12B}^{\phantom{12B}E_1}\A_{12}^{E_2F}(\A_{12}^{EF})^\ell
}
so that
\eqna{
b=1:\qquad&n_b=\ell+1,\qquad i_b=0,\qquad(\tCF{b}{k}{l}{m}{3}{4})_{DE''_2E''_1F''}=\A_{34DE''_1}\A_{34E''_2F''},\\
a=1:\qquad&n_a=\ell+1,\qquad i_a=0,\qquad(\tCF{a}{i}{j}{m}{1}{2})_B^{\phantom{B}E_1E_2F}=\A_{12B}^{\phantom{12B}E_1}\A_{12}^{E_2F}.
}
It is evident that we do not need to extract any indices, and we may therefore bypass the diagrammatic notation altogether.  From \eqref{EqPemple1}, we see that there are six different contributions involved in the decomposition \eqref{EqPExp} of the projection operator.  In consequence, we have six contributions to the conformal blocks \eqref{EqCB4}, namely
\eqna{
\mathscr{G}_{(1|1]}^{ij|m+\ell|kl}&=\frac{2\ell!}{(\ell+2)(d/2)_\ell}\left(C_\ell^{(d/2)}(X)\right)_{s_{(1|1)}^1}-\frac{2\ell!}{(\ell+2)(d/2+1)_{\ell-1}}\left(C_{\ell-1}^{(d/2+1)}(X)\right)_{s_{(1|1)}^2}\\
&\phantom{=}\qquad-\frac{2\ell!}{(\ell+2)(d/2+1)_{\ell-1}}\left(C_{\ell-1}^{(d/2+1)}(X)\right)_{s_{(1|1)}^3}\\
&\phantom{=}\qquad+\frac{2\ell!(\ell+d/2)(d+\ell-1)}{(\ell+2)(d+\ell-2)(d/2)_\ell}\left(C_{\ell-1}^{(d/2)}(X)\right)_{s_{(1|1)}^4}\\
&\phantom{=}\qquad-\frac{2\ell!(\ell+d/2)}{(\ell+2)(d+\ell-2)(d/2+1)_{\ell-1}}\left(C_{\ell-2}^{(d/2+1)}(X)\right)_{s_{(1|1)}^5}\\
&\phantom{=}\qquad+\frac{\ell!}{(\ell+2)(d/2+1)_{\ell-1}}\left(C_{\ell-2}^{(d/2+1)}(X)\right)_{s_{(1|1)}^6},
}
with the substitutions
\eqna{
s_{(1|1)}^1:\alpha_2^{s_2}\alpha_3^{s_3}\alpha_4^{s_4}x_3^{r_3}x_4^{r_4}&\to-(\mathcal{S}\cdot\A_{34})_{B[D}(\bar{\eta}_2\cdot\mathcal{S}\cdot\A_{34})_{F'']}\left(G_{(0,1,1,1,-1)}^{ij|m+\ell|kl}\right)^{F''},\\
s_{(1|1)}^2:\alpha_2^{s_2}\alpha_3^{s_3}\alpha_4^{s_4}x_3^{r_3}x_4^{r_4}&\to-\frac{1}{2}(\mathcal{S}\cdot\A_{34})_{B[D}(\bar{\eta}_2\cdot\mathcal{S}\cdot\A_{34})_{F'']}(\bar{\eta}_2\cdot\mathcal{S}\cdot\A_{34})_{F''}\left(G_{(-1,1,2,2,-2)}^{ij|m+\ell|kl}\right)^{F''^2},\\
s_{(1|1)}^3:\alpha_2^{s_2}\alpha_3^{s_3}\alpha_4^{s_4}x_3^{r_3}x_4^{r_4}&\to(\mathcal{S}\cdot\A_{34})_{B[D}\A_{34F'']F''}\left(G_{(1,1,4,2,-2)}^{ij|m+\ell|kl}\right)^{F''^2}\\
&\phantom{\to}\qquad-\frac{1}{2}(\bar{\eta}_2\cdot\mathcal{S}\cdot\A_{34})_{[D}\A_{34F'']F''}\left(G_{(-1,1,4,2,-2)}^{ij|m+\ell|kl}\right)_B^{F''^2},\\
s_{(1|1)}^4:\alpha_2^{s_2}\alpha_3^{s_3}\alpha_4^{s_4}x_3^{r_3}x_4^{r_4}&\to(\mathcal{S}\cdot\A_{34})_{B[D}\A_{34F'']F''}\left(G_{(1,1,4,2,-2)}^{ij|m+\ell|kl}\right)^{F''^2}\\
&\phantom{\to}\qquad-\frac{1}{2}(\bar{\eta}_2\cdot\mathcal{S}\cdot\A_{34})_{[D}\A_{34F'']F''}\left(G_{(-1,1,4,2,-2)}^{ij|m+\ell|kl}\right)_B^{F''^2},\\
s_{(1|1)}^5:\alpha_2^{s_2}\alpha_3^{s_3}\alpha_4^{s_4}x_3^{r_3}x_4^{r_4}&\to(\mathcal{S}\cdot\A_{34})_{BF''}(\bar{\eta}_2\cdot\mathcal{S}\cdot\A_{34})_{[D}\A_{34F'']F''}\left(G_{(0,1,5,3,-3)}^{ij|m+\ell|kl}\right)^{F''^3}\\
&\phantom{\to}\qquad-\frac{1}{2}(\bar{\eta}_2\cdot\mathcal{S}\cdot\A_{34})_{[D}\A_{34F'']F''}(\bar{\eta}_2\cdot\mathcal{S}\cdot\A_{34})_{F''}\left(G_{(-2,1,5,3,-3)}^{ij|m+\ell|kl}\right)_B^{F''^3},\\
s_{(1|1)}^6:\alpha_2^{s_2}\alpha_3^{s_3}\alpha_4^{s_4}x_3^{r_3}x_4^{r_4}&\to4(\mathcal{S}\cdot\A_{34})_{B[D}(\bar{\eta}_2\cdot\mathcal{S}\cdot\A_{34})_{F'']}\left(G_{(0,1,1,1,-1)}^{ij|m+\ell|kl}\right)^{F''},
}
from \eqref{EqS}.  Here, we replaced $\A_{12}^{E_2F}\to-\bar{\eta}_2^{E_2}\bar{\eta}_1^F$ and performed the contractions using the contiguous relations \eqref{EqContG}.  For example, we used
\eqn{\bar{\eta}_2\cdot\mathcal{S}\cdot\A_{13}\cdot\mathcal{S}^T\cdot\bar{\eta}_2=-2G_{(2,0,2,0,0)}^{ij|m+\ell|kl},\qquad(\mathcal{S}\cdot\A_{13}\cdot\mathcal{S}^T\cdot\bar{\eta}_2)_B\to-G_{(0,0,2,0,0)B}^{ij|m+\ell|kl},}
where the replacements are warranted by the contractions with the half-projectors in \eqref{Eq4ptind}.

To complete our analysis of $\vev{SVSV}$, we need to also compute the rotation matrix for the exchanged quasi-primary operators in the $\boldsymbol{e}_2+\ell\boldsymbol{e}_1$ representation.  Using \eqref{EqCB3} directly with the tensor structures yields
\eqna{
(R_{ij,m+\ell}^{-1})_{1,1}&=2{}_1\kappa_{(0,0,0,0,2,0,0,1,0)}+2{}_1\kappa_{(0,0,0,1,2,0,0,0,0)}+{}_1\kappa_{(0,0,1,0,1,0,0,2,0)}-\frac{1}{2}{}_1\kappa_{(0,0,1,0,1,0,1,1,0)}\\
&\phantom{=}\qquad-\frac{1}{2}{}_1\kappa_{(0,0,1,0,1,0,1,1,1)}+{}_1\kappa_{(0,0,1,0,1,1,0,1,0)}+2{}_1\kappa_{(0,0,1,0,2,0,0,0,0)}+\frac{1}{2}{}_1\kappa_{(0,0,1,1,1,0,0,1,0)}\\
&\phantom{=}\qquad-\frac{1}{4}{}_1\kappa_{(0,0,1,1,1,0,1,0,0)}-\frac{1}{4}{}_1\kappa_{(0,0,1,1,1,0,1,0,1)}+\frac{1}{2}{}_1\kappa_{(0,0,1,1,1,1,0,0,0)}+\frac{1}{3}{}_1\kappa_{(0,0,1,2,1,0,0,0,0)}\\
&\phantom{=}\qquad+\frac{1}{2}{}_1\kappa_{(0,0,2,0,1,0,0,1,0)}-\frac{1}{2}{}_1\kappa_{(0,0,2,0,1,0,1,0,0)}-\frac{1}{2}{}_1\kappa_{(0,0,2,0,1,0,1,0,1)}+{}_1\kappa_{(0,0,2,0,1,1,0,0,0)}\\
&\phantom{=}\qquad+\frac{1}{3}{}_1\kappa_{(0,0,2,1,1,0,0,0,0)}+2{}_1\kappa_{(0,1,0,0,2,0,0,0,0)}+\frac{1}{4}{}_1\kappa_{(0,1,1,0,1,0,0,1,0)}+\frac{1}{6}{}_1\kappa_{(0,1,1,1,1,0,0,0,0)}\\
&\phantom{=}\qquad+\frac{1}{3}{}_1\kappa_{(0,1,2,0,1,0,0,0,0)}+{}_1\kappa_{(1,0,0,0,1,0,0,1,0)}+\frac{2}{3}{}_1\kappa_{(1,0,0,1,1,0,0,0,0)}+{}_1\kappa_{(1,0,1,0,1,0,0,0,0)},
}
where we have taken into account the antisymmetry of the pair $E_1$, $E_2$.

Proceeding further, we find that the case $\vev{SSVV}$ is much simpler, as there is only one infinite tower of exchanged quasi-primary operators involved, namely $\ell\boldsymbol{e}_1$.  The tensor structures are given by
\eqna{
b=1:\qquad&(\FCF{b}{k}{l}{,m+\ell}{3}{4})_{\{cC\}\{dD\}\{e''E''\}}=(\A_{34}\cdot\bar{\bar{\eta}}_2)_C(\A_{34}\cdot\bar{\bar{\eta}}_2)_D[(\A_{34}\cdot\bar{\bar{\eta}}_2)_{E''}]^\ell\\
&\qquad\to(\tCF{b}{k}{l}{,m+\ell}{3}{4})_{\{cC\}\{dD\}\{e''E''\}\{F''\}}=\A_{34CF''}\A_{34DF''}(\A_{34E''F''})^\ell,\\
b=2:\qquad&(\FCF{b}{k}{l}{,m+\ell}{3}{4})_{\{cC\}\{dD\}\{e''E''\}}=\A_{34CD}[(\A_{34}\cdot\bar{\bar{\eta}}_2)_{E''}]^\ell\\
&\qquad\to(\tCF{b}{k}{l}{,m+\ell}{3}{4})_{\{cC\}\{dD\}\{e''E''\}\{F''\}}=\A_{34CD}(\A_{34E''F''})^\ell,\\
b=3:\qquad&(\FCF{b}{k}{l}{,m+\ell}{3}{4})_{\{cC\}\{dD\}\{e''E''\}}=\A_{34CE''_1}(\A_{34}\cdot\bar{\bar{\eta}}_2)_D[(\A_{34}\cdot\bar{\bar{\eta}}_2)_{E''}]^{\ell-1}\\
&\qquad\to(\tCF{b}{k}{l}{,m+\ell}{3}{4})_{\{cC\}\{dD\}\{e''E''\}\{F''\}}=\A_{34CE''_1}\A_{34DF''}(\A_{34E''F''})^{\ell-1},\\
b=4:\qquad&(\FCF{b}{k}{l}{,m+\ell}{3}{4})_{\{cC\}\{dD\}\{e''E''\}}=(\A_{34}\cdot\bar{\bar{\eta}}_2)_C\A_{34DE''_1}[(\A_{34}\cdot\bar{\bar{\eta}}_2)_{E''}]^{\ell-1}\\
&\qquad\to(\tCF{b}{k}{l}{,m+\ell}{3}{4})_{\{cC\}\{dD\}\{e''E''\}\{F''\}}=\A_{34CF''}\A_{34DE''_1}(\A_{34E''F''})^{\ell-1}\\
b=5:\qquad&(\FCF{b}{k}{l}{,m+\ell}{3}{4})_{\{cC\}\{dD\}\{e''E''\}}=\A_{34CE''_2}\A_{34DE''_2}[(\A_{34}\cdot\bar{\bar{\eta}}_2)_{E''}]^{\ell-2}\\
&\qquad\to(\tCF{b}{k}{l}{,m+\ell}{3}{4})_{\{cC\}\{dD\}\{e''E''\}\{F''\}}=\A_{34CE''_2}\A_{34DE''_2}(\A_{34E''F''})^{\ell-2},
}
which imply
\eqna{
b=1:\qquad&n_b=\ell+2,\qquad i_b=0,\qquad(\tCF{b}{k}{l}{m}{3}{4})_{CDF''^2}=\A_{34CF''}\A_{34DF''},\\
b=2:\qquad&n_b=\ell,\qquad i_b=0,\qquad(\tCF{b}{k}{l}{m}{3}{4})_{CDF''^2}=\A_{34CD},\\
b=3:\qquad&n_b=\ell,\qquad i_b=1,\qquad(\tCF{b}{k}{l}{,m+1}{3}{4})_{CDE''_1F''}=\A_{34CE''_1}\A_{34DF''},\\
b=4:\qquad&n_b=\ell,\qquad i_b=1,\qquad(\tCF{b}{k}{l}{,m+1}{3}{4})_{CDE''_1F''}=\A_{34CF''}\A_{34DE''_1},\\
b=5:\qquad&n_b=\ell-2,\qquad i_b=2,\qquad(\tCF{b}{k}{l}{,m+2}{3}{4})_{CDE''_2E''_1}=\A_{34CE''_1}\A_{34DE''_2}.
}
Given these, we find that we therefore need to extract zero, zero, one, one, and two $E''$-indices, respectively, which results in the diagrams
\eqna{
b\in\{1,2\}:\qquad&\hat{\mathcal{P}}_{13|d}^{\ell\boldsymbol{e}_1}=\Diag{1}{0}{0}{0}{0}{0}{0}{0},\\
b\in\{3,4\}:\qquad&\hat{\mathcal{P}}_{13|d}^{\ell\boldsymbol{e}_1}=\Diag{1}{0}{0}{0}{0}{0}{0}{1}+\Diag{1}{0}{0}{0}{0}{0}{1}{0},\\
b=5:\qquad&\hat{\mathcal{P}}_{13|d}^{\ell\boldsymbol{e}_1}=\Diag{1}{0}{0}{0}{0}{0}{0}{2}+\Diag{1}{0}{0}{0}{0}{0}{1}{1}+\Diag{1}{0}{0}{0}{0}{0}{2}{0}+\Diag{1}{0}{0}{0}{0}{1}{0}{0}.
}
The associated extended partitions allow us to straightforwardly write the conformal blocks as
\eqna{
b\in\{1,2\}:\qquad&\mathscr{G}_{(1|b]}^{ij|m+\ell|kl}=\frac{\ell!}{(d/2-1)_\ell}\left(C_\ell^{(d/2-1)}(X)\right)_{s_{(1|b)}^1},\\
b\in\{3,4\}:\qquad&\mathscr{G}_{(1|b]}^{ij|m+\ell|kl}=-\frac{(\ell-1)!}{(d/2)_{\ell-1}}\left(C_{\ell-1}^{(d/2)}(X)\right)_{s_{(1|b)}^1}+\frac{(\ell-1)!}{(d/2)_{\ell-1}}\left(C_{\ell-2}^{(d/2)}(X)\right)_{s_{(1|b)}^2},\\
b=5:\qquad&\mathscr{G}_{(1|b]}^{ij|m+\ell|kl}=\frac{(\ell-2)!}{(d/2+1)_{\ell-2}}\left(C_{\ell-2}^{(d/2+1)}(X)\right)_{s_{(1|b)}^1}-\frac{2(\ell-2)!}{(d/2+1)_{\ell-2}}\left(C_{\ell-3}^{(d/2+1)}(X)\right)_{s_{(1|b)}^2}\\
&\phantom{\mathscr{G}_{(1|b]}^{ij|m+\ell|kl}=}\qquad+\frac{(\ell-2)!}{(d/2+1)_{\ell-2}}\left(C_{\ell-4}^{(d/2+1)}(X)\right)_{s_{(1|b)}^3}+\frac{(\ell-2)!}{(d/2)_{\ell-1}}\left(C_{\ell-2}^{(d/2)}(X)\right)_{s_{(1|b)}^4},
}
using \eqref{EqCB4}.  Moreover, we can easily extract the substitution rules \eqref{EqS} for each block.  We find these to be 
\eqna{
s_{(1|1)}^1:\alpha_2^{s_2}\alpha_3^{s_3}\alpha_4^{s_4}x_3^{r_3}x_4^{r_4}&\to G_{(0,0,4,2,-2)CD}^{ij|m+\ell|kl},\\
s_{(1|2)}^1:\alpha_2^{s_2}\alpha_3^{s_3}\alpha_4^{s_4}x_3^{r_3}x_4^{r_4}&\to g_{CD}G_{(0,0,0,0,0)}^{ij|m+\ell|kl},\\
s_{(1|3)}^1:\alpha_2^{s_2}\alpha_3^{s_3}\alpha_4^{s_4}x_3^{r_3}x_4^{r_4}&\to(\bar{\eta}_2\cdot\mathcal{S})_CG_{(-1,0,1,1,-1)D}^{ij|m+\ell|kl},\\
s_{(1|3)}^2:\alpha_2^{s_2}\alpha_3^{s_3}\alpha_4^{s_4}x_3^{r_3}x_4^{r_4}&\to G_{(0,0,4,2,-2)CD}^{ij|m+\ell|kl},\\
s_{(1|4)}^1:\alpha_2^{s_2}\alpha_3^{s_3}\alpha_4^{s_4}x_3^{r_3}x_4^{r_4}&\to(\bar{\eta}_2\cdot\mathcal{S})_DG_{(-1,0,1,1,-1)C}^{ij|m+\ell|kl},\\
s_{(1|4)}^2:\alpha_2^{s_2}\alpha_3^{s_3}\alpha_4^{s_4}x_3^{r_3}x_4^{r_4}&\to G_{(0,0,4,2,-2)CD}^{ij|m+\ell|kl},\\
s_{(1|5)}^1:\alpha_2^{s_2}\alpha_3^{s_3}\alpha_4^{s_4}x_3^{r_3}x_4^{r_4}&\to(\bar{\eta}_2\cdot\mathcal{S})_C(\bar{\eta}_2\cdot\mathcal{S})_DG_{(-2,0,-2,0,0)}^{ij|m+\ell|kl},\\
s_{(1|5)}^2:\alpha_2^{s_2}\alpha_3^{s_3}\alpha_4^{s_4}x_3^{r_3}x_4^{r_4}&\to(\bar{\eta}_2\cdot\mathcal{S})_{(C}G_{(-1,0,1,1,-1)D)}^{ij|m+\ell|kl},\\
s_{(1|5)}^3:\alpha_2^{s_2}\alpha_3^{s_3}\alpha_4^{s_4}x_3^{r_3}x_4^{r_4}&\to G_{(0,0,4,2,-2)CD}^{ij|m+\ell|kl},\\
s_{(1|5)}^4:\alpha_2^{s_2}\alpha_3^{s_3}\alpha_4^{s_4}x_3^{r_3}x_4^{r_4}&\to g_{CD}G_{(0,0,0,0,0)}^{ij|m+\ell|kl}.
}
Again, here we replaced all metrics of the type $\A_{34CF''}$ by $g_{CF''}$ without loss of generality.

Although written differently, we have checked that all the conformal blocks above match the ones found in \cite{Fortin:2019gck}.


\subsubsection{\texorpdfstring{$\vev{SFSF}$}{SFSF} and \texorpdfstring{$\vev{SSFF}$}{SSFF}}

We next consider some examples involving fermions, namely $\vev{SFSF}$ and $\vev{SSFF}$.  Technically, for fermionic representations we should in principle consider odd and even dimensions separately.  However, a study of the tensor structures shows that the even-dimensional case corresponds to half of the odd-dimensional case.  We may therefore restrict attention to fermions in odd dimensions in our analysis, since the even-dimensional case may be straightforwardly derived from these results.

We first analyze the $\vev{SFSF}$ conformal blocks.  For these, the sole exchanged quasi-primary operators are in the $\boldsymbol{e}_r+\ell\boldsymbol{e}_1$ irreducible representation, with projection operators \eqref{EqPerple1odd}.  There are thus two terms in the sum over $t$, each with nontrivial special parts and shifted projection operators.

Moreover, in our simple basis, the tensor structures are given by
\eqna{
b=1:\qquad&(\FCF{b}{k}{l}{,m+\ell}{3}{4})_{\{cC\}\{dD\}\{e''E''\}}=(C_\Gamma^{-1})_{de''}[(\A_{34}\cdot\bar{\bar{\eta}}_2)_{E''}]^\ell\\
&\qquad\to(\tCF{b}{k}{l}{,m+\ell}{3}{4})_{\{cC\}\{dD\}\{e''E''\}\{F''\}}=(C_\Gamma^{-1})_{de''}(\A_{34E''F''})^\ell,\\
b=2:\qquad&(\FCF{b}{k}{l}{,m+\ell}{3}{4})_{\{cC\}\{dD\}\{e''E''\}}=(\bar{\bar{\eta}}_2\cdot\Gamma_{34}C_\Gamma^{-1})_{de''}[(\A_{34}\cdot\bar{\bar{\eta}}_2)_{E''}]^\ell\\
&\qquad\to(\tCF{b}{k}{l}{,m+\ell}{3}{4})_{\{cC\}\{dD\}\{e''E''\}\{F''\}}=(\Gamma_{34F''}C_\Gamma^{-1})_{de''}(\A_{34E''F''})^\ell,\\
a=1:\qquad&(\FCF{a}{i}{j}{,m+\ell}{1}{2})_{\{aA\}\{bB\}\{eE\}}=(C_\Gamma^{-1})_{be}[(\A_{12}\cdot\bar{\eta}_3)_E]^\ell\\
&\qquad\to(\tOPE{a}{i}{j}{,m+\ell}{1}{2})_{\{aA\}\{bB\}}^{\phantom{\{aA\}\{bB\}}\{Ee\}\{F\}}=\delta_b^{\phantom{b}e}(\A_{12}^{EF})^\ell,\\
a=2:\qquad&(\FCF{a}{i}{j}{,m+\ell}{1}{2})_{\{aA\}\{bB\}\{eE\}}=(\bar{\eta}_3\cdot\Gamma_{12}C_\Gamma^{-1})_{be}[(\A_{12}\cdot\bar{\eta}_3)_E]^\ell\\
&\qquad\to(\tOPE{a}{i}{j}{,m+\ell}{1}{2})_{\{aA\}\{bB\}}^{\phantom{\{aA\}\{bB\}}\{Ee\}\{F\}}=(\Gamma_{12}^F)_b^{\phantom{b}e}(\A_{12}^{EF})^\ell,
}
so that
\eqna{
b=1:\qquad&n_b=\ell,\qquad i_b=0,\qquad(\tCF{b}{k}{l}{m}{3}{4})_{de''}=(C_\Gamma^{-1})_{de''},\\
b=2:\qquad&n_b=\ell+1,\qquad i_b=0,\qquad(\tCF{b}{k}{l}{m}{3}{4})_{de''F''}=(\Gamma_{34F''}C_\Gamma^{-1})_{de''},\\
a=1:\qquad&n_a=\ell,\qquad i_a=0,\qquad(\tCF{a}{i}{j}{m}{1}{2})_b^{\phantom{d}e}=\delta_b^{\phantom{b}e},\\
a=2:\qquad&n_a=\ell+1,\qquad i_a=0,\qquad(\tCF{a}{i}{j}{m}{1}{2})_b^{\phantom{d}eF}=(\Gamma_{12}^F)_b^{\phantom{b}e}.
}
Since $i_a=i_b=0$ for all tensor structures, we do not need to extract any indices from the projection operators in this case.  Diagramatically, we therefore have
\eqn{\forall\,a,b:\qquad\mathscr{A}_1\hat{\mathcal{Q}}_{13|1}^{\boldsymbol{e}_r}\hat{\mathcal{P}}_{13|d+2}^{\ell\boldsymbol{e}_1}+\mathscr{A}_2\hat{\mathcal{Q}}_{13|2}^{\boldsymbol{e}_r+\boldsymbol{e}_1}\hat{\mathcal{P}}_{13|d+2}^{(\ell-1)\boldsymbol{e}_1}=\mathscr{A}_1\hat{\mathcal{Q}}_{13|1}^{\boldsymbol{e}_r}\times\Diag{1}{0}{0}{0}{0}{0}{0}{0}+\mathscr{A}_2\hat{\mathcal{Q}}_{13|2}^{\boldsymbol{e}_r+\boldsymbol{e}_1}\times\Diag{1}{0}{0}{0}{0}{0}{0}{0},}
and the four different conformal blocks have the same form when expressed in terms of Gegenbauer polynomials, namely
\eqn{\forall\,a,b:\qquad\mathscr{G}_{(a|b]}^{ij|m+\ell|kl}=\frac{\ell!}{(d/2)_\ell}\left(C_\ell^{(d/2)}(X)\right)_{s_{(a|b)}^1}+\frac{\ell!}{2(d/2)_\ell}\left(C_{\ell-1}^{(d/2)}(X)\right)_{s_{(a|b)}^2},}
with the explicit values for $\mathscr{A}_{t=1,2}$ \eqref{EqPerple1odd}.  This is however not the case for their associated substitutions, which are all different due to the tensor structures and the different values of $n_a$ and $n_b$.  They are
\eqna{
s_{(1|1)}^1:\alpha_2^{s_2}\alpha_3^{s_3}\alpha_4^{s_4}x_3^{r_3}x_4^{r_4}&\to-(\Gamma_{F''}\bar{\eta}_3\cdot\Gamma\,\Gamma_{F''}C_\Gamma^{-T})_{bd}\left(G_{(0,0,4,3,-1)}^{ij|m+\ell|kl}\right)^{F''^2}\\
&=-2(\Gamma_{F''}C_\Gamma^{-T})_{bd}\left(G_{(0,0,2,1,-1)}^{ij|m+\ell|kl}\right)^{F''},\\
s_{(1|1)}^2:\alpha_2^{s_2}\alpha_3^{s_3}\alpha_4^{s_4}x_3^{r_3}x_4^{r_4}&\to-(\Gamma_{F''}\bar{\eta}_3\cdot\Gamma\,\bar{\eta}_2\cdot\mathcal{S}\cdot\Gamma_{13}\Gamma_{134F''}\Gamma_{F''}C_\Gamma^{-T})_{bd}\left(G_{(-1,0,5,4,-2)}^{ij|m+\ell|kl}\right)^{F''^3},\\
s_{(1|2)}^1:\alpha_2^{s_2}\alpha_3^{s_3}\alpha_4^{s_4}x_3^{r_3}x_4^{r_4}&\to-(\Gamma_{F''}\bar{\eta}_3\cdot\Gamma\,\Gamma_{F''}(\Gamma_{34F''}C_\Gamma^{-1})^T)_{bd}\left(G_{(0,0,6,4,-2)}^{ij|m+\ell|kl}\right)^{F''^3}\\
&=-2(\Gamma_{F''}(\Gamma_{34F''}C_\Gamma^{-1})^T)_{bd}\left(G_{(0,0,4,2,-2)}^{ij|m+\ell|kl}\right)^{F''^2},\\
s_{(1|2)}^2:\alpha_2^{s_2}\alpha_3^{s_3}\alpha_4^{s_4}x_3^{r_3}x_4^{r_4}&\to-(\Gamma_{F''}\bar{\eta}_3\cdot\Gamma\,\bar{\eta}_2\cdot\mathcal{S}\cdot\Gamma_{13}\Gamma_{134F''}\Gamma_{F''}(\Gamma_{34F''}C_\Gamma^{-1})^T)_{bd}\left(G_{(-1,0,6,5,-3)}^{ij|m+\ell|kl}\right)^{F''^4},\\
s_{(2|1)}^1:\alpha_2^{s_2}\alpha_3^{s_3}\alpha_4^{s_4}x_3^{r_3}x_4^{r_4}&\to-(\Gamma_{12}^F\Gamma_{F''}\bar{\eta}_3\cdot\Gamma\,\Gamma_{F''}C_\Gamma^{-T})_{bd}\left(G_{(0,1,5,3,-1)}^{ij|m+\ell|kl}\right)_F^{F''^2}\\
&=-2(\Gamma_{12}^F\Gamma_{F''}C_\Gamma^{-T})_{bd}\left(G_{(0,1,3,1,-1)}^{ij|m+\ell|kl}\right)_F^{F''},\\
s_{(2|1)}^2:\alpha_2^{s_2}\alpha_3^{s_3}\alpha_4^{s_4}x_3^{r_3}x_4^{r_4}&\to-(\Gamma_{12}^F\Gamma_{F''}\bar{\eta}_3\cdot\Gamma\,\bar{\eta}_2\cdot\mathcal{S}\cdot\Gamma_{13}\Gamma_{134F''}\Gamma_{F''}C_\Gamma^{-T})_{bd}\left(G_{(-1,1,6,4,-2)}^{ij|m+\ell|kl}\right)_F^{F''^3},\\
s_{(2|2)}^1:\alpha_2^{s_2}\alpha_3^{s_3}\alpha_4^{s_4}x_3^{r_3}x_4^{r_4}&\to-(\Gamma_{12}^F\Gamma_{F''}\bar{\eta}_3\cdot\Gamma\,\Gamma_{F''}(\Gamma_{34F''}C_\Gamma^{-1}))_{bd}\left(G_{(0,1,7,4,-2)}^{ij|m+\ell|kl}\right)_F^{F''^3}\\
&=-2(\Gamma_{12}^F\Gamma_{F''}(\Gamma_{34F''}C_\Gamma^{-1})^T)_{bd}\left(G_{(0,1,5,2,-2)}^{ij|m+\ell|kl}\right)_F^{F''^2},\\
s_{(2|2)}^2:\alpha_2^{s_2}\alpha_3^{s_3}\alpha_4^{s_4}x_3^{r_3}x_4^{r_4}&\to-(\Gamma_{12}^F\Gamma_{F''}\bar{\eta}_3\cdot\Gamma\,\bar{\eta}_2\cdot\mathcal{S}\cdot\Gamma_{13}\Gamma_{134F''}\Gamma_{F''}(\Gamma_{34F''}C_\Gamma^{-1})^T)_{bd}\left(G_{(-1,1,8,5,-3)}^{ij|m+\ell|kl}\right)_F^{F''^4},
}
upon invoking \eqref{EqS}.  Here, we do not need to distinguish the $F''$-indices, as they are fully symmetrized.  Moreover, for some of the substitutions, we have simplified the result using the usual $\Gamma$-matrix algebra.

Finally, by straightforward substitution in \eqref{EqCB3} (with $r_0=r_3=0$ but possible sums on $s_0$, $s_3$ and $t$) we obtain the three-point conformal blocks which then lead to the rotation matrix
\eqna{
(R_{ij,m+\ell}^{-1})_{1,1}&=0,\\
(R_{ij,m+\ell}^{-1})_{1,2}&=(-1)^{r+1}{}_1\kappa_{(0,0,1,0,0,0,0,0,0)}^{ij|m+\ell},\\
(R_{ij,m+\ell}^{-1})_{2,1}&=(-1)^r\left[{}_2\kappa_{(0,0,1,0,0,0,0,1,0)}^{ij|m+\ell}-{}_2\kappa_{(0,0,1,0,0,0,1,0,0)}^{ij|m+\ell}-{}_2\kappa_{(0,0,1,0,0,0,1,0,1)}^{ij|m+\ell}+{}_2\kappa_{(0,0,1,1,0,0,0,0,0)}^{ij|m+\ell}\right.\\
&\phantom{=}\qquad\left.+d{}_2\kappa_{(0,0,2,0,0,0,0,0,0)}^{ij|m+\ell}+{}_2\kappa_{(0,2,0,0,0,0,0,0,0)}^{ij|m+\ell}+{}_2\kappa_{(1,0,0,0,0,0,0,0,0)}^{ij|m+\ell}\right],\\
(R_{ij,m+\ell}^{-1})_{2,2}&=0,
}
as in \eqref{EqRM}.  Here $r$ is the rank of the Lorentz group, and we have again applied the $\Gamma$-matrix algebra to simplify the rotation matrix.

We next examine $\vev{SSFF}$ conformal blocks.  Now, there are four different tensor structures given by
\eqna{
b=1:\qquad&(\FCF{b}{k}{l}{,m+\ell}{3}{4})_{\{cC\}\{dD\}\{e''E''\}}=(C_\Gamma^{-1})_{cd}[(\A_{34}\cdot\bar{\bar{\eta}}_2)_{E''}]^\ell\\
&\qquad\to(\tCF{b}{k}{l}{,m+\ell}{3}{4})_{\{cC\}\{dD\}\{e''E''\}\{F''\}}=(C_\Gamma^{-1})_{cd}(\A_{34E''F''})^\ell,\\
b=2:\qquad&(\FCF{b}{k}{l}{,m+\ell}{3}{4})_{\{cC\}\{dD\}\{e''E''\}}=(\bar{\bar{\eta}}_2\cdot\Gamma_{34}C_\Gamma^{-1})_{cd}[(\A_{34}\cdot\bar{\bar{\eta}}_2)_{E''}]^\ell\\
&\qquad\to(\tCF{b}{k}{l}{,m+\ell}{3}{4})_{\{cC\}\{dD\}\{e''E''\}\{F''\}}=(\Gamma_{34F''}C_\Gamma^{-1})_{cd}(\A_{34E''F''})^\ell,\\
b=3:\qquad&(\FCF{b}{k}{l}{,m+\ell}{3}{4})_{\{cC\}\{dD\}\{e''E''\}}=(\Gamma_{34E''_1}C_\Gamma^{-1})_{cd}[(\A_{34}\cdot\bar{\bar{\eta}}_2)_{E''}]^{\ell-1}\\
&\qquad\to(\tCF{b}{k}{l}{,m+\ell}{3}{4})_{\{cC\}\{dD\}\{e''E''\}\{F''\}}=(\Gamma_{34E''_1}C_\Gamma^{-1})_{cd}(\A_{34E''F''})^{\ell-1}\\
b=4:\qquad&(\FCF{b}{k}{l}{,m+\ell}{3}{4})_{\{cC\}\{dD\}\{e''E''\}}=(\bar{\bar{\eta}}_2\cdot\Gamma_{34}\Gamma_{34E''_1}C_\Gamma^{-1})_{cd}[(\A_{34}\cdot\bar{\bar{\eta}}_2)_{E''}]^{\ell-1}\\
&\qquad\to(\tCF{b}{k}{l}{,m+\ell}{3}{4})_{\{cC\}\{dD\}\{e''E''\}\{F''\}}=(\Gamma_{34F''}\Gamma_{34E''_1}C_\Gamma^{-1})_{cd}(\A_{34E''F''})^{\ell-1},
}
so that
\eqna{
b=1:\qquad&n_b=\ell,\qquad i_b=0,\qquad(\tCF{b}{k}{l}{m}{3}{4})_{cd}=(C_\Gamma^{-1})_{cd},\\
b=2:\qquad&n_b=\ell+1,\qquad i_b=0,\qquad(\tCF{b}{k}{l}{m}{3}{4})_{cdF''}=(\Gamma_{34F''}C_\Gamma^{-1})_{cd},\\
b=3:\qquad&n_b=\ell-1,\qquad i_b=1,\qquad(\tCF{b}{k}{l}{,m+1}{3}{4})_{cdE''_1}=(\Gamma_{34E''_1}C_\Gamma^{-1})_{cd},\\
b=4:\qquad&n_b=\ell,\qquad i_b=1,\qquad(\tCF{b}{k}{l}{,m+1}{3}{4})_{cdE''_1F''}=(\Gamma_{34F''}\Gamma_{34E''_1}C_\Gamma^{-1})_{cd}.
}
From the form of the first two tensor structures, it is apparent that no indices need to be extracted from the projection operators.  Meanwhile, it is necessary to extract one $E''$-index for the last two tensor structures.  Diagrammatically, we thus have
\eqna{
b\in\{1,2\}:\qquad&\hat{\mathcal{P}}_{13|d}^{\ell\boldsymbol{e}_1}=\Diag{1}{0}{0}{0}{0}{0}{0}{0},\\
b\in\{3,4\}:\qquad&\hat{\mathcal{P}}_{13|d}^{\ell\boldsymbol{e}_1}=\Diag{1}{0}{0}{0}{0}{0}{0}{1}+\Diag{1}{0}{0}{0}{0}{0}{1}{0},
}
which give the conformal blocks
\eqna{
b\in\{1,2\}:\qquad&\mathscr{G}_{(1|b]}^{ij|m+\ell|kl}=\frac{\ell!}{(d/2-1)_\ell}\left(C_\ell^{(d/2-1)}(X)\right)_{s_{(1|b)}^1},\\
b\in\{3,4\}:\qquad&\mathscr{G}_{(1|b]}^{ij|m+\ell|kl}=-\frac{(\ell-1)!}{(d/2)_{\ell-1}}\left(C_{\ell-1}^{(d/2)}(X)\right)_{s_{(1|b)}^1}+\frac{(\ell-1)!}{(d/2)_{\ell-1}}\left(C_{\ell-2}^{(d/2)}(X)\right)_{s_{(1|b)}^2},
}
in terms of Gegenbauer polynomials \eqref{EqCB4}.  From the partitions associated with the diagrams, the substitutions \eqref{EqS} for each block are easily found to be
\eqna{
s_{(1|1)}^1:\alpha_2^{s_2}\alpha_3^{s_3}\alpha_4^{s_4}x_3^{r_3}x_4^{r_4}&\to(C_\Gamma^{-1})_{cd}G_{(0,0,0,0,0)}^{ij|m+\ell|kl},\\
s_{(1|2)}^1:\alpha_2^{s_2}\alpha_3^{s_3}\alpha_4^{s_4}x_3^{r_3}x_4^{r_4}&\to(\Gamma_{34F''}C_\Gamma^{-1})_{cd}\left(G_{(0,0,2,1,-1)}^{ij|m+\ell|kl}\right)^{F''},\\
s_{(1|3)}^1:\alpha_2^{s_2}\alpha_3^{s_3}\alpha_4^{s_4}x_3^{r_3}x_4^{r_4}&\to(\bar{\eta}_2\cdot\mathcal{S}\cdot\Gamma_{34}C_\Gamma^{-1})_{cd}G_{(-1,0,-1,0,0)}^{ij|m+\ell|kl},\\
s_{(1|3)}^2:\alpha_2^{s_2}\alpha_3^{s_3}\alpha_4^{s_4}x_3^{r_3}x_4^{r_4}&\to(\Gamma_{34E''}C_\Gamma^{-1})_{cd}\left(G_{(0,0,2,1,-1)}^{ij|m+\ell|kl}\right)^{E''},\\
s_{(1|4)}^1:\alpha_2^{s_2}\alpha_3^{s_3}\alpha_4^{s_4}x_3^{r_3}x_4^{r_4}&\to(\Gamma_{34F''}\bar{\eta}_2\cdot\mathcal{S}\cdot\Gamma_{34}C_\Gamma^{-1})_{cd}\left(G_{(-1,0,1,1,-1)}^{ij|m+\ell|kl}\right)^{F''},\\
s_{(1|4)}^2:\alpha_2^{s_2}\alpha_3^{s_3}\alpha_4^{s_4}x_3^{r_3}x_4^{r_4}&\to(\Gamma_{34F''}\Gamma_{34E''}C_\Gamma^{-1})_{cd}\left(G_{(0,0,4,2,-2)}^{ij|m+\ell|kl}\right)^{E''F''}=-2(C_\Gamma^{-1})_{cd}G_{(0,0,0,0,0)}^{ij|m+\ell|kl},
}
where we simplified whenever possible.

As mentioned above, the even-dimensional case may be straightforwardly derived from the above results.  Indeed, a comparison of \eqref{EqPerple1odd} and \eqref{EqPerple1even} shows little difference between the projection operators in odd and even dimensions.  Hence, for the even-dimensional case, the form of the conformal blocks in terms of Gegenbauer polynomials is equivalent to the odd-dimensional one.  The same statement does not apply to the tensor structures, however.  Since there are two different spinor representations in even dimensions, namely $F$ and $\tilde{F}$, not all tensor structures exist for each of the four possible pairs of fermions $FF$, $F\tilde{F}$, $\tilde{F}F$ and $\tilde{F}\tilde{F}$.  An inspection of the tensor structures shows that only half of these are possible for a given fermion pair (depending on the rank and the exchanged fermion, either the half with an even number of $\Gamma$-matrices, or the half with an odd number of $\Gamma$-matrices, but not both).  In this way, conformal blocks for fermions in even dimensions can be seen as the appropriate half of the conformal blocks for fermions in odd dimensions.


\section{Conclusion}\label{SecConc}

In this work, we have established a set of highly efficient rules for determining all possible four-point conformal blocks in terms of fundamental group theoretic quantities, namely the projection operators of the external and exchanged quasi-primary operators.  Once known, these projection operators imply two sets of tensor structures, one for the left and right OPE at the origin of the conformal blocks.  With the knowledge of the projection operators and the tensor structures in hand, the rules introduced here allow us to seamlessly generate any conformal block of interest.

For infinite towers of exchanged quasi-primary operators in irreducible representations $\boldsymbol{N}_m+\ell\boldsymbol{e}_1$, the results summarized in Section \ref{SecSum} lead to simple conformal blocks expressed in terms of linear combinations of Gegenbauer polynomials in a specific variable $X$, coupled with associated substitutions.  The attractive simplicity of the blocks has its origin in the embedding space OPE formalism applied in the mixed basis of tensor structures.

Although the blocks feature the simplest available form in the mixed basis, it is in our best interest to derive their corresponding form in the pure three-point basis, given our hope of ultimately implementing the conformal bootstrap program.  Obtaining the conformal blocks in a pure basis, either the OPE or the three-point one, necessitates the computation of rotation matrices.  These are obtained from the three-point correlation functions and are summarized in Section \ref{SecSum}.

In this work, we also introduce a convenient diagrammatic notation in order to easily determine the appropriate linear combination of Gegenbauer polynomials appearing in a specific conformal block.  The rules are quite straightforward to apply.  To illustrate their utility in action, we have applied them explicitly across a range of examples involving quasi-primary operators in scalar, vector and fermion irreducible representations.

Our results make it transparent that all one requires in order to compute conformal blocks are the projection operators for the infinite towers of exchanged irreducible representations.  We have conveniently expressed these group theoretic objects in terms of shifted projection operators for $\ell\boldsymbol{e}_1$, \textit{i.e.}\ projection operators with an unnatural spacetime dimension.  As a consequence, these shifted projection operators are not traceless.  Nevertheless, the original projection operators for the infinite tower of irreducible representations $\boldsymbol{N}_m+\ell\boldsymbol{e}_1$ are most directly useful in the computation of conformal blocks when cast in terms of these shifted projection operators.  Moreover, the shifted projection operators satisfy several interesting properties that will be described in an upcoming work.

A salient feature of the form of the blocks presented in this work is the ubiquitous presence of the Gegenbauer polynomials.  This aspect is not surprising, as we expect Gegenbauer polynomials to appear for any tower of conformal blocks with exchanged quasi-primary operators in $\boldsymbol{N}_m+\ell\boldsymbol{e}_1$.  The existence of such a form raises the question: Is there another closed form expression that we may write down, which may effectively enable us to remove the multiple finite sums arising here and replace them by a smaller number of sums?  Motivated by the well known closed form expressions for the $\ell\boldsymbol{e}_1$ exchange blocks in $d=2$ and $d=4$ spacetime dimensions in scalar four-point functions in terms of specific linear combinations of products of hypergeometric functions, we may hope to determine a suitable generalization of such expressions for conformal blocks for nontrivial Lorentz representations in arbitrary spacetime dimensions.

With the rules laid out in this paper, the next logical step is to study correlation functions of the energy-momentum tensors.  Indeed, the energy-momentum tensor is the only nontrivial local quasi-primary operator present in all CFTs.  However, even when all the appropriate projection operators are known, it is still necessary to understand conserved currents within the context of the present formalism.  The analysis of conserved currents in the embedding space OPE formalism will be the subject of a forthcoming publication.

\ack{
The work of JFF and VP is supported by NSERC and FRQNT.  The work of WJM is supported by the Chinese Scholarship Council and in part by NSERC and FRQNT.
}


\setcounter{section}{0}
\renewcommand{\thesection}{\Alph{section}}

\section{Projection Operators}\label{SecProj}

In this appendix, we list the projection operators needed to compute the infinite towers of conformal blocks for the examples presented in Section \ref{SecEx}.  The projection operators are first expressed in terms of the usual $\ell$-dependent sums over traces with coefficients related to
\eqn{a_i(d,\ell)=\frac{(-\ell)_{2i}}{2^{2i}i!(-\ell+2-d/2)_i},}
as in the shifted projection operators \eqref{EqPShift}.  They are then re-expressed in terms of finite $\ell$-independent sums in these same shifted projection operators \eqref{EqPShift}, as in \eqref{EqPExp}.


\subsection{Projection Operator in the \texorpdfstring{$\ell\boldsymbol{e}_1$}{le1} Irreducible Representation}

The projection operator in the $\ell\boldsymbol{e}_1$ irreducible representation is well known.  It is given by
\eqn{(\hat{\mathcal{P}}^{\ell\boldsymbol{e}_1})_{\mu_\ell\cdots\mu_1}^{\phantom{\mu_\ell\cdots\mu_1}\mu'_1\cdots\mu'_\ell}=\sum_{i=0}^{\lfloor\ell/2\rfloor}a_i(d,\ell)g_{(\mu_1\mu_2}g^{(\mu'_1\mu'_2}\cdots g_{\mu_{2i-1}\mu_{2i}}g^{\mu'_{2i-1}\mu'_{2i}}g_{\mu_{2i+1}}^{\phantom{\mu_{2i+1}}\mu'_{2i+1}}\cdots g_{\mu_\ell)}^{\phantom{\mu_\ell)}\mu'_\ell)}.}
Since it is already written in terms of the shifted projection operators, for our purposes the $\ell\boldsymbol{e}_1$ projection operator is simply
\eqn{(\hat{\mathcal{P}}^{\ell\boldsymbol{e}_1})_{\mu_\ell\cdots\mu_1}^{\phantom{\mu_\ell\cdots\mu_1}\mu'_1\cdots\mu'_\ell}=(\hat{\mathcal{P}}_d^{\ell\boldsymbol{e}_1})_{\mu^\ell}^{\phantom{\mu^\ell}\mu'^\ell},}[EqPle1]
since the $\mu$-indices (and also the $\mu'$-indices) are symmetrized, with
\eqn{
\begin{array}{|cccc|}\hline
t & (d_t,\ell_t) & \mathscr{A}_t(d,\ell) & \hat{\mathcal{Q}}_{13|t}\\\hline
1 & (0,0) & 1 & 1\\
\hline
\end{array}
}
in the form \eqref{EqPExp}.


\subsection{Projection Operator in the \texorpdfstring{$\boldsymbol{e}_r+\ell\boldsymbol{e}_1$}{er+le1} Irreducible Representation}

For fermionic irreducible representation representations, the projection operators depend on the spacetime dimensions.

In odd dimensions, there is only one fermionic representation, given by $\boldsymbol{e}_r$, and the associated $\boldsymbol{e}_r+\ell\boldsymbol{e}_1$ projection operator is
\eqna{
&(\hat{\mathcal{P}}^{\boldsymbol{e}_r+\ell\boldsymbol{e}_1})_{\alpha\mu_\ell\cdots\mu_1}^{\phantom{\alpha\mu_\ell\cdots\mu_1}\mu'_1\cdots\mu'_\ell\alpha'}=\sum_{i=0}^{\lfloor\ell/2\rfloor}a_i(d+2,\ell)g_{(\mu_1\mu_2}g^{(\mu'_1\mu'_2}\cdots g_{\mu_{2i-1}\mu_{2i}}g^{\mu'_{2i-1}\mu'_{2i}}g_{\mu_{2i+1}}^{\phantom{\mu_{2i+1}}\mu'_{2i+1}}\cdots g_{\mu_\ell)}^{\phantom{\mu_\ell)}\mu'_\ell)}\delta_\alpha^{\phantom{\alpha}\alpha'}\\
&\quad+\sum_{i=0}^{\lfloor(\ell-1)/2\rfloor}\frac{\ell a_i(d+2,\ell-1)}{2(-\ell+1-d/2)}g_{(\mu_1\mu_2}g^{(\mu'_1\mu'_2}\cdots g_{\mu_{2i-1}\mu_{2i}}g^{\mu'_{2i-1}\mu'_{2i}}g_{\mu_{2i+1}}^{\phantom{\mu_{2i+1}}\mu'_{2i+1}}\cdots g_{\mu_{\ell-1}}^{\phantom{\mu_{\ell-1}}\mu'_{\ell-1}}(\gamma_{\mu_\ell)}\gamma^{\mu'_\ell)})_\alpha^{\phantom{\alpha}\alpha'}.
}
This result is obtained by combining allowed objects (among metrics, epsilon tensors and $\gamma$-matrices) in all possible ways consistent with the symmetry properties of the irreducible representation, demanding tracelessness, and enforcing the projection property $\hat{\mathcal{P}}^2=\hat{\mathcal{P}}$.

It can be rewritten in terms of the shifted projection operators as
\eqn{(\hat{\mathcal{P}}^{\boldsymbol{e}_r+\ell\boldsymbol{e}_1})_{\alpha\mu_\ell\cdots\mu_1}^{\phantom{\alpha\mu_\ell\cdots\mu_1}\mu'_1\cdots\mu'_\ell\alpha'}=\delta_\alpha^{\phantom{\alpha}\alpha'}(\hat{\mathcal{P}}_{d+2}^{\ell\boldsymbol{e}_1})_{\mu^\ell}^{\phantom{\mu^\ell}\mu'^\ell}+\frac{\ell}{2(-\ell+1-d/2)}(\gamma_{(\mu}\gamma^{(\mu'})_\alpha^{\phantom{\alpha}\alpha'}(\hat{\mathcal{P}}_{d+2}^{(\ell-1)\boldsymbol{e}_1})_{\mu^{\ell-1})}^{\phantom{\mu^{\ell-1})}\mu'^{\ell-1})}.}[EqPerple1odd]
Here, the fact that the projection operator accompanying $\delta_\alpha^{\phantom{\alpha}\alpha'}$ is shifted implies that the first term is not traceless by itself, and a second term is therefore necessary.  Thus there are two terms, given by
\eqn{
\begin{array}{|cccc|}\hline
t & (d_t,\ell_t) & \mathscr{A}_t(d,\ell) & \hat{\mathcal{Q}}_{13|t}\\\hline
1 & (2,0) & 1 & \delta_\alpha^{\phantom{\alpha}\alpha'}\\
2 & (2,1) & \frac{\ell}{2(-\ell+1-d/2)} & (\gamma_\mu\gamma^{\mu'})_\alpha^{\phantom{\alpha}\alpha'}\\
\hline
\end{array}
}
in the decomposition \eqref{EqPExp} of \eqref{EqPerple1odd}.

In even dimensions, there are two irreducible fermionic representations, namely $\boldsymbol{e}_{r-1}$ and $\boldsymbol{e}_r$.  However, their respective associated projection operators $\boldsymbol{e}_{r-1}+\ell\boldsymbol{e}_1$ and $\boldsymbol{e}_r+\ell\boldsymbol{e}_1$ are straightforwardly obtained from the equivalent projection operator in odd dimensions \eqref{EqPerple1odd}.  Indeed, they are given by
\eqna{
(\hat{\mathcal{P}}^{\boldsymbol{e}_{r-1}+\ell\boldsymbol{e}_1})_{\alpha\mu_\ell\cdots\mu_1}^{\phantom{\alpha\mu_\ell\cdots\mu_1}\mu'_1\cdots\mu'_\ell\alpha'}&=\delta_\alpha^{\phantom{\alpha}\alpha'}(\hat{\mathcal{P}}_{d+2}^{\ell\boldsymbol{e}_1})_{\mu^\ell}^{\phantom{\mu^\ell}\mu'^\ell}+\frac{\ell}{2(-\ell+1-d/2)}(\gamma_{(\mu}\tilde{\gamma}^{(\mu'})_\alpha^{\phantom{\alpha}\alpha'}(\hat{\mathcal{P}}_{d+2}^{(\ell-1)\boldsymbol{e}_1})_{\mu^{\ell-1})}^{\phantom{\mu^{\ell-1})}\mu'^{\ell-1})},\\
(\hat{\mathcal{P}}^{\boldsymbol{e}_r+\ell\boldsymbol{e}_1})_{\tilde{\alpha}\mu_\ell\cdots\mu_1}^{\phantom{\tilde{\alpha}\mu_\ell\cdots\mu_1}\mu'_1\cdots\mu'_\ell\tilde{\alpha}'}&=\delta_{\tilde{\alpha}}^{\phantom{\tilde{\alpha}}\tilde{\alpha}'}(\hat{\mathcal{P}}_{d+2}^{\ell\boldsymbol{e}_1})_{\mu^\ell}^{\phantom{\mu^\ell}\mu'^\ell}+\frac{\ell}{2(-\ell+1-d/2)}(\tilde{\gamma}_{(\mu}\gamma^{(\mu'})_{\tilde{\alpha}}^{\phantom{\tilde{\alpha}}\tilde{\alpha}'}(\hat{\mathcal{P}}_{d+2}^{(\ell-1)\boldsymbol{e}_1})_{\mu^{\ell-1})}^{\phantom{\mu^{\ell-1})}\mu'^{\ell-1})},
}[EqPerple1even]
and therefore have the same expansion according to \eqref{EqPExp} as the one in odd dimensions.


\subsection{Projection Operators in \texorpdfstring{$\boldsymbol{e}_m+\ell\boldsymbol{e}_1$}{em+le1} Irreducible Representations}

For the $m$-index antisymmetric irreducible representations of the type $\boldsymbol{e}_m+\ell\boldsymbol{e}_1$, where in terms of Dynkin indices one has more precisely
\eqn{
\begin{gathered}
\boldsymbol{e}_m\in\{\boldsymbol{e}_2,\boldsymbol{e}_3,\ldots,\boldsymbol{e}_{r-1},2\boldsymbol{e}_r\}\text{ in odd dimensions},\\
\boldsymbol{e}_m\in\{\boldsymbol{e}_2,\boldsymbol{e}_3,\ldots,\boldsymbol{e}_{r-2},\boldsymbol{e}_{r-1}+\boldsymbol{e}_r\}\text{ in even dimensions},
\end{gathered}
}
for $m$ from $2$ to $r$ (represented by $2\boldsymbol{e}_r$ Dynkin indices) in odd dimensions and $m$ from $2$ to $r-1$ (represented by $\boldsymbol{e}_{r-1}+\boldsymbol{e}_r$ Dynkin indices) in even dimensions, we find that the projection operators are
\eqna{
&(\hat{\mathcal{P}}^{\boldsymbol{e}_m+\ell\boldsymbol{e}_1})_{\nu_m\cdots\nu_1\mu_\ell\cdots\mu_1}^{\phantom{\nu_m\cdots\nu_1\mu_\ell\cdots\mu_1}\mu'_1\cdots\mu'_\ell\nu'_1\cdots\nu'_m}\\
&\quad=\sum_{i=0}^{\lfloor\ell/2\rfloor}a_i^mg_{[\nu_1}^{\phantom{[\nu_1}\nu'_1}\cdots g_{\nu_m]}^{\phantom{\nu_m]}\nu'_m}g_{(\mu_1\mu_2}g^{(\mu'_1\mu'_2}\cdots g_{\mu_{2i-1}\mu_{2i}}g^{\mu'_{2i-1}\mu'_{2i}}g_{\mu_{2i+1}}^{\phantom{\mu_{2i+1}}\mu'_{2i+1}}\cdots g_{\mu_\ell)}^{\phantom{\mu_\ell)}\mu'_\ell)}\\
&\quad\phantom{=}+\sum_{i=0}^{\lfloor(\ell-1)/2\rfloor}b_i^mg_{[\nu_1}^{\phantom{[\nu_1}[\nu'_1}\cdots g_{\nu_{m-1}}^{\phantom{\nu_{m-1}}\nu'_{m-1}}g_{\nu_m]}^{\phantom{\nu_m]}(\mu'_1}g_{(\mu_1}^{\phantom{(\mu_1}\nu'_m]}g_{\mu_2\mu_3}g^{\mu'_2\mu'_3}\cdots g_{\mu_{2i}\mu_{2i+1}}g^{\mu'_{2i}\mu'_{2i+1}}g_{\mu_{2i+2}}^{\phantom{\mu_{2i+2}}\mu'_{2i+2}}\cdots g_{\mu_\ell)}^{\phantom{\mu_\ell)}\mu'_\ell)}\\
&\quad\phantom{=}+\sum_{i=0}^{\lfloor(\ell-1)/2\rfloor}c_i^mg_{[\nu_1}^{\phantom{[\nu_1}[\nu'_1}\cdots g_{\nu_{m-1}}^{\phantom{\nu_{m-1}}\nu'_{m-1}}g_{\nu_m](\mu_1}g^{\nu'_m](\mu'_1}g_{\mu_2\mu_3}g^{\mu'_2\mu'_3}\cdots g_{\mu_{2i}\mu_{2i+1}}g^{\mu'_{2i}\mu'_{2i+1}}g_{\mu_{2i+2}}^{\phantom{\mu_{2i+2}}\mu'_{2i+2}}\cdots g_{\mu_\ell)}^{\phantom{\mu_\ell)}\mu'_\ell)}\\
&\quad\phantom{=}+\sum_{i=0}^{\lfloor(\ell-2)/2\rfloor}d_i^mg_{[\nu_1}^{\phantom{[\nu_1}[\nu'_1}\cdots g_{\nu_{m-2}}^{\phantom{\nu_{m-2}}\nu'_{m-2}}g_{\nu_{m-1}(\mu_1}g^{\nu'_{m-1}(\mu'_1}g_{\nu_m]}^{\phantom{\nu_m]}\mu'_2}g_{\mu_2}^{\phantom{\mu_2}\nu'_m]}\\
&\quad\phantom{=}\times g_{\mu_3\mu_4}g^{\mu'_3\mu'_4}\cdots g_{\mu_{2i+1}\mu_{2i+2}}g^{\mu'_{2i+1}\mu'_{2i+2}}g_{\mu_{2i+3}}^{\phantom{\mu_{2i+3}}\mu'_{2i+3}}\cdots g_{\mu_\ell)}^{\phantom{\mu_\ell)}\mu'_\ell)}\\
&\qquad\phantom{=}+\sum_{i=0}^{\lfloor(\ell-2)/2\rfloor}e_i^mg_{[\nu_1}^{\phantom{[\nu_1}[\nu'_1}\cdots g_{\nu_{m-2}}^{\phantom{\nu_{m-2}}\nu'_{m-2}}\left(g_{\nu_{m-1}(\mu_1}g_{\nu_m]}^{\phantom{\nu_m]}\nu'_{m-1}}g_{\mu_2}^{\phantom{\mu_2}\nu'_m]}g^{(\mu'_1\mu'_2}+g^{\nu'_{m-1}(\mu'_1}g_{\nu_{m-1}}^{\phantom{\nu_{m-1}}\nu'_m]}g_{\nu_m]}^{\phantom{\nu_m]}\mu'_2}g_{(\mu_1\mu_2}\right)\\
&\quad\phantom{=}\times g_{\mu_3\mu_4}g^{\mu'_3\mu'_4}\cdots g_{\mu_{2i+1}\mu_{2i+2}}g^{\mu'_{2i+1}\mu'_{2i+2}}g_{\mu_{2i+3}}^{\phantom{\mu_{2i+3}}\mu'_{2i+3}}\cdots g_{\mu_\ell)}^{\phantom{\mu_\ell)}\mu'_\ell)},
}
with
\eqn{
\begin{gathered}
a_i^m=\frac{m}{\ell+m}a_i(d+2,\ell),\qquad b_i^m=(\ell-2i)a_i^m,\\
c_i^m=\frac{(\ell-2i)[(d+\ell-m)i+(m+1)d/2+m(\ell-1)]}{(-\ell+1-d/2+i)(d+\ell-m)}a_i^m,\\
d_i^m=-\frac{4(m-1)(i+1)(-\ell-d/2)}{d+\ell-m}a_{i+1}^m,\qquad e_i^m=-2(i+1)a_{i+1}^m.
\end{gathered}
}
Here, the $\mu^\ell$ indices are the $\ell\boldsymbol{e}_1$ symmetrized indices while the $\nu^m$ indices are the $\boldsymbol{e}_m$ antisymmetrized indices.\footnote{The case $\boldsymbol{e}_m=\boldsymbol{e}_2$ was already found in \cite{Rejon-Barrera:2015bpa}.}

The above projection operators can be recast in terms of the shifted projection operators as
\eqna{
&(\hat{\mathcal{P}}^{\boldsymbol{e}_m+\ell\boldsymbol{e}_1})_{\nu_m\cdots\nu_1\mu_\ell\cdots\mu_1}^{\phantom{\nu_m\cdots\nu_1\mu_\ell\cdots\mu_1}\mu'_1\cdots\mu'_\ell\nu_1'\cdots\nu_m'}\\
&\qquad=\frac{m}{\ell+m}g_{[\nu_1}^{\phantom{[\nu_1}\nu'_1}\cdots g_{\nu_m]}^{\phantom{\nu_m]}\nu'_m}(\hat{\mathcal{P}}_{d+2}^{\ell\boldsymbol{e}_1})_{\mu^\ell}^{\phantom{\mu^\ell}\mu'^\ell}\\
&\qquad\phantom{=}+\frac{m\ell}{\ell+m}g_{[\nu_1}^{\phantom{[\nu_1}[\nu'_1}\cdots g_{\nu_{m-1}}^{\phantom{\nu_{m-1}}\nu'_{m-1}}g_{\nu_m]}^{\phantom{\nu_m]}(\mu'}g_{(\mu}^{\phantom{(\mu}\nu'_m]}(\hat{\mathcal{P}}_{d+4}^{(\ell-1)\boldsymbol{e}_1})_{\mu^{\ell-1})}^{\phantom{\mu^{\ell-1})}\mu'^{\ell-1})}\\
&\qquad\phantom{=}+\frac{m\ell}{\ell+m}g_{[\nu_1}^{\phantom{[\nu_1}[\nu'_1}\cdots g_{\nu_{m-1}}^{\phantom{\nu_{m-1}}\nu'_{m-1}}g_{\nu_m](\mu}g^{\nu'_m](\mu'}\\
&\qquad\phantom{=}\times\left[(\hat{\mathcal{P}}_{d+4}^{(\ell-1)\boldsymbol{e}_1})_{\mu^{\ell-1})}^{\phantom{\mu^{\ell-1})}\mu'^{\ell-1})}-\frac{(-\ell-d/2)(d+\ell-1)}{(-\ell+1-d/2)(d+\ell-m)}(\hat{\mathcal{P}}_{d+2}^{(\ell-1)\boldsymbol{e}_1})_{\mu^{\ell-1})}^{\phantom{\mu^{\ell-1})}\mu'^{\ell-1})}\right]\\
&\qquad\phantom{=}-\frac{m(m-1)\ell(\ell-1)(-\ell-d/2)}{(\ell+m)(-\ell+1-d/2)(d+\ell-m)}g_{[\nu_1}^{\phantom{[\nu_1}[\nu'_1}\cdots g_{\nu_{m-2}}^{\phantom{\nu_{m-2}}\nu'_{m-2}}\\
&\qquad\phantom{=}\times g_{\nu_{m-1}(\mu}g^{\nu'_{m-1}(\mu'}g_{\nu_m]}^{\phantom{\nu_m]}\mu'}g_\mu^{\phantom{\mu}\nu'_m]}(\hat{\mathcal{P}}_{d+4}^{(\ell-2)\boldsymbol{e}_1})_{\mu^{\ell-2})}^{\phantom{\mu^{\ell-2})}\mu'^{\ell-2})}\\
&\qquad\phantom{=}-\frac{m\ell(\ell-1)}{2(\ell+m)(-\ell+1-d/2)}g_{[\nu_1}^{\phantom{[\nu_1}[\nu'_1}\cdots g_{\nu_{m-2}}^{\phantom{\nu_{m-2}}\nu'_{m-2}}\\
&\qquad\phantom{=}\times\left(g_{\nu_{m-1}(\mu}g_{\nu_m]}^{\phantom{\nu_m]}\nu'_{m-1}}g_\mu^{\phantom{\mu}\nu'_m]}g^{(\mu'\mu'}+g^{\nu'_{m-1}(\mu'}g_{\nu_{m-1}}^{\phantom{\nu_{m-1}}\nu'_m]}g_{\nu_m]}^{\phantom{\nu_m]}\mu'}g_{(\mu\mu}\right)(\hat{\mathcal{P}}_{d+4}^{(\ell-2)\boldsymbol{e}_1})_{\mu^{\ell-2})}^{\phantom{\mu^{\ell-2})}\mu'^{\ell-2})},
}[EqPemple1]
which corresponds to
\eqn{
\begin{array}{|cccc|}\hline
t & (d_t,\ell_t) & \mathscr{A}_t(d,\ell) & \hat{\mathcal{Q}}_{13|t}\\\hline
1 & (2,0) & \frac{m}{\ell+m} & g_{[\nu_1}^{\phantom{[\nu_1}\nu'_1}\cdots g_{\nu_m]}^{\phantom{\nu_m]}\nu'_m}\\
2 & (4,1) & \frac{m\ell}{\ell+m} & g_{[\nu_1}^{\phantom{[\nu_1}[\nu'_1}\cdots g_{\nu_{m-1}}^{\phantom{\nu_{m-1}}\nu'_{m-1}}g_{\nu_m]}^{\phantom{\nu_m]}\mu'}g_{\mu}^{\phantom{\mu}\nu'_m]}\\
3 & (4,1) & \frac{m\ell}{\ell+m} & g_{[\nu_1}^{\phantom{[\nu_1}[\nu'_1}\cdots g_{\nu_{m-1}}^{\phantom{\nu_{m-1}}\nu'_{m-1}}g_{\nu_m]\mu}g^{\nu'_m]\mu'}\\
4 & (2,1) & -\frac{m\ell(-\ell-d/2)(d+\ell-1)}{(\ell+m)(-\ell+1-d/2)(d+\ell-m)} & g_{[\nu_1}^{\phantom{[\nu_1}[\nu'_1}\cdots g_{\nu_{m-1}}^{\phantom{\nu_{m-1}}\nu'_{m-1}}g_{\nu_m]\mu}g^{\nu'_m]\mu'}\\
5 & (4,2) & -\frac{m(m-1)\ell(\ell-1)(-\ell-d/2)}{(\ell+m)(-\ell+1-d/2)(d+\ell-m)} & g_{[\nu_1}^{\phantom{[\nu_1}[\nu'_1}\cdots g_{\nu_{m-2}}^{\phantom{\nu_{m-2}}\nu'_{m-2}}g_{\nu_{m-1}\mu}g^{\nu'_{m-1}\mu'}g_{\nu_m]}^{\phantom{\nu_m]}\mu'}g_\mu^{\phantom{\mu}\nu'_m]}\\
6 & (4,2) & -\frac{m\ell(\ell-1)}{2(\ell+m)(-\ell+1-d/2)} & \substack{g_{[\nu_1}^{\phantom{[\nu_1}[\nu'_1}\cdots g_{\nu_{m-2}}^{\phantom{\nu_{m-2}}\nu'_{m-2}}\\\times\left(g_{\nu_{m-1}\mu}g_{\nu_m]}^{\phantom{\nu_m]}\nu'_{m-1}}g_\mu^{\phantom{\mu}\nu'_m]}g^{\mu'\mu'}+g^{\nu'_{m-1}\mu'}g_{\nu_{m-1}}^{\phantom{\nu_{m-1}}\nu'_m]}g_{\nu_m]}^{\phantom{\nu_m]}\mu'}g_{\mu\mu}\right)}\\
\hline
\end{array}
}
in the decomposition \eqref{EqPExp}.

The projection operators \eqref{EqPemple1} can be obtained directly by combining the allowed objects in the most general way satisfying the symmetry properties of the irreducible representations, leading to the five terms above (symmetry under the exchange of the primed and unprimed indices implies two contributions to the last term).  Then, the projection property $\hat{\mathcal{P}}^{\boldsymbol{N}}\cdot\hat{\mathcal{P}}^{\boldsymbol{N}'}=\delta_{\boldsymbol{N}'\boldsymbol{N}}\hat{\mathcal{P}}^{\boldsymbol{N}}$ for mixed symmetry and tracelessness relates $\mathscr{A}_2$ to $\mathscr{A}_1$.  Using again the projection property, tracelessness, and the Fock condition fixes $\mathscr{A}_1$.  Finally, the tracelessness condition determines the remaining $\mathscr{A}_{t=3,4,5,6}$.


\newpage
\bibliography{DiagrammaticRules}

\end{document}